\documentclass[11pt]{article}
\usepackage{amsfonts,amsmath,amsthm}
\usepackage{algorithm2e,framed,algorithmic}
\RequirePackage{graphicx}
\RequirePackage{subfigure}

\newcommand{\mbf}{\mathbf}
\newcommand{\sbf}{\boldsymbol}
\newtheorem{definition}{Definition}
\newtheorem{assumption}{Assumption}
\newtheorem{lemma}{Lemma}

\newcommand{\Diagnl}[1]{\mbox{}{Diag}\left\{#1\right\}}
\newcommand{\Trace }[1]{\mbox{}{\bf{Tr}}\left(#1\right)}
\newcommand{\Probab}[1]{\mbox{}{\bf{Pr}}\left[#1\right]}
\newcommand{\Expect}[1]{\mbox{}{\bf{E}}\left[#1\right]}
\newcommand{\CExpect}[1]{\mbox{}{\bf{E}_{w}}\left[#1\right]}
\newcommand{\Varnce}[1]{\mbox{}{\bf{Var}}\left[#1\right]}
\newcommand{\CVarnce}[1]{\mbox{}{\bf{Var}_{w}}\left[#1\right]}
\newlength{\defbaselineskip}
\begin{document}

\title{A Statistical Perspective on Algorithmic Leveraging}

\author{
Ping Ma
\thanks{
Department of Statistics,
University of Illinois at Urbana-Champaign,
Champaign, IL 61820.
Email: \texttt{pingma@illinois.edu}.
}
\and
Michael W.~Mahoney
\thanks{
Department of Mathematics,
Stanford University,
Stanford, CA 94305.
Email: \texttt{mmahoney@cs.stanford.edu}.
}
\and
Bin Yu
\thanks{
Department of Statistics,
University of California at Berkeley,
Berkeley, CA 94720.
Email: \texttt{binyu@stat.berkeley.edu}.
}
}

\date{}
\maketitle

\begin{abstract}
\noindent
One popular method for dealing with large-scale data sets is sampling.
For example, by using the empirical statistical leverage scores as an 
importance sampling distribution, the method of \emph{algorithmic leveraging} 
samples and rescales rows/columns of data matrices to reduce the data size 
before performing computations on the subproblem.
This method has been successful in improving computational efficiency of 
algorithms for matrix problems such as least-squares approximation, least 
absolute deviations approximation, and low-rank matrix approximation.
Existing work has focused on algorithmic issues such as worst-case running 
times and numerical issues associated with providing high-quality 
implementations, but none of it addresses statistical aspects of this 
method.

In this paper, we provide a simple yet effective framework to evaluate the statistical 
properties of algorithmic leveraging in the context of estimating parameters 
in a linear regression model with a fixed number of predictors.
In particular, for several versions of leverage-based sampling, we derive 
results for the bias and variance, both conditional and unconditional on 
the observed data.
We show that from the statistical perspective of bias and variance, neither 
leverage-based sampling nor uniform sampling dominates the other.
This result is particularly striking, given the well-known result that, 
from the algorithmic perspective of worst-case 
analysis, leverage-based sampling provides uniformly superior worst-case 
algorithmic results, when compared with uniform sampling.

Based on these theoretical results, we propose and analyze two new
leveraging algorithms: one constructs a smaller least-squares problem 
with ``shrinked'' leverage scores (SLEV), and the other solves a smaller 
and unweighted (or biased) least-squares problem (LEVUNW).
A detailed empirical evaluation of existing leverage-based methods as well 
as these two new methods is carried out on both synthetic and real data sets.
The empirical results indicate that our theory is a good predictor of
practical performance of existing and new leverage-based algorithms and that the new algorithms achieve improved performance.
For example, with the same computation reduction as in the original 
algorithmic leveraging approach, our proposed SLEV typically leads to 
improved biases and variances both unconditionally and conditionally (on the 
observed data), and our proposed LEVUNW typically yields improved 
unconditional biases and variances.
\end{abstract}

\newpage
\section{Introduction}
\label{sxn:intro}

One popular method for dealing with large-scale data sets is sampling.
In this approach, one first chooses a small portion of the full data, and
then one uses this sample as a surrogate to carry out computations of 
interest for the full data.
For example, one might randomly sample a small number of rows from an input
matrix and use those rows to construct a low-rank approximation to the
original matrix, or one might randomly sample a small number of constraints
or variables in a regression problem and then perform a regression 
computation on the subproblem thereby defined.
For many problems, it is very easy to construct ``worst-case'' input for
which \emph{uniform} random sampling will perform very poorly.
Motivated by this, there has been a great deal of work on developing
algorithms for matrix-based machine learning and data analysis problems
that construct the random sample in a \emph{nonuniform} data-dependent
fashion~\cite{Mah-mat-rev_BOOK}.

Of particular interest here is when that data-dependent sampling process
selects rows or columns from the input matrix according to a probability 
distribution that depends on the empirical statistical leverage scores of 
that matrix.
This recently-developed approach of \emph{algorithmic leveraging} has been 
applied to matrix-based problems that are of interest in large-scale data 
analysis, e.g., least-squares approximation~\cite{DMM06,DMMS07_FastL2_NM10}, 
least absolute deviations regression~\cite{CDMMMW12_TR,MM12_TR}, and 
low-rank matrix approximation~\cite{CUR_PNAS,CW12sparse_TR}. 
Typically, the leverage scores are computed
approximately~\cite{DMMW12_JMLR,CDMMMW12_TR}, or otherwise a random 
projection~\cite{AC10,CDMMMW12_TR} is used to precondition by approximately 
uniformizing them~\cite{DMMS07_FastL2_NM10,AMT10,MSM11_TR}.
A detailed discussion of this approach can be found in the recent review
monograph on randomized algorithms for matrices and matrix-based data
problems~\cite{Mah-mat-rev_BOOK}.

This algorithmic leveraging paradigm has already yielded impressive 
algorithmic benefits: by preconditioning with a high-quality
numerical implementation of a Hadamard-based random projection, the
Blendenpik code of~\cite{AMT10} ``beats \textsc{Lapack}'s%
\footnote{\textsc{Lapack} (short for Linear Algebra PACKage) is a
high-quality and widely-used software library of numerical routines for
solving a wide range of numerical linear algebra problems.}
direct dense least-squares solver by a large margin on
essentially any dense tall matrix;'' the LSRN algorithm of \cite{MSM11_TR}
preconditions with a high-quality numerical implementation of a normal
random projection in order to solve large over-constrained least-squares
problems on clusters with high communication cost, e.g., on Amazon
Elastic Cloud Compute clusters; the solution to the $\ell_1$ regression or 
least absolute deviations problem as well as to quantile regression problems 
can be approximated for problems with 
billions of constraints~\cite{CDMMMW12_TR,YMM13_TR}; 
and CUR-based low-rank matrix approximations~\cite{CUR_PNAS} have been used 
for structure extraction in DNA SNP matrices of size thousands of 
individuals by hundreds of thousands of SNPs~\cite{Paschou07b,Paschou10b}.
In spite of these impressive \emph{algorithmic} results, none of this recent 
work on leveraging or leverage-based sampling addresses \emph{statistical} 
aspects of this approach.
This is in spite of the central role of statistical leverage, a traditional 
concept from regression diagnostics~\cite{HW78,CH86,VW81}.

In this paper, we bridge that gap by providing the first statistical
analysis of the algorithmic leveraging paradigm.
We do so in the context of parameter estimation in fitting linear regression
models for large-scale data---where, by ``large-scale,'' we mean that the
data define a high-dimensional problem in terms of sample size $n$, as
opposed to the dimension $p$ of the parameter space.
Although $n\gg p$ is the classical regime in theoretical statistics, it is a 
relatively new phenomenon that in practice we routinely see a sample size $n$ 
in the hundreds of thousands or millions or more. 
This is a size regime where sampling methods such as algorithmic leveraging 
are indispensable to meet computational constraints. 

Our main theoretical contribution is to provide an analytic framework for 
evaluating the statistical properties of algorithmic leveraging.
This involves performing a Taylor series analysis around the ordinary 
least-squares solution to approximate the subsampling estimators as linear 
combinations of random sampling matrices.
Within this framework, we consider biases and variances, both conditioned as 
well as not conditioned on the data, for several versions of the basic 
algorithmic leveraging procedure. 
We show that both leverage-based sampling and uniform sampling are unbiased
to leading order;
and that while leverage-based sampling improves the ``size-scale'' of the 
variance, relative to uniform sampling, the presence of very small leverage 
scores can inflate the variance considerably.
It is well-known that, from the algorithmic perspective of worst-case 
analysis, leverage-based sampling provides uniformly superior worst-case 
algorithmic results, when compared with uniform sampling.
However, our statistical analysis here reveals that from the statistical 
perspective of bias and variance, neither leverage-based sampling nor uniform 
sampling dominates the~other.

Based on these theoretical results, we propose and analyze two new
leveraging algorithms designed to improve upon vanilla leveraging
and uniform sampling algorithms in terms of bias and variance.
The first of these (denoted SLEV below) involves increasing the probability 
of low-leverage samples, and thus it also has the effect of ``shrinking'' 
the effect of large leverage scores.
The second of these (denoted LEVUNW below) constructs an unweighted version 
of the leverage-subsampled problem; and thus for a given data set it 
involves solving a biased subproblem.
In both cases, we obtain the algorithmic benefits of leverage-based 
sampling, while achieving improved statistical performance.

Our main empirical contribution is to provide a detailed evaluation of the 
statistical properties of these algorithmic leveraging estimators on both 
synthetic and real data sets.
These empirical results indicate that our theory is a good predictor of 
practical performance for both existing algorithms and
our two new leveraging algorithms as well as that our two new algorithms lead to improved performance.
In addition, we show that using shrinked leverage scores typically leads to 
improved conditional and unconditional biases and variances; and that 
solving a biased subproblem typically yields improved unconditional biases 
and variances. 
By using a recently-developed algorithm of~\cite{DMMW12_JMLR} to compute fast approximations to 
the statistical leverage scores, we also demonstrate a regime for large data 
where our shrinked leveraging procedure is better algorithmically, in the 
sense of computing an answer more quickly than the usual black-box 
least-squares solver, as well as statistically, in the sense of having 
smaller mean squared error than na\"{i}ve uniform sampling.
Depending on whether one is interested in results unconditional on the data
(which is more traditional from a statistical perspective) or conditional on 
the data (which is more natural from an algorithmic perspective), we 
recommend the use of SLEV or LEVUNW, respectively, in the future.

The remainder of this article is organized as follows.
We will start in Section~\ref{sxn:background} with a brief review of linear
models, the algorithmic leveraging approach, and related work.
Then, in Section~\ref{sxn:theory}, we will present our main theoretical
results for bias and variance of leveraging estimators.
This will be followed in Sections~\ref{sxn:empirical} and~\ref{sxn:moreempirical} by a detailed empirical 
evaluation on a wide range of synthetic and several real data sets. 
Then, in Section~\ref{sxn:conc}, we will conclude with a brief discussion of
our results in a broader context.
Appendix~\ref{sxn:interlude} will describe our results from the perspective 
of asymptotic relative efficiency and will consider several toy data sets 
that illustrate various aspects of algorithmic leveraging; and 
Appendix~\ref{sxn:app-proofs} will provide the proofs of our main 
theoretical~results.

\section{Background, Notation, and Related Work}
\label{sxn:background}

In this section, we will provide a brief review of relevant background,
including our notation for linear models, an overview of the algorithmic
leveraging approach, and a review of related work in statistics and computer 
science.

\subsection{Least-squares and Linear Models}
\label{sxn:background-ls}

We start with relevant background and notation.
Given an $n \times p$ matrix $X$ and an $n$-dimensional vector $\sbf{y}$,
the least-squares (LS) problem is to~solve
\begin{equation}
\label{eqn:ls}
 \text{argmin}_{\beta \in \mathbb{R}^p}||\sbf{y}-X\sbf{\beta}||^2  ,
\end{equation}
where $||\cdot||$ represents the Euclidean norm on $\mathbb{R}^n$.
Of interest is both a vector exactly or approximately optimizing
Problem~(\ref{eqn:ls}), as well as the value of the objective function at
the optimum.
Using one of several related methods~\cite{GVL96}, this LS problem can be
solved exactly in $O(np^2)$ time (but, as we will discuss in
Section~\ref{sxn:background-al}, it can be solved approximately in $o(np^2)$ 
time% 
\footnote{Recall that, formally, $f(n) = o(g(n))$ as $n \rightarrow \infty$ 
means that for every positive constant $\epsilon$ there exists a constant 
$N$ such that $|f(n)| \le \epsilon|g(n)|$, for all $n \ge N$.  Informally, 
this means that $f(n)$ grows more slowly than $g(n)$.  Thus, if the running 
time of an algorithm is $o(np^2)$ time, then it is asymptotically faster 
than any (arbitrarily small) constant times $np^2$.}%
).
For example, LS can be solved using the singular value decomposition (SVD):
the so-called \emph{thin SVD} of $X$ can be written as $ X=U \Lambda V^{T} $,
where $U$ is an $n \times p$ orthogonal matrix whose columns contain the
left singular vectors of $X$, $V$ is an $p \times p$ orthogonal matrix
whose columns contain the right singular vectors of $X$, and the 
$p \times p$ matrix $\Lambda =\Diagnl{\lambda_i}$, where $\lambda_i$, 
$ i=1,\ldots, p$, are the singular values of $X$.
In this case, $\hat{\sbf{\beta}}_{ols}=V\Lambda^{-1}U^{T}\sbf{y}$.

We consider the use of LS for parameter estimation in a
Gaussian linear regression model.
Consider the model
\begin{equation}
\label{eqn:reg}
\sbf{y}=X\sbf{\beta}_0+\sbf{\epsilon} ,
\end{equation}
where $\sbf{y}$ is an $n \times 1$ response vector, $X$ is an $n \times p$
\emph{fixed} predictor or design matrix, $\sbf{\beta}_0$ is a $p \times 1$
coefficient vector, and the noise vector
$\sbf{\epsilon} \sim N(0, \sigma^2I)$.
In this case, the unknown coefficient $\sbf{\beta}_0$ can be estimated via maximum-likelihood estimation as
\begin{equation}
\label{eqn:lsq}
\hat{\sbf{\beta}}_{ols}= \text{argmin}_{\sbf{\beta}}||\sbf{y}-X\sbf{\beta}||^2=(X^{T}X)^{-1}X^{T}\sbf{y}  ,
\end{equation}
in which case the predicted response vector is $\hat{\sbf{y}} = H \sbf{y}$, 
where $H=X(X^{T}X)^{-1}X^{T}$ is the so-called Hat Matrix, which is of 
interest in classical regression diagnostics~\cite{HW78,CH86,VW81}.
The $i^{th}$ diagonal element of $H$, 
$h_{ii}=\sbf{x}_i^{T}(X^{T}X)^{-1}\sbf{x}_i$, where $\sbf{x}_i^{T}$ is the
$i^{th}$ row of $X$, is the \emph{statistical leverage} of $i^{th}$
observation or sample.
The statistical leverage scores have been used historically to quantify the 
extent to which an observation is an outlier~\cite{HW78,CH86,VW81}, and they
will be important for our main results below.
Since $H$ can alternatively be expressed as $H=UU^{T}$, where $U$ is any 
orthogonal basis for the column space of $X$, e.g., the $Q$ matrix from a 
QR decomposition or the matrix of left singular vectors from the thin SVD, 
the leverage of the $i^{th}$ observation can also be expressed~as
\begin{equation}
\label{eqn:lev-scores}
h_{ii}=\sum_{j=1}^{p}U_{ij}^2=||\sbf{u}_i||^2 ,
\end{equation}
where $\sbf{u}_i^{T}$ is the $i^{th}$ row of $U$.
Using Eqn.~(\ref{eqn:lev-scores}), the exact computation of $h_{ii}$, for 
$i\in[n]$, requires $O(np^2)$ time~\cite{GVL96} (but, as we 
will discuss in Section~\ref{sxn:background-al}, they can be approximated 
in $o(np^2)$ time).

For an estimate $\hat{\sbf{\beta}}$ of $\sbf{\beta}$, the MSE (mean squared 
error) associated with the prediction error is defined to be
\begin{align}
\label{eqn:MSE}
MSE(\hat{\sbf{\beta}}) 
   &= \frac{1}{n}\Expect{(X\sbf{\beta}_0-X\hat{\sbf{\beta}})^{T}(X\sbf{\beta}_0-X\hat{\sbf{\beta}})}\\\notag
   &= \frac{1}{n}\Trace{\Varnce{X\hat{\sbf{\beta}}}}
    + \frac{1}{n}\Expect{(X\hat{\sbf{\beta}}- X{\sbf{\beta}}_0)^T(X\hat{\sbf{\beta}}- X{\sbf{\beta}}_0)}\\\notag
   &= \frac{1}{n}\Trace{\Varnce{X\hat{\sbf{\beta}}}}
    + \frac{1}{n}[\textbf{bias}(X\hat{\sbf{\beta}})]^T[\textbf{bias}(X\hat{\sbf{\beta}})]
\end{align}
where ${\sbf{\beta}}_0$ is the true value of ${\sbf{\beta}}$.
The MSE provides a benchmark to compare the different subsampling
estimators, and we will be interested in both the bias and variance components.

\subsection{Algorithmic Leveraging for Least-squares Approximation}
\label{sxn:background-al}

Here, we will review relevant work on random sampling algorithms for 
computing approximate solutions to the general overconstrained LS 
problem~\cite{DMM06,Mah-mat-rev_BOOK,DMMW12_JMLR}.
These algorithms choose (in general, nonuniformly) a subsample of the data, 
e.g., a small number of rows of $X$ and the corresponding elements of 
$\sbf{y}$, and then they perform (typically weighted) LS on the subsample.
Importantly, these algorithms make no assumptions on the input data $X$ and 
$\sbf{y}$, except that~$n \gg p$.

A prototypical example of this approach is given by the following 
meta-algorithm~\cite{DMM06,Mah-mat-rev_BOOK,DMMW12_JMLR}, which we call 
\texttt{SubsampleLS}, and which takes as input an $n \times p$ matrix $X$, 
where $n \gg p$, a vector $\sbf{y}$, and a probability distribution 
$\{\pi_i\}_{i=1}^{n}$, and which returns as output an approximate solution 
$\tilde{\sbf{\beta}}_{ols}$, which is an estimate of 
$\hat{\sbf{\beta}}_{ols}$ of Eqn.~(\ref{eqn:lsq}).

\begin{itemize}
\item
Randomly sample $r > p$ constraints, i.e., rows of $X$ and the corresponding 
elements of $\sbf{y}$, using $\{\pi_i\}_{i=1}^{n}$ as an importance sampling 
distribution.
\item
Rescale each sampled row/element by $1/(r\pi_{i})$ to form a weighted LS 
subproblem.
\item
Solve the weighted LS subproblem, formally given in Eqn.~(\ref{lsq-sample}) 
below, and then return the solution $\tilde{\sbf{\beta}}_{ols}$.
\end{itemize}

\noindent
It is convenient to describe \texttt{SubsampleLS} in terms of a random 
``sampling matrix'' $S_X^T$ and a random diagonal ``rescaling matrix'' (or ``reweighting matrix'') $D$,
in the following manner.
If we draw $r$ samples (rows or constraints or data points) with 
replacement, then define an $r \times n$ sampling matrix, $S_X^T$, where 
each of the $r$ rows of $S_X^T$ has one non-zero element indicating which 
row of $X$ (and element of $\sbf{y}$) is chosen in a given random trial.
That is, if the $k^{th}$ data unit (or observation) in the original data set 
is chosen in the $i^{th}$ random trial, then the $i^{th}$ row of $S_X^T$ 
equals $\mbf{e}_k$; and thus $S_X^T$ is a random matrix that describes the 
process of sampling \emph{with} replacement.
As an example of applying this sampling matrix, when the sample size $n=6$
and the subsample size $r=3$, then premultiplying~by
$$
S_X^T=\left( \begin{array}{cccccc}
                   0 & 1 & 0 & 0 & 0 & 0 \\
                   0 & 0 & 0 & 1 & 0 & 0 \\
                   0 & 0 & 0 & 1 & 0 & 0 \\
              \end{array}
        \right)
$$
represents choosing the second, fourth, and fourth data points or samples.
The resulting subsample of $r$ data points can be denoted as
$(X^{*},\sbf{y}^{*})$, where $X^{*}_{r\times p}=S_X^{T}X$ and
$\sbf{y}^{*}_{r\times 1}=S_X^{T}\sbf{y}$.
In this case, an $r\times r$ diagonal rescaling matrix $D$ can be defined so
that $i^{th}$ diagonal element of $D$ equals $1/\sqrt{r\pi_k}$ if the 
$k^{th}$ data point is chosen in the $i^{th}$ random trial (meaning, in
particular, that every diagonal element of $D$ equals $\sqrt{n/r}$ for 
uniform sampling).
With this notation, \texttt{SubsampleLS} constructs and solves the 
\emph{weighted LS estimator}:
\begin{equation}
\label{lsq-sample}
\text{argmin}_{\beta \in \mathbb{R}^p}||DS_X^{T}\sbf{y}-DS_X^{T}X\sbf{\beta}||^2  .
\end{equation}

Since \texttt{SubsampleLS} samples constraints and not variables, the 
dimensionality of the vector $\tilde{\sbf{\beta}}_{ols}$ that solves the 
(still overconstrained, but smaller) weighted LS subproblem 
is the same as that of the vector $\hat{\sbf{\beta}}_{ols}$ that solves the 
original LS problem.
The former may thus be taken as an approximation of the latter, where,
of course, the quality of the approximation depends critically on the 
choice of $\{\pi_i\}_{i=1}^{n}$. 
There are several distributions that have been considered 
previously~\cite{DMM06,Mah-mat-rev_BOOK,DMMW12_JMLR}. 

\begin{itemize}
\item
\textbf{Uniform Subsampling.}
Let $\pi_i=1/n$, for all $i\in[n]$, i.e., draw the sample uniformly at random.
\item
\textbf{Leverage-based Subsampling.}
Let $\pi_i = h_{ii}/\sum_{i=1}^{n}h_{ii} = h_{ii}/p$ be the normalized
statistical leverage scores of Eqn.~(\ref{eqn:lev-scores}), i.e., draw the
sample according to an importance sampling distribution that is proportional
to the statistical leverage scores of the data matrix $X$.
\end{itemize}

\noindent
Although Uniform Subsampling (with or without replacement) is very simple to 
implement, it is easy to construct examples where it will perform very 
poorly (e.g., see below or see~\cite{DMM06,Mah-mat-rev_BOOK}).
On the other hand, it has been shown that, for a parameter 
$\gamma \in (0,1]$ to be tuned, if
\begin{equation}
\label{eqn:approx-lev-score-probs}
\pi_i \ge \gamma \frac{h_{ii}}{p}, \;\;\mbox{and }
r=O(p \log(p) / (\gamma \epsilon) ),
\end{equation}
then the following relative-error bounds hold:
\begin{eqnarray}
\label{eqn:ls-bound-eq1}
||\sbf{y}-X\tilde{\sbf{\beta}}_{ols}||_2 
   &\leq& (1+\epsilon) ||\sbf{y}-X\hat{\sbf{\beta}}_{ols}||_2  \hspace{2mm} \mbox{ and } \\
\label{eqn:ls-bound-eq2}
||\hat{\sbf{\beta}}_{ols} - \tilde{\sbf{\beta}}_{ols}||_2
   &\leq& \sqrt{\epsilon} \left( \kappa(X)\sqrt{\xi^{-2}-1} \right) ||\hat{\sbf{\beta}}_{ols}||_2  ,
\end{eqnarray}
where $\kappa(X)$ is the condition number of $X$ and where 
$\xi = ||UU^T\sbf{y}||_2/||\sbf{y}||_2$ is a parameter defining the 
amount of the mass of $\sbf{y}$ inside the column space of 
$X$~\cite{DMM06,Mah-mat-rev_BOOK,DMMW12_JMLR}. 

Due to the crucial role of the statistical leverage scores in 
Eqn.~(\ref{eqn:approx-lev-score-probs}), we refer to algorithms of the 
form of \texttt{SubsampleLS} as the \emph{algorithmic leveraging} approach 
to approximating LS approximation.
Several versions of the \texttt{SubsampleLS} algorithm are of particular 
interest to us in this paper.
We start with two versions that have been studied in the past.
\begin{itemize}
\item
\textbf{Uniform Sampling Estimator (UNIF)} is the estimator resulting from
\emph{uniform subsampling} and \emph{weighted LS estimation}, i.e., where 
Eqn.~(\ref{lsq-sample}) is solved,
where both the sampling and rescaling/reweighting are done with the uniform 
sampling probabilities.
(Note that when the weights are uniform, then the weighted LS estimator of 
Eqn.~(\ref{lsq-sample}) leads to the same solution as same as the unweighted 
LS estimator of Eqn.~(\ref{unlsq-sample}) below.)
This version corresponds to vanilla uniform sampling, and it's solution will
be denoted by $\tilde{\sbf{\beta}}_{UNIF}$.
\item
\textbf{Basic Leveraging Estimator (LEV)} is the estimator resulting from
\emph{exact leverage-based sampling} and \emph{weighted LS estimation}, i.e., 
where Eqn.~(\ref{lsq-sample}) is solved,
where both the sampling and rescaling/reweighting are done with 
the leverage-based sampling probabilities given in 
Eqn.~(\ref{eqn:approx-lev-score-probs}).
This is the basic algorithmic leveraging algorithm that was originally
proposed in~\cite{DMM06}, where the exact empirical statistical leverage 
scores of $X$ were first used to construct the subsample and reweight the
subproblem, and it's solution will be denoted by $\tilde{\sbf{\beta}}_{LEV}$.
\end{itemize}

\noindent
Motivated by our statistical analysis (to come later in the paper), we will introduce two variants of 
\texttt{SubsampleLS}; since these are new to this paper, we also describe 
them here.
\begin{itemize}
\item
\textbf{Shrinked Leveraging Estimator (SLEV)} is the estimator resulting
from a \emph{shrinked leverage-based sampling} and \emph{weighted LS
estimation}.
By shrinked leverage-based sampling, we mean that we will sample according 
to a distribution that is a convex combination of a leverage score 
distribution and the uniform distribution, thereby obtaining the benefits of 
each; and the rescaling/reweighting is done according to the same distribution.
That is, if $\pi^{Lev}$ denotes a distribution defined by the normalized
leverage scores and $\pi^{Unif}$ denotes the uniform distribution, then the 
sampling and reweighting probabilities for SLEV are of the~form 
\begin{equation}
\label{eqn:slev_probs}
\pi_i = \alpha \pi^{Lev}_{i} + (1-\alpha) \pi^{Unif}_{i}  ,
\end{equation}
where $\alpha\in(0,1)$.
Thus, with SLEV, Eqn.~(\ref{lsq-sample}) is solved, where both the sampling 
and rescaling/reweighting are done with the probabilities given in 
Eqn.~(\ref{eqn:slev_probs}).
This estimator will be denoted by $\tilde{\sbf{\beta}}_{SLEV}$, and to our
knowledge it has not been explicitly considered previously.
\item
\textbf{Unweighted Leveraging Estimator (LEVUNW)} is the estimator resulting
from a \emph{leverage-based sampling} and \emph{unweighted LS estimation}.
That is, after the samples have been selected with leverage-based sampling probabilities, rather than solving the 
unweighted LS estimator of (\ref{lsq-sample}), we will compute the solution 
of the \emph{unweighted LS estimator}:
\begin{equation}
\label{unlsq-sample}
 \text{argmin}_{\beta\in \mathbb{R}^p}||S_X^{T}\sbf{y}-S_X^{T}X\sbf{\beta}||^2 .
\end{equation}
Whereas the previous estimators all follow the basic framework of sampling 
and rescaling/reweighting according to the same distribution (which is used in 
worst-case analysis to control the properties of both eigenvalues and 
eigenvectors and provide unbiased estimates of certain quantities within the 
analysis~\cite{DMM06,Mah-mat-rev_BOOK,DMMW12_JMLR}), with LEVUNW they are 
essentially done according to two different distributions---the reason being 
that not rescaling leads to the same solution as rescaling with the uniform 
distribution.
This estimator will be denoted by $\tilde{\sbf{\beta}}_{LEVUNW}$, and to our
knowledge it has not been considered previously.
\end{itemize}
These methods can all be used to estimate the coefficient vector $\sbf{\beta}$, 
and we will analyze---both theoretically and empirically---their statistical 
properties in terms of bias and variance.

\subsection{Running Time Considerations}
\label{sxn:background-running}

Although it is not our main focus, the running time for leverage-based 
sampling algorithms is of interest.
The running times of these algorithms depend on both the time to construct 
the probability distribution, $\{\pi_i\}_{i=1}^{n}$, and the time to solve 
the subsampled problem.
For UNIF, the former is trivial and the latter depends on the size of the 
subproblem.
For estimators that depend on the exact or approximate (recall the 
flexibility in Eqn.~(\ref{eqn:approx-lev-score-probs}) provided by $\gamma$) 
leverage scores, the running time is dominated by the exact or approximate 
computation of those scores.
A na\"{i}ve algorithm involves using a QR decomposition or the thin SVD of 
$X$ to obtain the exact leverage scores.  
Unfortunately, this exact algorithm takes $O(np^2)$ time and is thus no 
faster than solving the original LS problem exactly.
Of greater interest is the algorithm of~\cite{DMMW12_JMLR} that computes 
relative-error approximations to all of the leverage scores of $X$ in 
$o(np^2)$ time.

In more detail, given as input an arbitrary $n \times p$ matrix $X$, with 
$n \gg p$, and an error parameter $\epsilon\in(0,1)$, the main algorithm 
of~\cite{DMMW12_JMLR} (described also in Section~\ref{sxn:empirical-fast} 
below) computes numbers $\tilde{\ell}_{i}$, for all $ i = 1,\ldots,n $, 
that are relative-error approximations to the leverage scores $h_{ii}$, 
in the sense that $ |h_{ii}- \tilde{\ell}_i | \le \epsilon h_{ii} $, for all
$ i = 1,\ldots,n $.
This algorithm runs in roughly $O(n p \log(p)/\epsilon)$ time,%
\footnote{
In more detail, the asymptotic running time of the main algorithm 
of~\cite{DMMW12_JMLR} is
$
O\left(np\ln\left(p\epsilon^{-1}\right) + np\epsilon^{-2}\ln n + p^3\epsilon^{-2}\left(\ln n\right)\left(\ln \left(p\epsilon^{-1}\right)\right)\right).
$
To simplify this expression, suppose that $p\leq n\leq e^p$ and treat 
$\epsilon$ as a constant; then, the asymptotic running time is
$
O\left(np\ln n + p^3 \left(\ln n\right) \left(\ln p\right)\right).
$
} 
which for appropriate parameter settings is $o(np^2)$ time~\cite{DMMW12_JMLR}.
Given the numbers $\tilde{\ell}_{i}$, for all $ i = 1,\ldots,n $, we can let 
$\pi_i = \tilde{\ell}_{i}/\sum_{i=1}^{n}\tilde{\ell}_{i}$, which then yields
probabilities of the form of Eqn.~(\ref{eqn:approx-lev-score-probs}) with 
(say) $\gamma=0.5$ or $\gamma=0.9$.
Thus, we can use these $\pi_i$ in place of $h_{ii}$ in BELV, SLEV, or LEVUNW,
thus providing a way to implement these procedures in $o(np^2)$ time.

The running time of the relative-error approximation algorithm 
of~\cite{DMMW12_JMLR} depends on the time needed to premultiply $X$ by a 
randomized Hadamard transform (i.e., a ``structured'' random projection).
Recently, high-quality numerical implementations of such random projections
have been provided; see, e.g., Blendenpik~\cite{AMT10}, as well as
LSRN~\cite{MSM11_TR}, which extends these implementations to large-scale
parallel environments.
These implementations demonstrate that, for matrices as small as several 
thousand by several hundred, leverage-based algorithms such as LEV and
SLEV can be better in terms of running time than the computation of QR 
decompositions or the SVD with, e.g., \textsc{Lapack}. 
See~\cite{AMT10,MSM11_TR} for details, and see~\cite{GM13_TR} for the 
application of these methods to the fast computation of leverage scores.
Below, we will evaluate an implementation of a variant of the main algorithm 
of~\cite{DMMW12_JMLR} in the software environment R.

\subsection{Additional Related Work}
\label{sxn:background-additional}

Our leverage-based methods for estimating $\sbf{\beta}$ are related to 
resampling methods such as the bootstrap \cite{efron:79}, and many of these 
resampling methods enjoy desirable asymptotic properties \cite{shaotu:95}.
Resampling methods in linear models were studied extensively
in \cite{Wu:jack:1986} and are related to the
jackknife \cite{millerbio:74,miller:74,jaeckel:72,EfronGong:83}.
They usually produce resamples at a similar size to that of the full data,
whereas algorithmic leveraging is primarily interested in constructing
subproblems that are much smaller than the full data.
In addition, the goal of resampling is traditionally to perform statistical 
inference  
and not to improve the running time of an algorithm, 
except in the very recent work \cite{Kleiner12_ICML}.
Additional related work in statistics
includes~\cite{hinkley:77,rubin:81,Liu:Chen:Wong:1998,bickel:97,Poli:Roma:Wolf:subs:1999}.

\section{Bias and Variance Analysis of Subsampling Estimators}
\label{sxn:theory}

In this section, we develop analytic methods to study the biases and
variances of the subsampling estimators described in
Section~\ref{sxn:background-al}.
Analyzing these subsampling methods is challenging for at least the
following two reasons:
first, there are two layers of randomness in the estimators, i.e., the
randomness inherent in the linear regression model as well as random
subsampling of a particular sample from the linear model; and
second, the estimators depends on random subsampling through the inverse of
random sampling matrix, which is a nonlinear function.
To ease the analysis, we will employ a Taylor series analysis to approximate 
the subsampling estimators as linear combinations of random sampling 
matrices, and we will consider biases and variances both conditioned as well 
as not conditioned on the data.
Here is a brief outline of the main results of this section.
\begin{itemize}
\item
We will start in Section~\ref{sxn:theory-weighted} with bias and variance
results for weighted LS estimators for general sampling/reweighting 
probabilities.
This will involve viewing the solution of the subsampled LS problem as a
function of the vector of sampling/reweighting probabilities and performing 
a Taylor series expansion of the solution to the subsampled LS problem 
around the expected value (where the expectation is taken with respect to 
the random choices of the algorithm) of that vector.
\item
Then, in Section~\ref{sxn:theory-levunif}, we will specialize these results
to leverage-based sampling and uniform sampling, describing their 
complementary properties.
We will see that, in terms of bias and variance, neither LEV nor UNIF is 
uniformly better than the other.
In particular, LEV has variance whose size-scale is better than the 
size-scale of UNIF; but UNIF does not have leverage scores in the 
denominator of its variance expressions, as does LEV, and thus the 
variance of UNIF is not inflated on inputs that have very small leverage 
scores.
\item
Finally, in Section~\ref{sxn:theory-novel}, we will propose and analyze two 
new leveraging algorithms that will address deficiencies of LEV and UNIF in 
two different ways.
The first, SLEV, constructs a smaller LS problem with ``shrinked'' leverage 
scores that are constructed as a convex combination of leverage score 
probabilities and uniform probabilities; and the second, LEVUNW, uses 
leverage-based sampling probabilities to construct
and solve an unweighted or biased LS problem.
\end{itemize}

\subsection{Traditional Weighted Sampling Estimators}
\label{sxn:theory-weighted}

We start with the bias and variance of the traditional weighted sampling 
estimator $\tilde{\sbf{\beta}}_{W}$, given in Eqn.~(\ref{lsq-wls}) below.
Recall that this estimator actually refers to a parameterized family of
estimators, parameterized by the sampling/rescaling probabilities.
The estimate obtained by solving the weighted LS problem
of~(\ref{lsq-sample}) can be represented~as
\begin{eqnarray}
\nonumber
\tilde{\sbf{\beta}}_{W}
   &=& (X^{T}S_X D^2S_X^{T}X)^{-1}X^{T}S_X^{T} D^2 S_X\sbf{y}  \\
\label{lsq-wls}
   &=& (X^{T}WX)^{-1}X^{T}W\sbf{y}   ,
\end{eqnarray}
where $W=S_X D^2S_X^{T}$ is an $r \times r$ diagonal random matrix, i.e., all 
off-diagonal elements are zeros, and where both $S_X$ and $D$ are defined in 
terms of the sampling/rescaling probabilities.
(In particular, $W$ describes the probability distribution with which to 
draw the sample \emph{and} with which to reweigh the subsample, where both 
are done according to the same distribution.
Thus, this section does \emph{not} apply to LEVUNW; see 
Section~\ref{sxn:theory-novel-levunw} for the extension to LEVUNW.)
Although our results hold more generally, we are most interested in UNIF, 
LEV, and SLEV, as described in Section~\ref{sxn:background-al}.

Clearly, the vector $\tilde{\sbf{\beta}}_{W}$ can be regarded as a function 
of the random weight vector $\sbf{w}=(w_1, w_2, \ldots, w_n)^{T}$, denoted as
$\tilde{\sbf{\beta}}_{W}(\sbf{w})$, where $(w_1, w_2, \ldots, w_n)$ are
diagonal entries of $W$. 
Since we are performing random sampling with replacement,
it is easy to see that $\sbf{w}=(w_1, w_2, \ldots, w_n)^{T}$ has a scaled
multinomial distribution,
\begin{equation*}
\label{lsq-multi}
\Probab{w_1=\frac{k_1}{r\pi_1},w_2=\frac{k_2}{r\pi_2}, \ldots, w_n=\frac{k_n}{r\pi_n} } = \frac{r!}{k_1! k_2!\ldots, k_n!}\pi_1^{k_1}\pi_2^{k_2}\cdots \pi_n^{k_n}  ,
\end{equation*}
and thus it can easily be shown that $\Expect{\sbf{w}}=\sbf{1}$.
By setting $\sbf{w}_0$, the vector around which we will perform our Taylor
series expansion, to be the all-ones vector, i.e., $\sbf{w}_0=\sbf{1}$, then
$\tilde{\sbf{\beta}}(\sbf{w})$ can be expanded around the full sample
ordinary LS estimate $\hat{\sbf{\beta}}_{ols}$, i.e.,
$\tilde{\sbf{\beta}}_{W}(\sbf{1})= \hat{\sbf{\beta}}_{ols}$.
From this, we can establish the following lemma, the proof of which may be
found in Section~\ref{sxn:app-proofs-lev-taylor}.

\begin{lemma}
\label{lem:taylor}
Let $\tilde{\sbf{\beta}}_{W}$ be the output of the \texttt{SubsampleLS}
Algorithm, obtained by solving the weighted LS problem of~(\ref{lsq-sample}).
Then, a Taylor expansion of $\tilde{\sbf{\beta}}_{W}$ around the point
$\sbf{w}_0=\sbf{1}$ yields
\begin{align}
\label{eqn:lev-taylor}
\tilde{\sbf{\beta}}_{W}
   = \hat{\sbf{\beta}}_{ols} + (X^{T}X)^{-1}X^{T}\Diagnl{\hat{\sbf{e}}}(\sbf{w}-\sbf{1}) + R_{W},
\end{align}
where $\hat{\sbf{e}}=\sbf{y}-X\hat{\sbf{\beta}}_{ols}$
is the LS residual vector,
and where $R_{W}$ is the Taylor expansion remainder.
\end{lemma}

\noindent
\textbf{Remark.}
The significance of Lemma~\ref{lem:taylor} is that, to leading
order, the vector $\sbf{w}$ that encodes information about the sampling
process and subproblem construction enters the estimator of
$\tilde{\sbf{\beta}}_{W}$ linearly.
The additional error, $R_{W}$ depends strongly on the details of the
sampling process, and in particular will be very different for UNIF, LEV,
and SLEV.

\noindent
\textbf{Remark.}
Our approximations hold when the Taylor series expansion is valid, i.e., 
when $R_{W}$ is ``small,'' e.g., $R_{W}=o_p(||\sbf{w}-\sbf{w}_0||)$, 
where $o_p(\cdot)$ means ``little o'' with high probability over the 
randomness in the random vector $\sbf{w}$.
Although we will evaluate the quality of our approximations empirically in 
Sections~\ref{sxn:empirical} and~\ref{sxn:moreempirical}, we currently do 
\emph{not} have a precise theoretical characterization of when this holds.
Here, we simply make two observations.
First, this expression will fail to hold if rank is lost in the sampling 
process.
This is because in general there will be a bias due to failing to capture 
information in the dimensions that are not represented in the sample.
(Recall that one may use the Moore-Penrose generalized inverse for inverting 
rank-deficient matrices.)
Second, this expression will tend to hold better as the subsample size $r$ is 
increased.
However, 
for a fixed value of 
$r$, the linear approximation regime will be larger when the sample is 
constructed using information in the leverage scores---since, among other 
things, using leverage scores in the sampling process is designed to 
preserve the rank of the subsampled 
problem~\cite{DMM06,Mah-mat-rev_BOOK,DMMW12_JMLR}.
A detailed discussion of this last point is available
in~\cite{Mah-mat-rev_BOOK};
and these observations will be confirmed empirically in 
Section~\ref{sxn:moreempirical}.

\noindent
\textbf{Remark.}
Since, essentially, LEVUNW involves sampling and reweighting according to 
two \emph{different} distributions%
\footnote{In this case, the latter distribution is the uniform distribution, 
where recall that reweighting uniformly leads to the same solution as not 
reweighting at all.},
the analogous expression for LEVUNW will be somewhat different, as will 
be discussed in Lemma~\ref{lem:unwl-taylor} in Section~\ref{sxn:theory-novel}.

Given Lemma~\ref{lem:taylor}, we can establish the following lemma, which
provides expressions for the conditional and unconditional expectations and
variances for the weighted sampling estimators.
The first two expressions in the lemma are conditioned on the data 
vector $\sbf{y}$%
\footnote{Here and below, the subscript $\bf{w}$ on $\bf{E}_{w}$ and 
$\bf{Var}_{w}$ refers to performing expectations and variances with respect to 
(just) the random weight vector $\sbf{w}$ and not the data.};
and the last two expressions in the lemma provide similar results, except
that they are not conditioned on the data vector $\sbf{y}$.
The proof of this lemma appears in Section~\ref{sxn:app-proofs-lev-bv}.

\begin{lemma}
\label{lem:bv}
The conditional expectation and conditional variance for the traditional
algorithmic leveraging procedure, i.e., when the subproblem solved is a
weighted LS problem of the form~(\ref{lsq-sample}), are given by:
\begin{eqnarray}
\label{eqn:slev-cond-bias}
\CExpect{\tilde{\sbf{\beta}}_{W} |\sbf{y}}
   \hspace{-4mm}
   &=&\hspace{-3mm} \hat{\sbf{\beta}}_{ols}
   +
   \CExpect{R_{W}}  ; \\
\label{eqn:slev-wls-var}
\CVarnce{\tilde{\sbf{\beta}}_{W} |\sbf{y} }
   \hspace{-4mm}
   &=&\hspace{-3mm} (X^{T}X)^{-1}X^{T}
     \left[\Diagnl{\hat{\sbf{e}}}\Diagnl{\frac{1}{r\sbf{\pi}}} \Diagnl{\hat{\sbf{e}}} \right]
     X(X^{T}X)^{-1}
   \hspace{-1mm}
   + 
   \hspace{-1mm}
   \CVarnce{R_{W}}  ,
\end{eqnarray}
where $W$ specifies the probability distribution used in the sampling
and rescaling steps.
The unconditional expectation and unconditional variance for the traditional
algorithmic leveraging procedure are given by:
\begin{eqnarray}
\label{eqn:slev-wls-bias}
\Expect{\tilde{\sbf{\beta}}_{W} }
   \hspace{-4mm}
   &=&\hspace{-3mm} \sbf{\beta}_0 + \Expect{R_{W}}  ;  \\
\label{eqn:slev-wls-var-r-d}
\Varnce{\tilde{\sbf{\beta}}_{W}}
   \hspace{-4mm}
   &=&\hspace{-3mm} \sigma^2(X^{T}X)^{-1} + \frac{\sigma^2}{r}(X^{T}X)^{-1}X^{T}\Diagnl{\frac{(1-h_{ii})^2}{\pi_{i}}} X(X^{T}X)^{-1} + \Varnce{R_{W}}  .
\end{eqnarray}
\end{lemma}

\noindent
\textbf{Remark.}
Eqn.~(\ref{eqn:slev-cond-bias}) states that, when the $\CExpect{R_{W}}$
term is negligible, i.e., when the linear approximation is valid, then, 
conditioning on the observed data $\sbf{y}$, the estimate 
$\tilde{\sbf{\beta}}_{W}$ is approximately unbiased, relative to the 
full sample ordinarily LS estimate $\hat{\sbf{\beta}}_{ols}$; and
Eqn.~(\ref{eqn:slev-wls-bias}) states that, when the $\Expect{R_{W}}$
term is negligible, then the estimate $\tilde{\sbf{\beta}}_{W}$ is
approximately unbiased, relative to the ``true'' value $\sbf{\beta}_0$ of
the parameter vector $\sbf{\beta}$.
That is, given a particular data set $(X, \sbf{y})$, the conditional
expectation result of Eqn.~(\ref{eqn:slev-cond-bias}) states that the
leveraging estimators can approximate well $\hat{\sbf{\beta}}_{ols}$;
and, as a statistical inference procedure for arbitrary data sets, the
unconditional expectation result of Eqn.~(\ref{eqn:slev-wls-bias}) states
that the leveraging estimators can infer well $\sbf{\beta}_0$.

\noindent
\textbf{Remark.}
Both the conditional variance of~Eqn.~(\ref{eqn:slev-wls-var}) and the
(second term of the) unconditional variance
of~Eqn.~(\ref{eqn:slev-wls-var-r-d}) are inversely proportional to the
subsample size $r$; and both contain a sandwich-type expression, the middle
of which depends on how the leverage scores interact with the sampling
probabilities.
Moreover, the first term of the unconditional variance,
$\sigma^2 (X^{T}X)^{-1}$, equals the variance of the
ordinary LS estimator; this implies, e.g., that the unconditional variance
of Eqn.~(\ref{eqn:slev-wls-var-r-d}) is larger than the variance of the
ordinary LS estimator, which is consistent with the Gauss-Markov theorem.

\subsection{Leverage-based Sampling and Uniform Sampling Estimators}
\label{sxn:theory-levunif}

Here, we specialize Lemma~\ref{lem:bv} by stating two lemmas that provide
the conditional and unconditional expectation and variance for LEV and UNIF,
and we will discuss the relative merits of each procedure.
The proofs of these two lemmas are immediate, given the proof of
Lemma~\ref{lem:bv}.
Thus, we omit the proofs, and instead discuss properties of the expressions
that are of interest in our empirical evaluation.

Our main conclusion here is that
Lemma~\ref{lem:lev-bv} and Lemma~\ref{lem:unif-bv} highlight that the 
statistical properties of the algorithmic leveraging method can be quite 
different than the algorithmic properties.
Prior work has adopted an \emph{algorithmic perspective} that has
focused on providing worst-case running time bounds for arbitrary input
matrices.
From this algorithmic perspective, leverage-based sampling (i.e., explicitly
or implicitly biasing toward high-leverage components, as is done in
particular with the LEV procedure) provides uniformly superior worst-case
algorithmic results, when compared with
UNIF~\cite{DMM06,Mah-mat-rev_BOOK,DMMW12_JMLR}.
Our analysis here reveals that, from a \emph{statistical perspective} 
where one is interested in the bias and variance properties of the 
estimators, the situation is considerably more subtle.
In particular, a key conclusion from Lemmas~\ref{lem:lev-bv}
and~\ref{lem:unif-bv} is that, with respect to their variance or MSE,
neither LEV nor UNIF is uniformly superior for all input.

We start with the bias and variance of the leverage subsampling estimator
$\tilde{\sbf{\beta}}_{LEV}$.

\begin{lemma}
\label{lem:lev-bv}
The conditional expectation and conditional variance for the LEV procedure
are given~by:
\begin{eqnarray*}
\CExpect{\tilde{\sbf{\beta}}_{LEV} |\sbf{y}}
\hspace{-4mm}
   &=&\hspace{-3mm} \hat{\sbf{\beta}}_{ols}+\CExpect{R_{LEV}}  ; \\
\CVarnce{\tilde{\sbf{\beta}}_{LEV} |\sbf{y} }
\hspace{-4mm}
   &=&\hspace{-3mm} \frac{p}{r}(X^{T}X)^{-1}X^{T}
       \left[\Diagnl{\hat{\sbf{e}}}\Diagnl{\frac{1}{h_{ii}}} \Diagnl{\hat{\sbf{e}}} \right]
       X(X^{T}X)^{-1}
     \hspace{-1mm}+\hspace{-1mm} \CVarnce{R_{LEV}}.
\end{eqnarray*}
The unconditional expectation and unconditional variance for the LEV procedure
are given by:
\begin{eqnarray}
\nonumber
\Expect{\tilde{\sbf{\beta}}_{LEV} }
   \hspace{-3mm}
   &=&\hspace{-2mm} \sbf{\beta}_0 + \Expect{R_{LEV}}  ;  \\
\nonumber
\Varnce{\tilde{\sbf{\beta}}_{LEV}}
   \hspace{-3mm}
   &=&\hspace{-2mm} \sigma^2(X^{T}X)^{-1} + \frac{p\sigma^2}{r}(X^{T}X)^{-1}X^{T}\Diagnl{\frac{(1-h_{ii})^2}{h_{ii}}} X(X^{T}X)^{-1} \\
   &+& \Varnce{R_{LEV}} .
\label{lsq-wls-var-r-d}
\end{eqnarray}
\end{lemma}

\noindent
\textbf{Remark.}
Two points are worth making.
First, the variance expressions for LEV depend on the size (i.e., the number 
of columns and rows) of the 
$n \times p$ matrix $X$ and the number of samples $r$ as $p/r$.
This variance size-scale many be made to be very small if
$p\ll r\ll n$.
Second, the sandwich-type expression depends on the leverage scores
as $1/h_{ii}$, implying that the variances could be inflated to arbitrarily
large values by very small leverage scores.
Both of these observations will be confirmed empirically in 
Section~\ref{sxn:empirical}.

We next turn to the bias and variance of the uniform subsampling estimator
$\tilde{\sbf{\beta}}_{UNIF}$.

\begin{lemma}
\label{lem:unif-bv}
The conditional expectation and conditional variance for the UNIF procedure
are given by:
\begin{eqnarray}
\nonumber
\CExpect{\tilde{\sbf{\beta}}_{UNIF} |\sbf{y}}
   \hspace{-3mm}
   &=&\hspace{-2mm} \hat{\sbf{\beta}}_{ols} + \CExpect{R_{UNIF}} \\
\label{uniform-CVar}
\CVarnce{\tilde{\sbf{\beta}}_{UNIF} |\sbf{y}}
   \hspace{-3mm}
   &=&\hspace{-2mm} \frac{n}{r} (X^{T}X)^{-1}X^{T} \left[ \Diagnl{\hat{\sbf{e}}}\Diagnl{\hat{\sbf{e}}} \right] X(X^{T}X)^{-1} + \CVarnce{R_{UNIF}} .
\end{eqnarray}
The unconditional expectation and unconditional variance for the UNIF procedure
are given by:
\begin{eqnarray}
\nonumber
\Expect{\tilde{\sbf{\beta}}_{UNIF}}
   \hspace{-3mm}
   &=&\hspace{-2mm} \sbf{\beta}_0 + \Expect{R_{UNIF}}  ; \\
\nonumber
\Varnce{\tilde{\sbf{\beta}}_{UNIF}}
   \hspace{-3mm}
   &=&\hspace{-2mm} \sigma^2(X^{T}X)^{-1}
   + \frac{n}{r} \sigma^2(X^{T}X)^{-1}X^{T} \Diagnl{ (1-h_{ii})^2 } X(X^{T}X)^{-1}  \\
\label{uniform-Var1}
   &+& \Varnce{R_{UNIF}}   .
\end{eqnarray}
\end{lemma}

\noindent
\textbf{Remark.}
Two points are worth making.
First, the variance expressions for UNIF depend on the size (i.e., the number 
of columns and rows) of the 
$n \times p$ matrix $X$ and the number of samples $r$ as $n/r$.
Since this variance size-scale is very large, e.g., compared to the $p/r$ 
from LEV, these variance expressions will be large unless $r$ is nearly 
equal to $n$.
Second, the sandwich-type expression is not inflated by very small 
leverage~scores.

\noindent
\textbf{Remark.}
Apart from a factor $n/r$, the conditional variance for UNIF, as given in 
Eqn.~(\ref{uniform-CVar}), is the same as Hinkley's weighted jackknife 
variance estimator~\cite{hinkley:77}.

\subsection{Novel Leveraging Estimators}
\label{sxn:theory-novel}

In view of Lemmas~\ref{lem:lev-bv} and~\ref{lem:unif-bv}, 
we consider several ways to take advantage of the complementary
strengths of the LEV and UNIF procedures.
Recall that we would like to sample with respect to probabilities that are 
``near'' those defined by the empirical statistical leverage scores.
We at least want to identify large leverage scores to preserve rank.
This helps ensure that the linear regime of the Taylor expansion is 
large, and it also helps ensure that the scale of the variance is $p/r$ 
and not $n/r$.
But we would like to avoid rescaling by $1/h_{ii}$ when certain leverage 
scores are extremely small, thereby avoiding inflated variance estimates.

\subsubsection{The Shrinked Leveraging (SLEV) Estimator}
\label{sxn:theory-novel-slev}

Consider first the SLEV procedure.
As described in Section~\ref{sxn:background-al}, this involves sampling and
reweighting with respect to a distribution that is a convex combination of the 
empirical leverage score distribution and the uniform distribution.
That is, let $\pi^{Lev}$ denote a distribution defined by the normalized
leverage scores (i.e., $\pi^{Lev}_{i}=h_{ii}/p$, or $\pi^{Lev}$ is 
constructed from the output of the algorithm of~\cite{DMMW12_JMLR} that 
computes relative-error approximations to the leverage scores), and let
$\pi^{Unif}$ denote the uniform distribution (i.e., $\pi^{Unif}_{i} = 1/n$, 
for all $i\in[n]$); then the sampling probabilities for the SLEV procedure 
are of the form 
\begin{equation}
\pi_i = \alpha \pi^{Lev}_{i} + (1-\alpha) \pi^{Unif}_{i}  ,
\label{eqn:slev-probs}
\end{equation}
where $\alpha\in(0,1)$.

Since SLEV involves solving a weighted LS problem of the form of
Eqn.~(\ref{lsq-sample}), expressions of the form provided by 
Lemma~\ref{lem:bv} hold immediately.
In particular, SLEV enjoys approximate unbiasedness, in the same sense that 
the LEV and UNIF procedures do.
The particular expressions for the higher order terms can be easily derived, 
but they are much messier and less transparent than the bounds provided by 
Lemmas~\ref{lem:lev-bv} and~\ref{lem:unif-bv} for LEV and UNIF, respectively.
Thus, rather than presenting them, we simply point out several aspects of 
the SLEV procedure that should be immediate, given our earlier theoretical 
discussion.

First, note that $\min_i \pi_{i} \ge (1-\alpha)/n$, with equality obtained when 
$h_{ii}=0$.
Thus, assuming that $1-\alpha$ is not extremely small, e.g., $1-\alpha=0.1$, 
then none of the SLEV sampling probabilities is too small, and thus the 
variance of the SLEV estimator does not get inflated too much, as it could 
with the LEV estimator.
Second, assuming that $1-\alpha$ is not too large, e.g., $1-\alpha=0.1$, then
Eqn.~(\ref{eqn:approx-lev-score-probs}) is satisfied with $\gamma=1.1$, and 
thus the amount of oversampling that is required, relative to the LEV 
procedure, is not much, e.g., $10\%$.
In this case, the variance of the SLEV procedure has a scale of $p/r$, as 
opposed to $n/r$ scale of UNIF, assuming that $r$ is increased by that 
$10\%$.
Third, since Eqn.~(\ref{eqn:slev-probs})
is still required to be a probability distribution, combining the leverage
score distribution with the uniform distribution has the effect of not only
increasing the very small scores, but it also has the effect of performing
shrinkage on the very large scores.
Finally, all of these observations also hold if, rather that using the exact 
leverage score distribution (which recall takes $O(np^2)$ time to compute), 
we instead use approximate leverage scores, as computed with the fast 
algorithm of~\cite{DMMW12_JMLR}.
For this reason, this approximate version of the SLEV procedure is the most 
promising for very large-scale applications.

\subsubsection{The Unweighted Leveraging (LEVUNW) Estimator}
\label{sxn:theory-novel-levunw}

Consider next the LEVUNW procedure.
As described in Section~\ref{sxn:background-al}, this estimator is different 
than the previous estimators, in that the sampling and reweighting are done 
according to different distributions.
(Since LEVUNW does \emph{not} sample and reweight according to the same 
probability distribution, our previous analysis does not apply.)
Thus, we shall examine the bias and variance of the unweighted leveraging 
estimator $\tilde{\sbf{\beta}}_{LEVUNW}$.
To do so, we first use a Taylor series expansion to get the following lemma,
the proof of which may be found in Section~\ref{sxn:app-proofs-unwl-taylor}.

\begin{lemma}
\label{lem:unwl-taylor}
Let $\tilde{\sbf{\beta}}_{LEVUNW}$ be the output of the modified
\texttt{SubsampleLS} Algorithm, obtained by solving the unweighted LS
problem of~(\ref{unlsq-sample}).
Then, a Taylor expansion of $\tilde{\sbf{\beta}}_{LEVUNW}$ around the point
$\sbf{w}_0=r\sbf{\pi}$ yields
\begin{align}
\label{eqn:unwl-taylor}
\tilde{\sbf{\beta}}_{LEVUNW}
   = \hat{\sbf{\beta}}_{wls} + (X^{T}W_0X)^{-1}X^{T}\Diagnl{\hat{\sbf{e}}_w}(\sbf{w}-r\sbf{\pi}) + R_{LEVUNW},
\end{align}
where $\hat{\sbf{\beta}}_{wls}= (X^{T}W_0X)^{-1}XW_0\sbf{y}$
is the full sample weighted LS estimator,
$\hat{\sbf{e}}_w=\sbf{y}-X\hat{\sbf{\beta}}_{wls}$
is the LS residual vector,
$W_0=\Diagnl{r\sbf{\pi}}= \Diagnl{r h_{ii}/p}$,
and $R_{LEVUNW}$ is the Taylor expansion remainder.
\end{lemma}

\noindent
\textbf{Remark.}
This lemma is analogous to Lemma~\ref{lem:taylor}.
Since the sampling and reweighting are performed according to different
distributions, however, the point about which the Taylor expansion is
performed, as well as the prefactors of the linear term, are somewhat
different.
In particular, here we expand around the point $\sbf{w}_0=r\sbf{\pi}$ 
since $\Expect{\sbf{w}}=r\sbf{\pi}$ when no reweighting takes place.

Given this Taylor expansion lemma, we can now establish the following
lemma for the mean and variance of LEVUNW, both conditioned and
unconditioned on the data $\sbf{y}$.
The proof of the following lemma may be found in
Section~\ref{sxn:app-proofs-unwl-bv}.

\begin{lemma}
\label{lem:unwl-bv}
The conditional expectation and conditional variance for the LEVUNW procedure
are given by:
\begin{eqnarray*}
\CExpect{\tilde{\sbf{\beta}}_{LEVUNW} |\sbf{y}}
   &=& \hat{\sbf{\beta}}_{wls}+ \CExpect{R_{LEVUNW}}  ;  \\
\CVarnce{\tilde{\sbf{\beta}}_{LEVUNW}|\sbf{y} }
   &=& (X^{T}W_0X)^{-1}X^{T}\Diagnl{\hat{\sbf{e}}_w}W_0\Diagnl{\hat{\sbf{e}}_w} X(X^{T}W_0X)^{-1} \\
  &+&  \CVarnce{R_{LEVUNW}}.
\end{eqnarray*}
where $W_0=\Diagnl{r\sbf{\pi}}$, and where
$\hat{\sbf{\beta}}_{wls}= (X^{T}W_0X)^{-1}XW_0\sbf{y}$
is the full sample weighted LS estimator.
The unconditional expectation and unconditional variance for the LEVUNW 
procedure are given by:
\begin{eqnarray}
\nonumber
\Expect{\tilde{\sbf{\beta}}_{LEVUNW}}
   &=& \sbf{\beta}_0 + \Expect{R_{LEVUNW}}  ; \\
\nonumber
\Varnce{\tilde{\sbf{\beta}}_{LEVUNW}}
   &=& \sigma^2(X^{T}W_0X)^{-1}X^{T}W_0^2X(X^{T}W_0X)^{-1}   \\
\nonumber
   &+& {\sigma^2}(X^{T}W_0X)^{-1}X^{T}\Diagnl{I-P_{X,W_0}}W_0 \Diagnl{I-P_{X,W_0}} X(X^{T}W_0 X)^{-1} \\
\label{levnoweightvar11}
   &+& \Varnce{R_{LEVUNW}}
\end{eqnarray}
where $P_{X,W_0}=X(X^{T}W_0X)^{-1}X^{T}W_0$.
\end{lemma}

\noindent
\textbf{Remark.}
The two expectation results in this lemma state:
(i), when $\CExpect{R_{LEVUNW}}$ is negligible, then, conditioning on the
observed data $\sbf{y}$, the estimator $\tilde{\sbf{\beta}}_{LEVUNW}$ is
approximately unbiased, relative to the full sample \emph{weighted} LS
estimator $\hat{\sbf{\beta}}_{wls}$; and (ii), when $\Expect{R_{LEVUNW}}$ is
negligible, then the estimator $\tilde{\sbf{\beta}}_{LEVUNW}$ is approximately
unbiased, relative to the ``true'' value $\sbf{\beta}_0$ of the parameter
vector $\sbf{\beta}$.
That is, if we apply LEVUNW to a given data set $N$ times, then the average
of the $N$ LEVUNW estimates are \emph{not} centered at the LS estimate, but 
instead are centered roughly at the weighted least squares estimate; while 
if we generate many data sets from the true model and apply LEVUNW to 
these data sets, then the average of these estimates is roughly centered 
around true value $\sbf{\beta}_0$.

\noindent
\textbf{Remark.}
As expected, when the leverage scores are all the same, the variance in 
Eqn.~(\ref{levnoweightvar11}) is the same as the variance of uniform random 
sampling.
This is expected since, when reweighting with respect to the uniform
distribution, one does not change the problem being solved, and thus the
solutions to the weighted and unweighted LS problems are identical.
More generally, the variance is not inflated by very small leverage scores, 
as it is with LEV.
For example, the conditional variance expression is also a sandwich-type
expression, the center of which is $W_0=\Diagnl{rh_{ii}/n}$, which is
not inflated by very small leverage scores.

\section{Main Empirical Evaluation}
\label{sxn:empirical}

In this section, we describe the main part of our empirical analysis of the 
behavior of the biases and variances of the subsampling estimators described 
in Section~\ref{sxn:background-al}.
Additional empirical results will be presented in 
Section~\ref{sxn:moreempirical}.
In these two sections,
we will consider both synthetic data as well as real data that have been 
chosen to illustrate the extreme properties of the subsampling methods in 
realistic settings.
We will use the MSE as a benchmark to compare the different subsampling 
estimators; but since we are interested in both the bias and variance 
properties of our estimates, we will present results for both the bias and 
variance separately.

Here is a brief outline of the main results of this section.
\begin{itemize}
\item
In Section~\ref{sxn:empirical-synthetic}, we will describe our synthetic 
data.
These data are drawn from three standard distributions, and they are 
designed to provide relatively-realistic synthetic examples where leverages 
scores are fairly uniform, moderately nonuniform, or very nonuniform.
\item
Then, in Section~\ref{sxn:empirical-levvunif}, we will summarize our results 
for the unconditional bias and variance for LEV and UNIF, when applied to 
the synthetic data.  
\item
Then, in Section~\ref{sxn:empirical-shrinkedunweight}, we will summarize our 
results for the unconditional bias and variance of SLEV and LEVUNW.
This will illustrate that both SLEV and LEVUNW can overcome some of the 
problems associated with LEV and UNIF.
\item
Finally, in Section~\ref{sxn:empirical-conditional}, we will present our results
for the conditional bias and variance of SLEV and LEVUNW (as well as LEV
and UNIF).
In particular, this will show that LEVUNW can incur substantial bias, 
relative to the other methods, when conditioning on a given data set.
\end{itemize}

\subsection{Description of Synthetic Data}
\label{sxn:empirical-synthetic}

We consider synthetic data of 1000 runs generated from 
$\sbf{y}=X\sbf{\beta}+\sbf{\epsilon}$, where
$\sbf{\epsilon} \sim N(0, 9I_n)$,
where several different values of $n$ and $p$, leading to both 
``very rectangular'' and ``moderately rectangular'' matrices $X$, are considered.
The design matrix $X$ is generated from one of three different classes of 
distributions introduced below.
These three distributions were chosen since the first has nearly uniform 
leverage scores, the second has mildly non-uniform leverage scores, and the 
third has very non-uniform leverage~scores.
\begin{itemize}
\item
\textbf{Nearly uniform leverage scores (GA).}
We generated an $n \times p$ matrix $X$ from multivariate normal
$N(\sbf{1}_{p},\Sigma)$, where the $(i, j)$th element of
$\Sigma_{ij}=2 \times 0.5^{|i-j|}$, and where we set
$\sbf{\beta}=(\sbf{1}_{10}, 0.1\sbf{1}_{p-20}, \sbf{1}_{10})^{T}$.
(Referred to as GA data.)
\item
\textbf{Moderately nonuniform leverage scores ($T_3$).}
We generated $X$ from multivariate $t$-distribution
with $3$ degree of freedom and covariance matrix $\Sigma$ as before.
(Referred to as $T_3$ data.)
\item
\textbf{Very nonuniform leverage scores ($T_1$).}
We generated $X$ from multivariate $t$-distribution with $1$
degree of freedom and covariance matrix $\Sigma$ as before.
(Referred to as $T_1$ data.) 
\end{itemize}
See Table~\ref{tab:synthetic-summary-stats} for a summary of the parameters 
for the synthetic data we considered and for basic summary statistics for 
the leverage scores probabilities (i.e., the leverage scores that have been 
normalized to sum to $1$ by dividing by $p$) of these data matrices.
The results reported in Table~\ref{tab:synthetic-summary-stats} are for 
leverage score statistics for a single fixed data matrix $X$ generated in 
the above manner (for each of the $3$ procedures and for each value of $n$ 
and $p$), but we have confirmed that similar results hold for other matrices 
$X$ generated in the same~manner.

\begin{table}[t]
\label{tab:synthetic-summary-stats}
\begin{center}
\begin{tabular}{lrrrrrrrrrr}
Dstbn & $n$ & $p$ & Min & Median & Max &  Mean & Std.Dev. & $\frac{\text{Max}}{\text{Min}}$ & $\frac{\text{Max}}{\text{Mean}}$ & $\frac{\text{Max}}{\text{Median}}$
\\ \hline \hline \\
GA  & 1K & 10  & 1.96e-4 & 9.24e-4 & 2.66e-3 & 1.00e-3 & 4.49e-4 & 13.5    & 2.66 & 2.88 \\
GA  & 1K & 50  & 4.79e-4 & 9.90e-4 & 1.74e-3 & 1.00e-3 & 1.95e-4 & 3.63    & 1.74 & 1.76 \\
GA  & 1K & 100 & 6.65e-4 & 9.94e-4 & 1.56e-3 & 1.00e-3 & 1.33e-4 & 2.35    & 1.56 & 1.57 \\
\hline 
GA  & 5K & 10  & 1.45e-5 & 1.88e-4 & 6.16e-4 & 2.00e-4 & 8.97e-5 & 42.4    & 3.08 & 3.28 \\
GA  & 5K & 50  & 9.02e-5 & 1.98e-4 & 3.64e-4 & 2.00e-4 & 3.92e-5 & 4.03    & 1.82 & 1.84 \\
GA  & 5K & 250 & 1.39e-4 & 1.99e-4 & 2.68e-4 & 2.00e-4 & 1.73e-5 & 1.92    & 1.34 & 1.34 \\
GA  & 5K & 500 & 1.54e-4 & 2.00e-4 & 2.48e-4 & 2.00e-4 & 1.20e-5 & 1.61    & 1.24 & 1.24 \\
\hline 
$T_3$  & 1K & 10  & 2.64e-5 & 4.09e-4 & 5.63e-2 & 1.00e-3 & 2.77e-3 & 2.13e+3 & 56.3 & 138 \\
$T_3$  & 1K & 50  & 6.57e-5 & 5.21e-4 & 1.95e-2 & 1.00e-3 & 1.71e-3 & 297     & 19.5 & 37.5 \\
$T_3$  & 1K & 100 & 7.26e-5 & 6.39e-4 & 9.04e-3 & 1.00e-3 & 1.06e-3 & 125     & 9.04 & 14.1 \\
\hline 
$T_3$  & 5K & 10  & 5.23e-6 & 7.73e-5 & 5.85e-2 & 2.00e-4 & 9.66e-4 & 1.12e+4 &  293 & 757 \\
$T_3$  & 5K & 50  & 9.60e-6 & 9.84e-5 & 1.52e-2 & 2.00e-4 & 4.64e-4 & 1.58e+3 & 76.0 & 154 \\
$T_3$  & 5K & 250 & 1.20e-5 & 1.14e-4 & 3.56e-3 & 2.00e-4 & 2.77e-4 & 296     & 17.8 & 31.2 \\
$T_3$  & 5K & 500 & 1.72e-5 & 1.29e-4 & 1.87e-3 & 2.00e-4 & 2.09e-4 & 108     & 9.34 & 14.5 \\
\hline 
$T_1$  & 1K & 10  & 4.91e-8 & 4.52e-6 & 9.69e-2 & 1.00e-3 & 8.40e-3 & 1.97e+6 & 96.9 & 2.14e+4 \\
$T_1$  & 1K & 50  & 2.24e-6 & 6.18e-5 & 2.00e-2 & 1.00e-3 & 3.07e-3 & 8.93e+3 &   20 & 323 \\
$T_1$  & 1K & 100 & 4.81e-6 & 1.66e-4 & 9.99e-3 & 1.00e-3 & 2.08e-3 & 2.08e+3 & 9.99 & 60.1 \\
\hline 
$T_1$  & 5K & 10  & 5.00e-9 & 6.18e-7 & 9.00e-2 & 2.00e-4 & 3.00e-3 & 1.80e+7 & 450 & 1.46e+5 \\
$T_1$  & 5K & 50  & 4.10e-8 & 2.71e-6 & 2.00e-2 & 2.00e-4 & 1.39e-3 & 4.88e+5 & 99.9 & 7.37e+3 \\
$T_1$  & 5K & 250 & 3.28e-7 & 1.50e-5 & 4.00e-3 & 2.00e-4 & 6.11e-4 & 1.22e+4 &   20 & 267 \\
$T_1$  & 5K & 500 & 1.04e-6 & 2.79e-5 & 2.00e-3 & 2.00e-4 & 4.24e-4 & 1.91e+3 &   10 & 71.6 \\
\hline \hline \\
\end{tabular}
\end{center}
\caption{Summary statistics for leverage-score probabilities (i.e., leverage scores divided by $p$) for the synthetic data sets.  }
\end{table}

Several observations are worth making about the summaries presented in 
Table~\ref{tab:synthetic-summary-stats}.
First, and as expected, the Gaussian data tend to have the most uniform 
leverage scores, the $T_3$ data are intermediate, and the $T_1$ data have 
the most nonuniform leverage scores, as measured by both the standard 
deviation of the scores as well as the ratio of maximum to minimum leverage 
score.
Second, the standard deviation of the leverage score distribution is 
substantially less sensitive to nonuniformities in the leverage scores than 
is the ratio of the maximum to minimum leverage score (or the maximum to the 
mean/median score, although all four measures exhibit the same qualitative 
trends).
Although we have not pursued it, this suggests that these latter measures 
will be more informative as to when leverage-based sampling might be 
necessary in a particular application.
Third, in all these cases, the variability trends are caused both by the 
large (in particular, the maximum) leverage scores increasing as well as the 
small (in particular, the minimum) leverage score decreasing.
Fourth, within a given type of distribution (i.e., GA or $T_3$ or $T_1$), 
leverage scores are more nonuniform when the matrix $X$ is more rectangular, 
and this is true both when $n$ is held fixed and when $p$ is held fixed.

\subsection{Leveraging Versus Uniform Sampling on Synthetic Data}
\label{sxn:empirical-levvunif}

\begin{figure}[t]
  \centering
   \makebox{\includegraphics[scale=1]{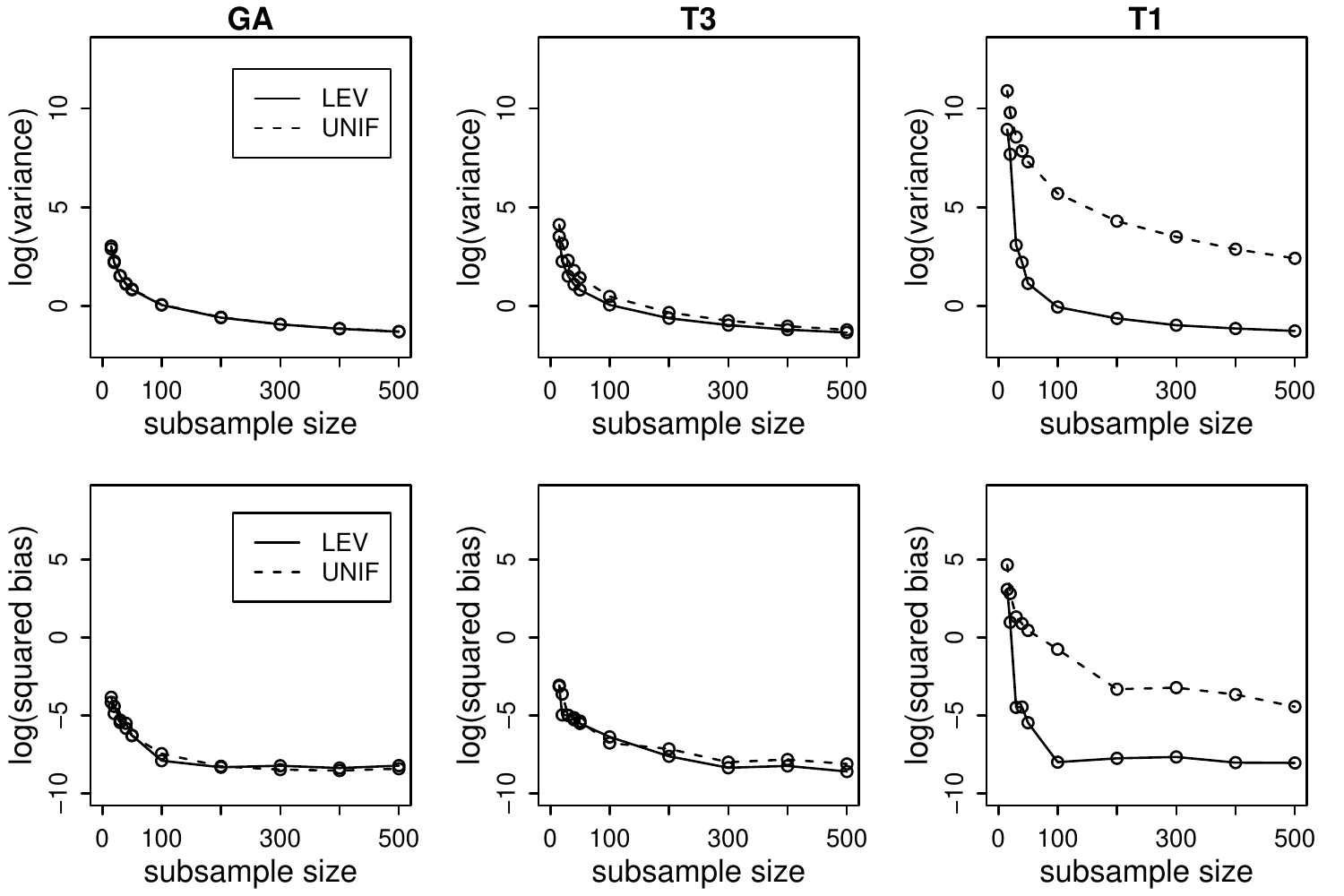}}
\vspace{-1em}
\caption{
(Leveraging Versus Uniform Sampling subsection.)
Comparison of variances and squared biases of the LEV and UNIF estimators 
in three data sets (GA, $T_3$, and $T_1$) for $n=1000$ and $p=10$.
Left panels are GA data;
Middle panels are $T_3$ data;
Right panels are $T_1$ data.
Upper panels are Logarithm of Variances; 
Lower panels are Logarithm of Squared bias.
Black lines are LEV; 
Dash lines are UNIF.
}
\label{fig:simuline1}
\end{figure}

\begin{figure}[t]
  \centering
  \makebox{\includegraphics[scale=1]{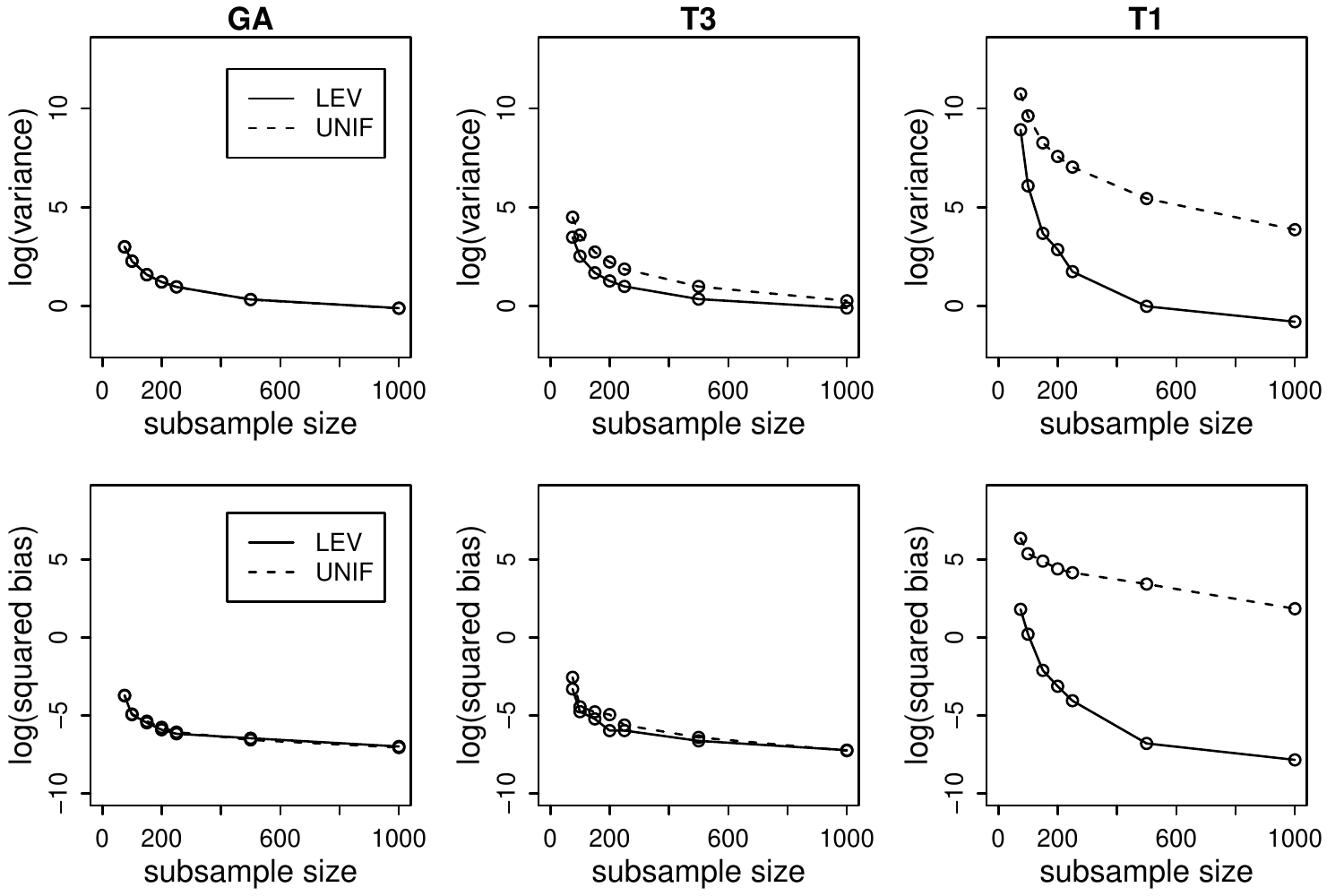}}
\vspace{-1em}
\caption{
(Leveraging Versus Uniform Sampling subsection.)
Same as Figure~\ref{fig:simuline1}, except that $n=1000$ and $p=50$.
}
\label{fig:simuline2}
\end{figure}

\begin{figure}[t]
  \centering
  \makebox{\includegraphics[scale=1]{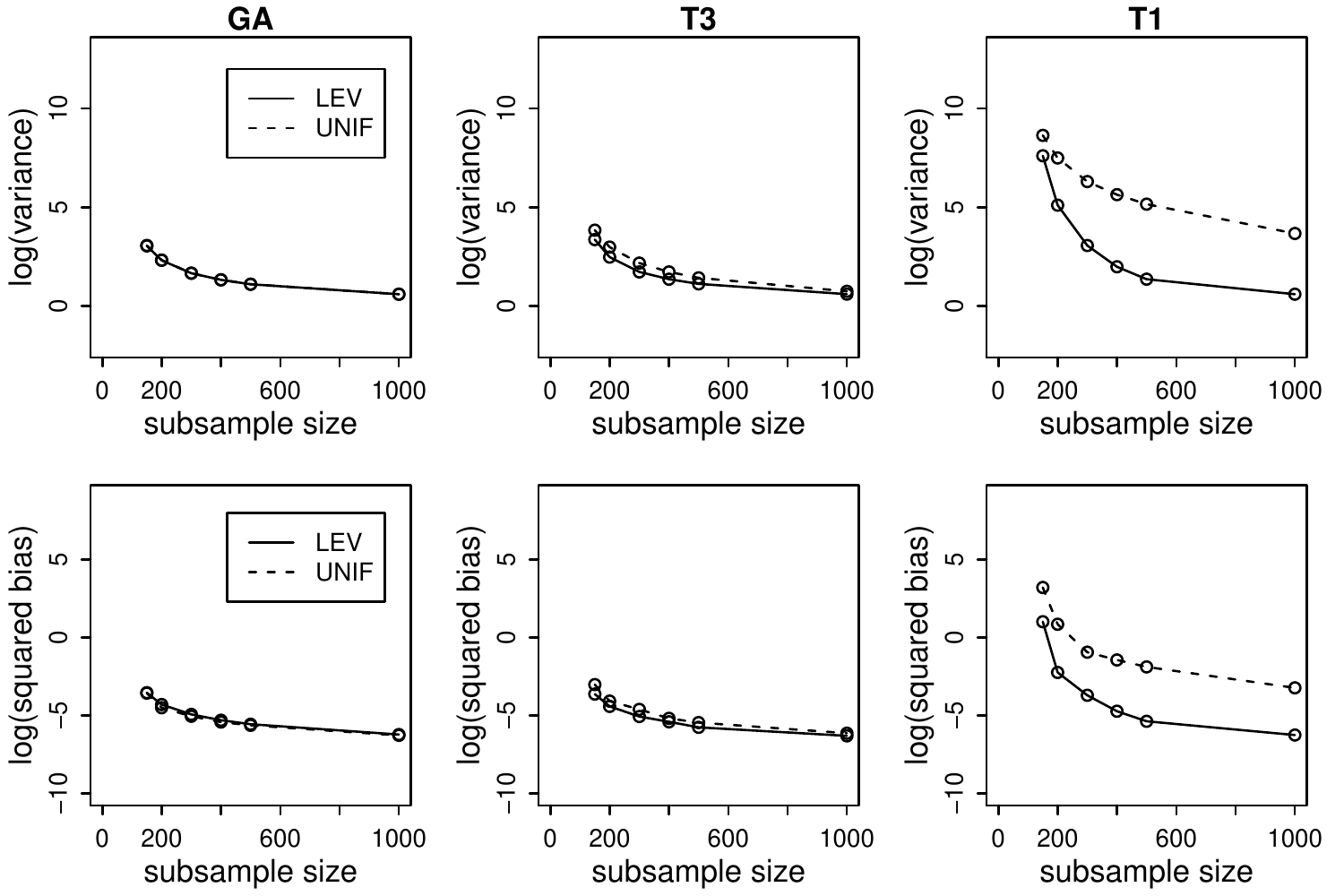}}
\vspace{-1em}
\caption{
(Leveraging Versus Uniform Sampling subsection.)
Same as Figure~\ref{fig:simuline1}, except that $n=1000$ and $p=100$.
}
\label{fig:simuline3}
\end{figure}

Here, we will describe the properties of LEV versus UNIF for synthetic data.
See Figures~\ref{fig:simuline1}, \ref{fig:simuline2}, 
and~\ref{fig:simuline3} for the results on data matrices with $n=1000$ and 
$p=10$, $50$, and $100$, respectively.
(The results for data matrices for $n=5000$ and other values of $n$ are 
similar.)
In each case, we generated a single matrix from that distribution (which we 
then fixed to generate the $\sbf{y}$ vectors) and $\sbf{\beta}_{0}$ was set 
to be the all-ones vector; and then 
we ran the sampling process multiple times, typically ca. $1000$ times, in 
order to obtain reliable estimates for the biases and variances.
In each of the Figures~\ref{fig:simuline1}, \ref{fig:simuline2},
and~\ref{fig:simuline3}, the top panel is the variance, the bottom panel is the 
squared bias; for both the bias and variance, we have plotted the results in 
log-scale; and, in each figure, the first column is the GA model, the middle 
column is the $T_3$ model, and the right column is the $T_1$ model.


The simulation results corroborate what we have learned from our theoretical 
analysis, and there are several things worth noting.
First, in general the squared bias is much less than the variance, even for
the $T_1$ data, suggesting that the solution is approximately unbiased, at 
least for the values of $r$ plotted here, in the sense quantified in 
Lemmas~\ref{lem:lev-bv} and~\ref{lem:unif-bv}.
Second, LEV and UNIF perform very similarly for GA, somewhat less 
similarly for $T_3$, and quite differently for $T_1$, consistent with the 
results in Table~\ref{tab:synthetic-summary-stats} that indicate that the
leverage scores are very uniform for GA and very nonuniform for $T_1$.
In addition, when they are different, LEV tends to perform better than 
UNIF, i.e., have a lower MSE for a fixed sampling complexity. 
Third, as the subsample size increases, the squared bias and variance tend 
to decrease monotonically.
In particular, the variance tends to decrease roughly as $1/r$, where $r$ is 
the size of the subsample, in agreement with Lemmas~\ref{lem:lev-bv} 
and~\ref{lem:unif-bv}.
Moreover, the decrease for UNIF is much slower, in a manner more consistent 
with the leading term of $n/r$ in Eqn.~(\ref{uniform-Var1}), than is the 
decrease for LEV, which by Eqn.~(\ref{lsq-wls-var-r-d}) has leading term 
$p/r$.
Fourth, for all three models, both the bias and variance tend to increase
when the matrix is less rectangular, e.g., as $p$ increases $10$ to $100$ 
for $n=1000$.
All in all, LEV is comparable to or outperforms UNIF, especially when the 
leverage scores are nonuniform.

\subsection{Improvements from Shrinked Leveraging and Unweighted Leveraging}
\label{sxn:empirical-shrinkedunweight}

Here, we will describe how our proposed SLEV and LEVUNW procedures can both 
lead to improvements relative to LEV and UNIF.
Recall that LEV can lead to large MSE by inflating very small leverage 
scores.
The SLEV procedure deals with this by considering a convex combination of 
the uniform distribution and the leverage score distribution, thereby 
providing a lower bound on the leverage scores; and the LEVUNW procedure 
deals with this by not rescaling the subproblem to be solved.

\begin{figure}[t]
  \centering
    \makebox{\includegraphics[scale=1]{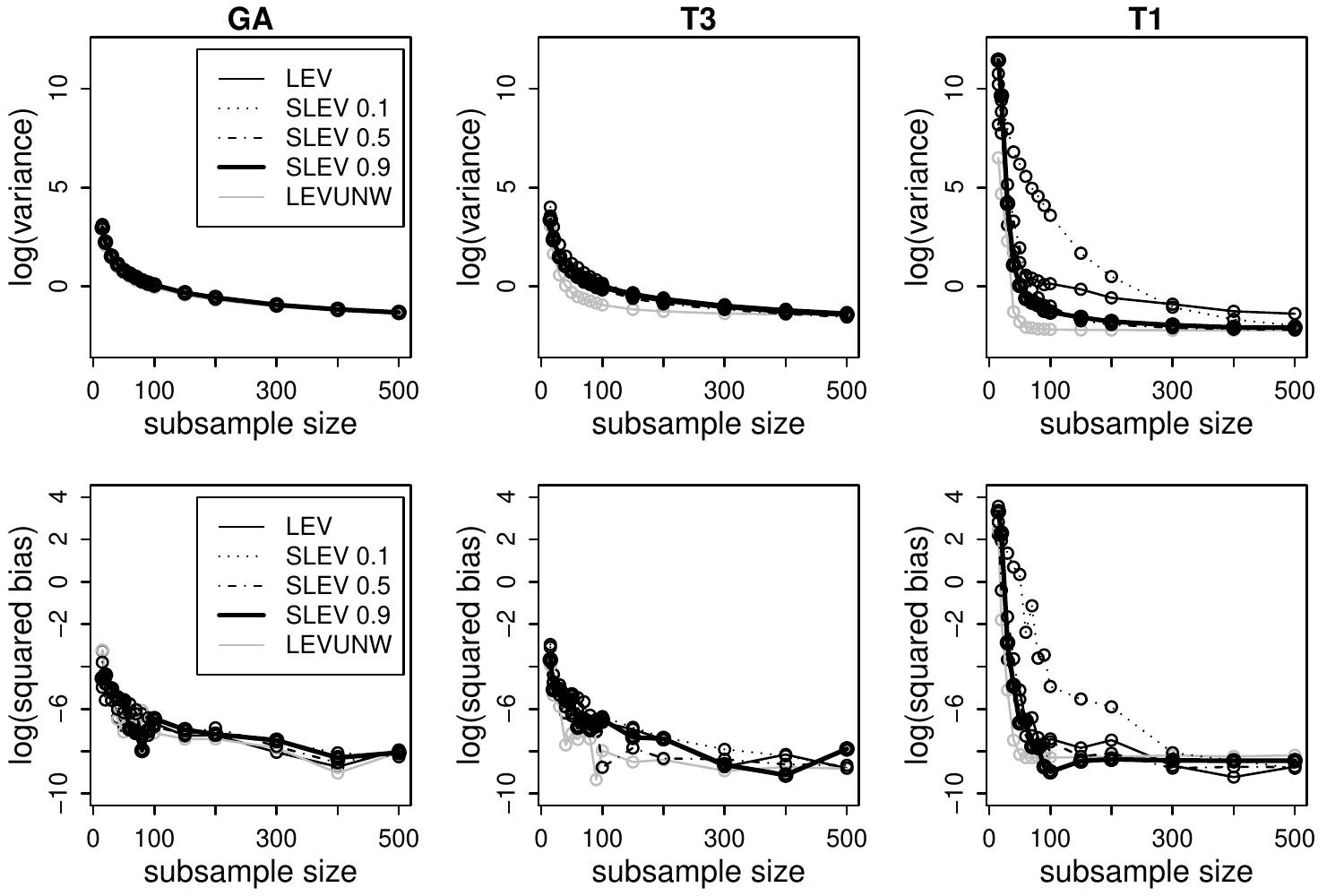}}
\vspace{-1em}
\caption{
(Improvements from SLEV and LEVUNW subsection.)
Comparison of variances and squared biases of the LEV, SLEV, and LEVUNW 
estimators in three data sets (GA, $T_3$, and $T_1$) for $n=1000$ and $p=10$.
Left panels are GA data;
Middle panels are $T_3$ data;
Right panels are $T_1$ data.
Grey lines are LEVUNW;
black lines are LEV;
dotted lines are SLEV with $\alpha=0.1$;
dotdashed lines are SLEV with $\alpha=0.5$;
thick black lines are SLEV with $\alpha=0.9$.
}
\label{fig:simual1}
\end{figure}
\begin{figure}[t]
  \centering
    \makebox{\includegraphics[scale=1]{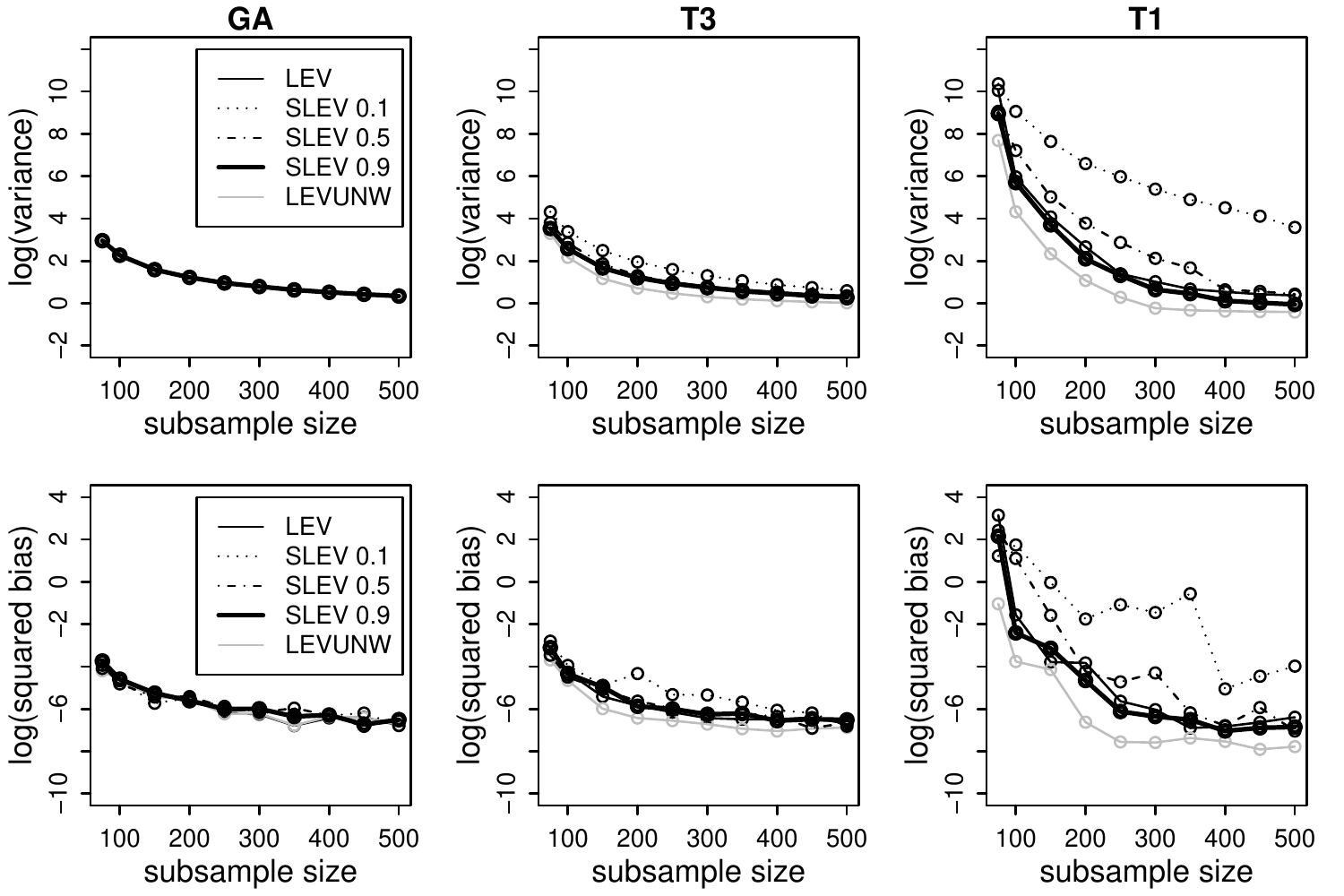}}
\vspace{-1em}
\caption{
(Improvements from SLEV and LEVUNW subsection.)
Same as Figure~\ref{fig:simual1}, except that $n=1000$ and $p=50$.
}
\label{fig:simual2}
\end{figure}
\begin{figure}[t]
  \centering
    \makebox{\includegraphics[scale=1]{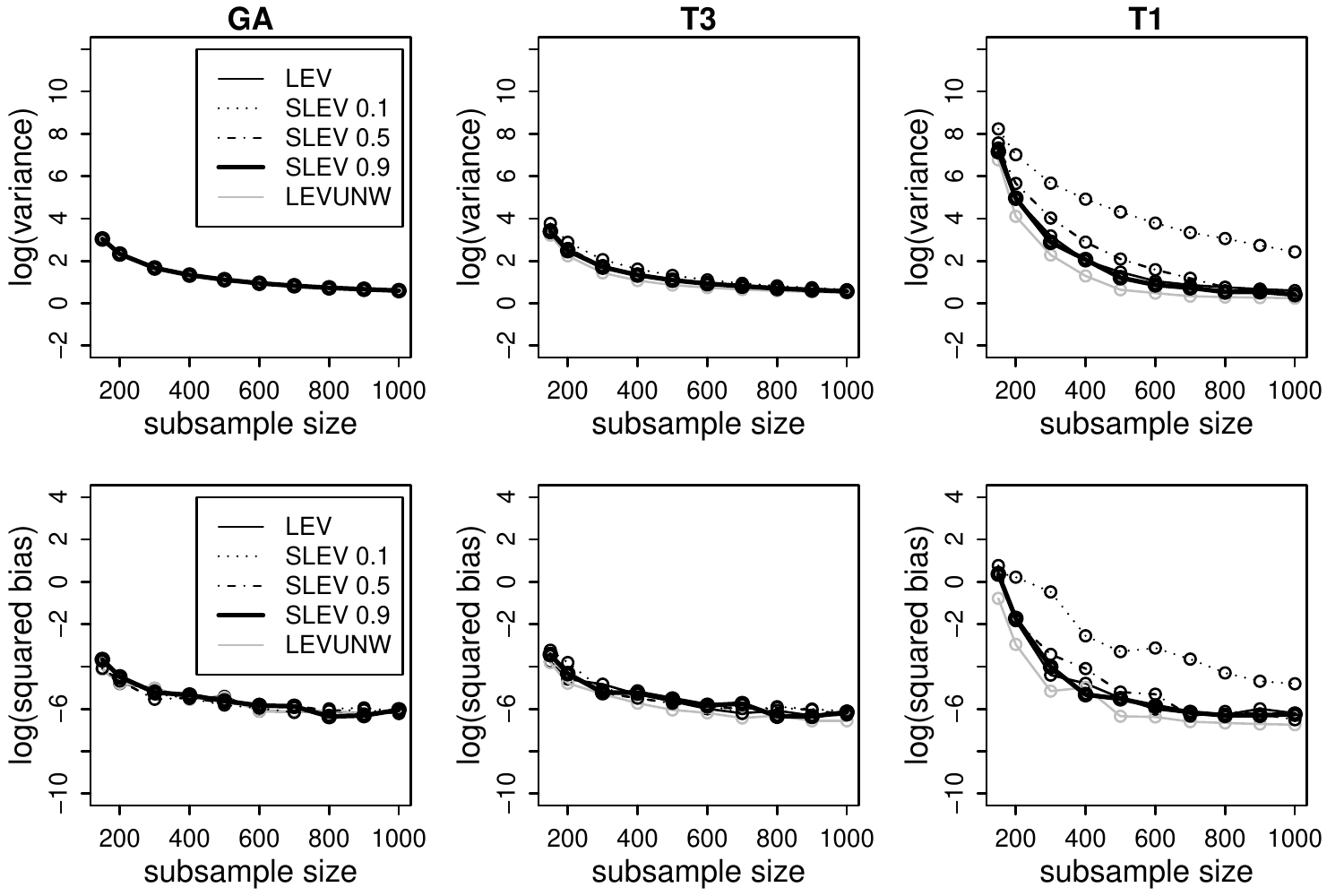}}
\vspace{-1em}
\caption{
(Improvements from SLEV and LEVUNW subsection.)
Same as Figure~\ref{fig:simual1}, except that $n=1000$ and $p=100$.
}
\label{fig:simual3}
\end{figure}

Consider Figures~\ref{fig:simual1}, \ref{fig:simual2}, and~\ref{fig:simual3}, 
which present the variance and bias for synthetic data matrices 
(for GA, $T_3$, and $T_1$ data) of size 
$n \times p$, where $n=1000$ and $p=10$, $50$, and $100$, respectively.
In each case, LEV, SLEV for three different values of the convex 
combination parameter $\alpha$, and LEVUNW were considered.
Several observations are worth making.
First of all, for GA data (left panel in these figures), all the results 
tend to be quite similar; but for $T_3$ data (middle panel) and even more so 
for $T_1$ data (right panel), differences appear.
Second, SLEV with $\alpha\simeq0.1$, i.e., when SLEV consists mostly of the 
uniform distribution, is notably worse in a manner similarly as with UNIF.
Moreover, there is a gradual decrease in both bias and variance for our 
proposed SLEV as $\alpha$ is increased; and when $\alpha\simeq0.9$ SLEV is 
slightly better than LEV.
Finally, our proposed LEVUNW often has the smallest bias and variance over a 
wide range of subsample sizes for both $T_3$ and $T_1$, although the effect 
is not major.
All in all, these observations are consistent with our main 
theoretical~results.

Consider next Figure~\ref{fig:simualev1}.
This figure examines the optimal convex combination choice for $\alpha$ in 
SLEV, and $\alpha$ is the x-axis in all the plots. 
Different column panels in Figure~\ref{fig:simualev1} correspond 
to different subsample sizes $r$.
Recall that there are two conflicting goals for SLEV: 
adding $(1-\alpha)/n$ to the small leverage scores will avoid 
substantially inflating the variance of the resulting estimate by samples 
with extremely small leverage scores; and doing so will 
lead to larger sample size $r$ in order to obtain bounds of the form 
Eqns.~(\ref{eqn:ls-bound-eq1}) and~(\ref{eqn:ls-bound-eq2}). 
Figure~\ref{fig:simualev1} plots the variance and bias for $T_1$ data for a 
range of parameter values and for a range of subsample sizes.
In general, one sees that using SLEV to increase the probability of choosing
small leverage components with $\alpha$ around $0.8 - 0.9$ (and relatedly 
shrinking the effect of large leverage components) has a beneficial effect 
on bias as well as variance.  
This is particularly true in two cases:
first, when the matrix is very rectangular, e.g., when the $p=10$, which is
consistent with the leverage score statistics from
Table~\ref{tab:synthetic-summary-stats}; and
second, when the subsample size $r$ is larger, as the results for $r=3p$ are 
much choppier (and for $r=2p$, they are still choppier).
As a rule of thumb, these plots suggest that choosing $\alpha=0.9$, and
thus using $\pi_i = \alpha \pi^{Lev}_{i} + (1-\alpha)/n$ as the importance 
sampling probabilities, strikes a balance between needing more samples and 
avoiding variance inflation.

One can also see in Figure~\ref{fig:simualev1} the grey lines, dots, and 
dashes, which correspond to LEVUNW for the corresponding values of $p$, 
that LEVUNW consistently has smaller variances than SLEV for all values of 
$\alpha$.
We should emphasize, though, that these are \emph{unconditional} 
biases and variances.  
Since LEVUNW is approximately unbiased relative to the full sample 
\emph{weighted} LS estimate $\hat{\sbf{\beta}}_{wls}$, however, there is a 
large bias away from the full sample \emph{unweighted} LS estimate 
$\hat{\sbf{\beta}}_{ols}$.  
This suggests that LEVUNW may be used when the primary goal is to infer the 
true $\sbf{\beta}_0$; but that when the primary goal is rather to 
approximate the full sample unweighted LS estimate, or when 
\emph{conditional} biases and variances are of interest, then SLEV may be 
more appropriate.  
We will discuss this in greater detail in 
Section~\ref{sxn:empirical-conditional} next. 

\begin{figure}[t]
  \centering
      \makebox{\includegraphics[scale=1]{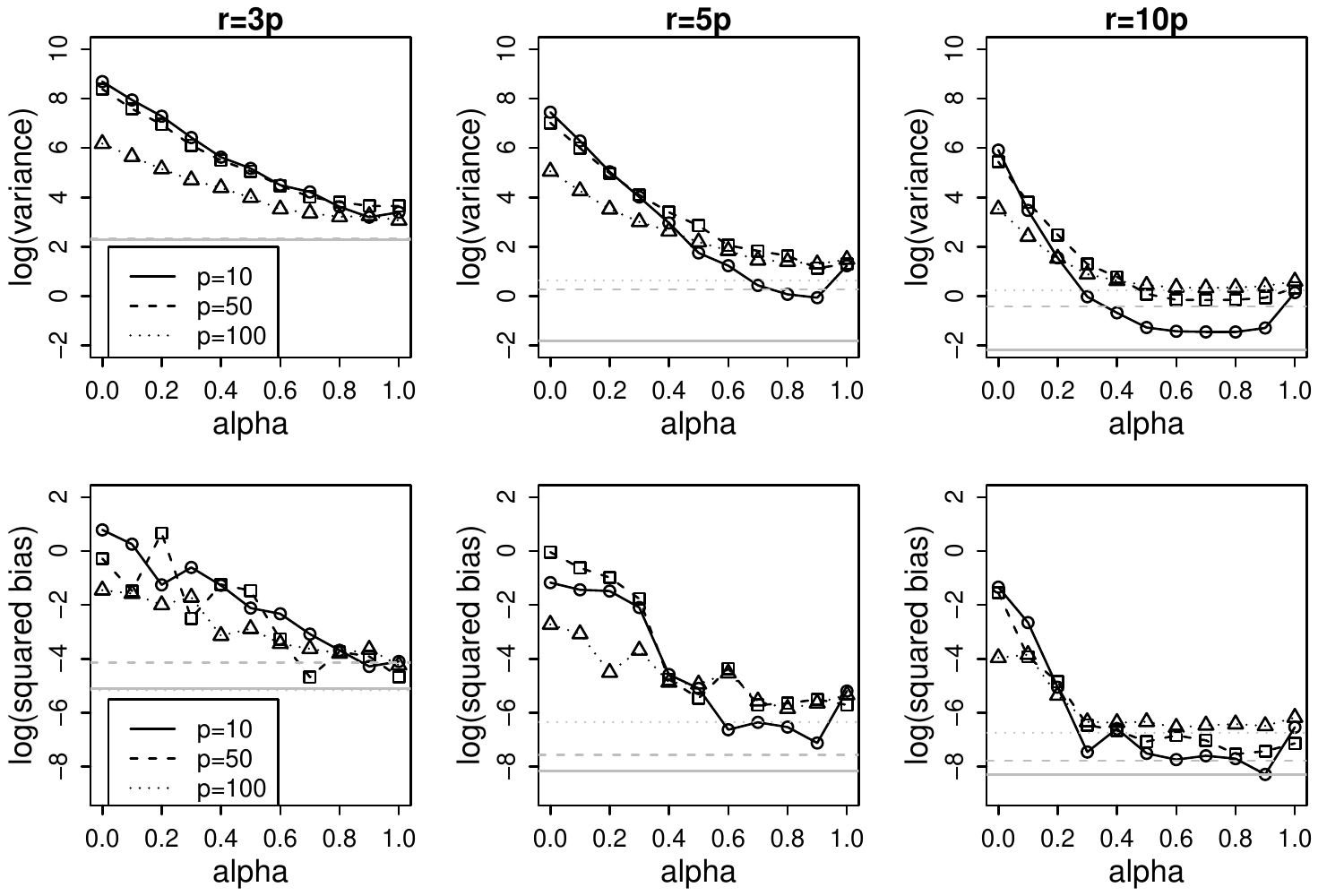}}
\vspace{-1em}
\caption{
(Improvements from SLEV and LEVUNW subsection.)
Varying $\alpha$ in SLEV.
Comparison of variances and squared biases of the SLEV estimator in 
data generated from $T_1$ with $n=1000$ and variable $p$.
Left panels are subsample size $r=3p$; 
Middle panels are $r=5p$; 
Right panels are $r=10p$.
Circles connected by black lines are $p=10$; 
squares connected by dash lines are $p=50$;
triangles connected by dotted lines are $p=100$.
Grey corresponds to the LEVUNW estimator.
}
\label{fig:simualev1}
\end{figure}

\subsection{Conditional Bias and Variance}
\label{sxn:empirical-conditional}

Here, we will describe the properties of the \emph{conditional} bias and 
variance under various subsampling estimators.
These will provide a more direct comparison with 
Eqns.~(\ref{eqn:slev-cond-bias}) and~(\ref{eqn:slev-wls-var}) from
Lemma~\ref{lem:bv} and the corresponding bounds from Lemma~\ref{lem:unwl-bv}.
These will also provide a more direct comparison with previous work that
has adopted an algorithmic perspective on algorithmic 
leveraging~\cite{DMM06,Mah-mat-rev_BOOK,DMMW12_JMLR}.

\begin{figure}[t]
  \centering
  \makebox{\includegraphics[scale=1]{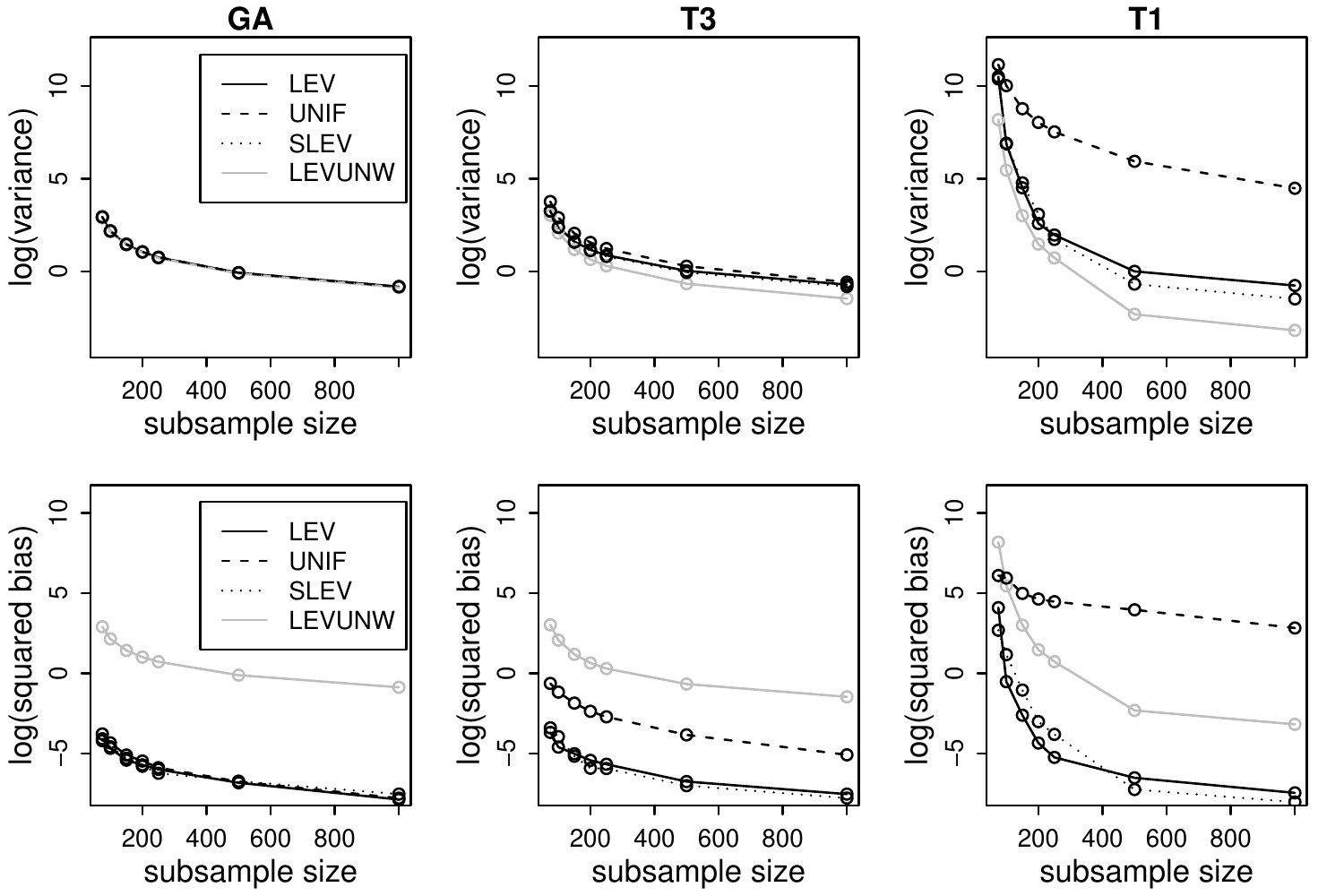}}
\vspace{-1em}
\caption{
(Conditional Bias and Variance subsection.)
Comparison of \emph{conditional} variances and squared biases of the 
LEV and UNIF estimators in three data sets (GA, $T_3$, and $T_1$) for 
$n=1000$ and $p=50$.
Left panels are GA data;
Middle panels are $T_3$ data;
Right panels are $T_1$ data.
Upper panels are Variances; 
Lower panels are Squared Bias.
Black lines for LEV estimate; 
dash lines for UNIF estimate; 
grey lines for LEVUNW estimate; 
dotted lines for SLEV estimate with~$\alpha=0.9$.
}
\label{fig:simuline2c}
\end{figure}

Consider Figure~\ref{fig:simuline2c}, which presents our main empirical 
results for conditional biases and variances.
As before, matrices were generated from GA, $T_3$ and $T_1$; and we 
calculated the empirical bias and variance of UNIF, LEV, SLEV with 
$\alpha=0.9$, and LEVUNW---in all cases, conditional on the empirical data 
$\sbf{y}$.
Several observations are worth making.
First, for GA the variances are all very similar the same; and the biases 
are also, with the exception of LEVUNW.
This is expected, since by the conditional expectation bounds from 
Lemma~\ref{lem:unwl-bv}, LEVUNW is approximately unbiased, relative to the 
full sample \emph{weighted} LS estimate $\hat{\sbf{\beta}}_{wls}$---and thus
there should be a large bias away from the full sample unweighted LS 
estimate.
Second, for $T_3$ and even more prominently for $T_1$, the variance of 
LEVUNW is less than that for the other estimators.
Third, when the leverage scores are very nonuniform, as with $T_1$, the 
relative merits of UNIF versus LEVUNW depend on the subsample size $r$.
In particular, the bias of LEVUNW is larger than that of even UNIF for very 
aggressive downsampling; but it is substantially less than UNIF for moderate 
to large sample sizes.

Based on these and our other results, our default recommendation is to 
use SLEV (with either exact or approximate leverage scores) with
$\alpha \approx 0.9$: it is no more than slightly worse than LEVUNW when 
considering unconditional biases and variances, and it can be much better 
than LEVUNW when considering conditional biases and variances.

\section{Additional Empirical Evaluation}
\label{sxn:moreempirical}

In this section, we provide additional empirical results (of a more 
specialized nature than those presented in Section~\ref{sxn:empirical}).
Here is a brief outline of the main results of this section.
\begin{itemize}
\item
In Section~\ref{sxn:empirical-singular}, we will consider the synthetic data, 
and we will describe what happens when the subsampled problem looses rank.
This can happen if one is \emph{extremely} aggressive in downsampling with 
SLEV; but it is much more common with UNIF, even if one samples many 
constraints.
In both cases, the behavior of bias and variance is very different than when 
rank is preserved.
\item
Then, in Section~\ref{sxn:empirical-fast}, we will summarize our results on synthetic data when 
the leverage scores are computed approximately with the fast approximation algorithm 
of~\cite{DMMW12_JMLR}.
Among other things, we will describe the running time of this algorithm, 
illustrating that it can solve larger problems than can be solved with 
traditional deterministic methods; and we will evaluate the unconditional 
bias and variance of SLEV when this algorithm is used to approximate the 
leverage scores.
\item
Finally, in Section~\ref{sxn:empirical-real}, we will consider real data, and we will present our results for 
the conditional bias and variance for two data sets that are drawn from 
our previous work in two genetics applications.
One of these has very uniform leverage scores, and the other has moderately 
nonuniform leverage scores; and our results from the synthetic data hold 
also in these realistic applications.
\end{itemize}

\subsection{Leveraging and Uniform Estimates for Singular Subproblems}
\label{sxn:empirical-singular}

Here, we will describe the properties of LEV versus UNIF for situations in 
which rank is lost in the construction of the subproblem.
That is, in some cases, the subsampled matrix, $X^{*}$, may have column rank 
that is smaller than the rank of the original matrix $X$, and this leads to 
a singular $X^{*T}X^{*}=X^TWX$.
Of course, the LS solution of the subproblem can still be solved, but there 
will be a ``bias'' due to the dimensions that are not represented in the 
subsample.
(We use the Moore-Penrose generalized inverse to compute the estimators when 
rank is lost in the construction of the subproblem.)
Before describing these results, recall that algorithmic leveraging (in 
particular, LEV, but it holds for SLEV as well) guarantees that this will 
\emph{not} happen in the following sense:
if roughly $O(p \log p)$ rows of $X$ are sampled using an importance sampling
distribution that approximates the leverage scores in the sense of 
Eqn.~(\ref{eqn:approx-lev-score-probs}), then with very high probability the 
matrix $X^{*}$ does not loose rank~\cite{DMM06,Mah-mat-rev_BOOK,DMMW12_JMLR}.
Indeed, this observation is crucial from the algorithmic perspective, 
i.e., in order to obtain relative-error bounds of the form of 
Eqns.~(\ref{eqn:ls-bound-eq1}) and~(\ref{eqn:ls-bound-eq2}), and thus it was 
central to the development of algorithmic leveraging.
On the other hand, if one downsamples more aggressively, e.g., if one samples
only, say, $p+100$ or $p+10$ rows, or if one uses uniform sampling when the 
leverage scores are very nonuniform, then it is possible to loose rank.
Here, we examine the statistical consequences of this.

\begin{figure}[t]
  \centering
    \makebox{\includegraphics[scale=0.9]{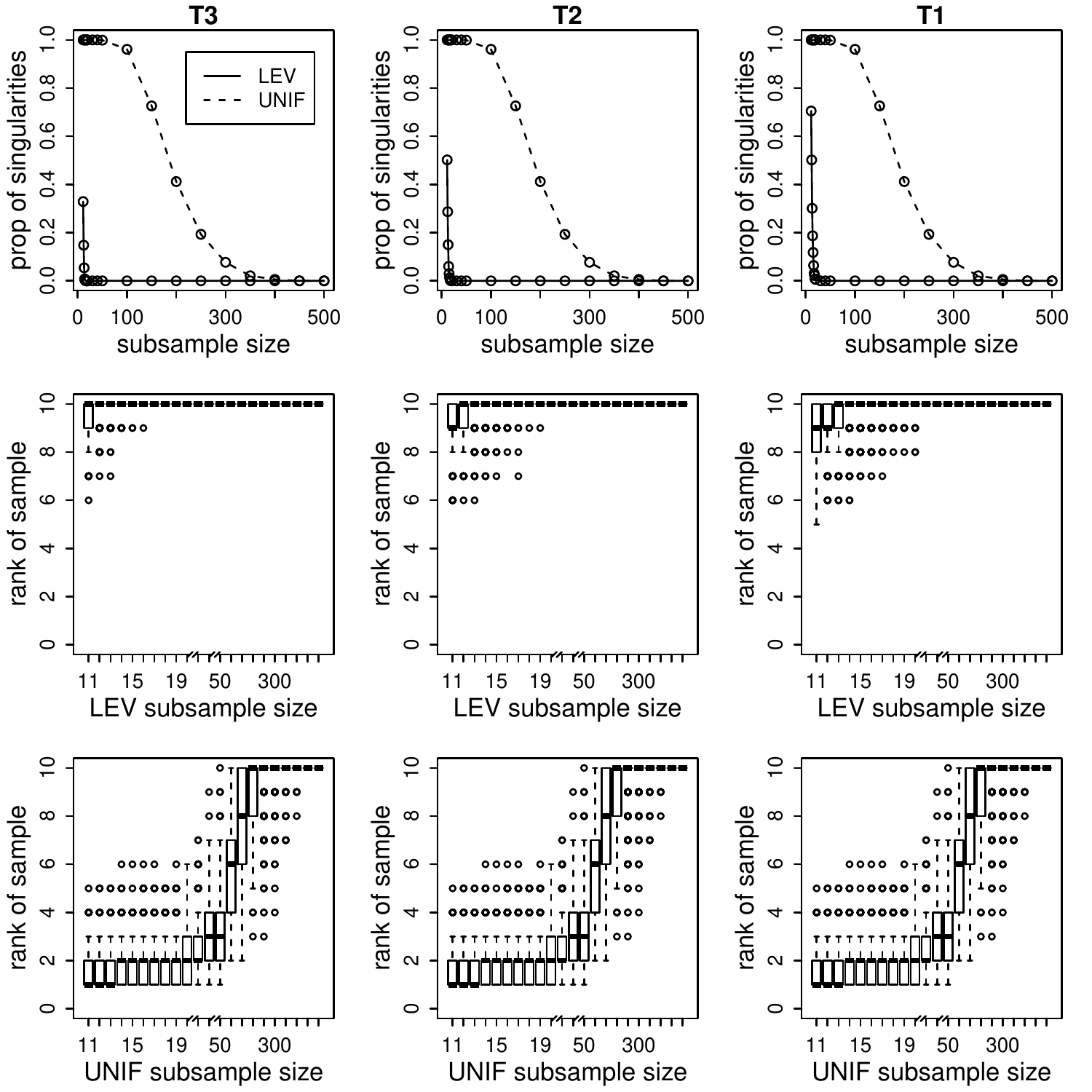}}
\vspace{-1em}
\caption{
Comparison of LEV and UNIF when rank is lost in the sampling process 
($n=1000$ and $p=10$ here). 
Left panels are $T_3$; 
Middle panels are $T_2$;
Right panels are $T_1$.
Upper panels are proportion of singular $X^{T}WX$, out of $500$ trials, 
for both LEV (solid lines) and UNIF (dashed lines); 
Middle panels are boxplots of ranks of $500$ LEV subsamples; 
Lower panels are boxplots of ranks of $500$ UNIF subsamples.
Note the nonstandard scaling of the X axis.
}
\label{fig:simurepeata}
\end{figure}

\begin{figure}[t]
  \centering
    \makebox{\includegraphics[scale=0.9]{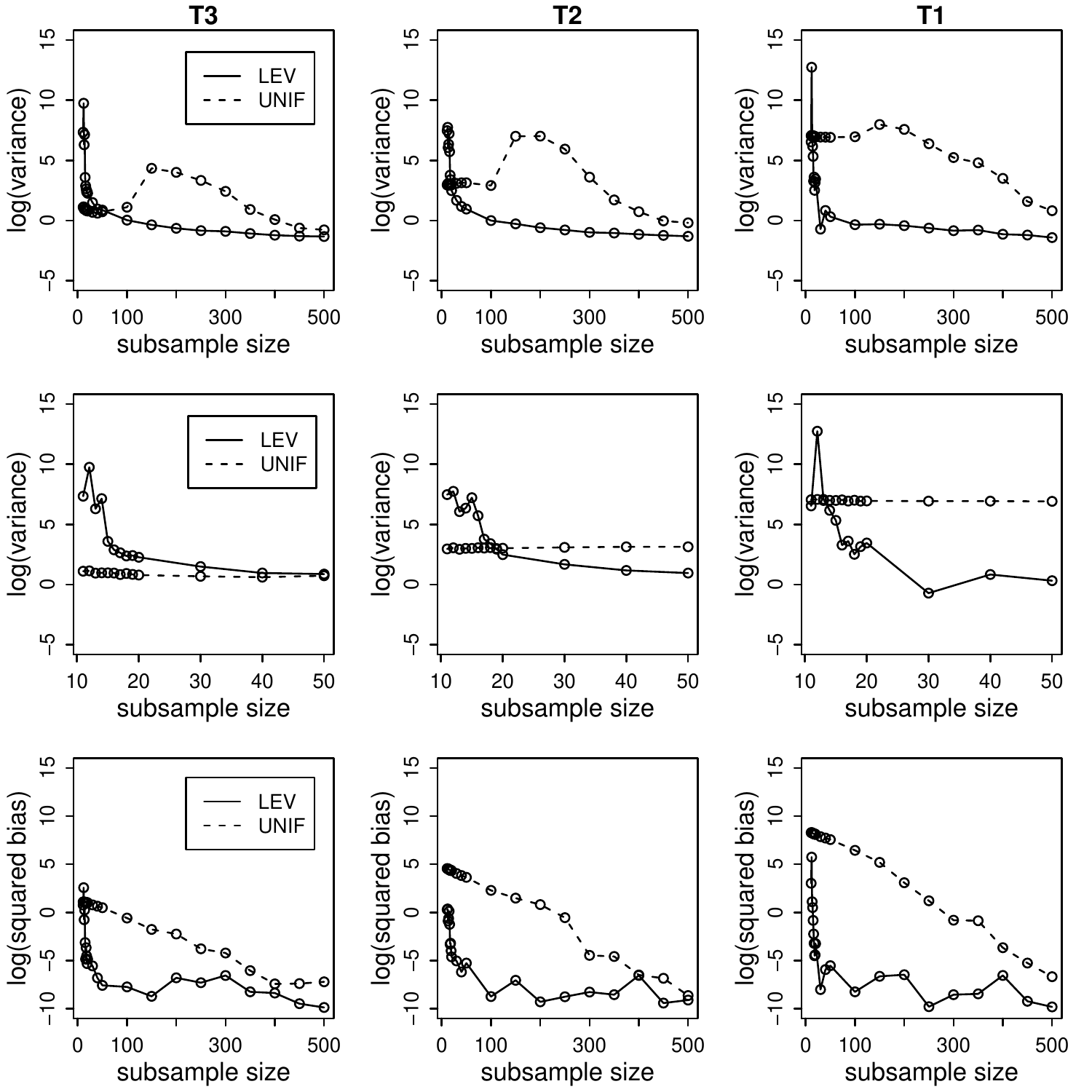}}
\vspace{-1em}
\caption{
Comparison of LEV and UNIF when rank is lost in the sampling process 
($n=1000$ and $p=10$ here). 
Left panel are $T_3$;
Middle panels are $T_2$;
Right panels are $T_1$. 
Upper panels are logarithm of variances of the estimates; 
Middle panels are logarithm of variances, zoomed-in on the X-axis;
Lower panels are logarithm of squared bias of the estimates.
Black line for LEV; 
Dash line for UNIF.
}
\label{fig:simurepeat}
\end{figure}

We have observed this phenomenon with the synthetic data for 
both UNIF as well as for leverage-based sampling procedures; but the 
properties are somewhat different depending on the sampling procedure.
To illustrate both of these with a single synthetic example, we first 
generated a $1000 \times 10$ matrix from multivariate $t$-distribution with 
$3$ (or $2$ or $1$, denoted $T_3$, $T_2$, and $T_1$, respectively) degrees of 
freedom and covariance matrix $\Sigma_{ij}=2 \times 0.5^{|i-j|} $;
we then calculated the leverage scores of all rows; and finally we formed 
the matrix $X$ was by keeping the $50$ rows with highest leverage scores and 
replicating $950$ times the row with the smallest leverage score.
(This is a somewhat more realistic version of the toy 
\textbf{Worst-case Matrix} that is described in 
Section~\ref{sxn:interlude-toy}.)
We then applied LEV and UNIF to the data sets with different subsample 
sizes, as we did for the results summarized in 
Section~\ref{sxn:empirical-levvunif}.
Our results are summarized in Figure~\ref{fig:simurepeata}
and~\ref{fig:simurepeat}.

The top row of Figure~\ref{fig:simurepeata} plots the fraction of singular 
$X^{T}WX$, out of $500$ trials, for both LEV and UNIF;
from left to right, results for $T_3$, $T_2$, and $T_1$ are shown.
Several points are worth emphasizing.
First, both LEV and UNIF loose rank if the downsampling is sufficiently 
aggressive.
Second, for LEV, as long as one chooses more than roughly $20$ (or less for 
$T_2$ and $T_1$), i.e., the ratio $r/p$ is at least roughly $2$, then rank 
is \emph{not} lost; but for uniform sampling, one must sample a \emph{much} 
larger fraction of the data.
In particular, when fewer than $r=100$ samples are drawn then nearly all of 
the subproblems constructed with the UNIF procedure are singular, and it is 
not until more than $r=300$ that nearly all of the subproblems are not 
singular.
Although these particular numbers depend on the particular data, needing to 
draw many more samples with UNIF than with LEV in order to preserve rank is 
a very general phenomenon.
The middle row of Figure~\ref{fig:simurepeata} shows the boxplots of rank 
for the subproblem for LEV for those $500$ tries; and the bottom row shows 
the boxplots of the rank of the subproblem for UNIF for those $500$ tries.
Note the unusual scale on the X-axis designed to highlight the lost rank
data for both LEV as well as UNIF.
These boxplots illustrate the sigmoidal distribution of ranks that obtained 
by UNIF as a function of the number of samples and the less severe beginning 
of the sigmoid for LEV; and they also show that when subproblems are 
singular, then often many dimensions fail to be captured.
All in all, LEV outperforms UNIF, especially when the 
leverage scores are nonuniform.

Figure~\ref{fig:simurepeat} illustrates the variance and bias of the 
corresponding estimators.
In particular, the upper panels plot the logarithm of variances;
the middle panels plot the same quantities, except that it is zoomed-in on 
the X-axis; and the lower panels plot the logarithm of squared bias.
As before, the left/middle/right panels present results for the 
$T_3$/$T_2$/$T_1$ data, respectively.
The behavior here is very different that that shown in 
Figures~\ref{fig:simuline1}, \ref{fig:simuline2}, and~\ref{fig:simuline3}; 
and several observations are worth making.
First, for all three models and for both LEV and UNIF, when the 
downsampling is very aggressive, e.g, $r=p+5$ or $r=p+10$, then the bias is 
comparable to the variance.
That is, since the sampling process has lost dimensions, the linear 
approximation implicit in our Taylor expansion is violated.
Second, both bias and variance are worse for $T_1$ than for $T_2$ than for 
$T_3$, which is consistent with Table~\ref{tab:synthetic-summary-stats}, but 
the effect is minor; and the bias and variance are generally much worse for 
UNIF than for LEV. 
Third, as $r$ increases, the variance for UNIF increases, hits a maximum 
and then decreases; and at the same time the bias for UNIF gradually 
decreases.
Upon examining the original data, the reason that there is very little 
variance initially is that most of the subsamples have rank $1$ or $2$; then 
the variance increases as the dimensionality of the subsamples increases; 
and then the variance decreases due to the $1/r$ scaling, as we saw in the 
plots in Section~\ref{sxn:empirical-levvunif}.
Fourth, as $r$ increases, both the variance and bias of LEV decrease, as we 
saw in Section~\ref{sxn:empirical-levvunif}; but in the aggressive 
downsampling regime, i.e., when $r$ is very small, the variance of LEV is 
particularly ``choppy,'' and is actually worse than that of UNIF, perhaps 
also due to rank deficiency issues.

\subsection{Approximate Leveraging via the Fast Leveraging Algorithm}
\label{sxn:empirical-fast}

Here, we will describe using the fast randomized algorithm 
from~\cite{DMMW12_JMLR} to compute approximations to the leverage scores 
of $X$, to be used in place of the exact leverage scores in LEV, SLEV, and
LEVUNW.
To start, we provide a brief description of the algorithm 
of~\cite{DMMW12_JMLR}, which takes as input an arbitrary $n \times p$ 
matrix $X$. 
\begin{itemize}
\item
Generate an $r_1 \times n$ random matrix $\Pi_1$ 
and a $p \times r_2$ random matrix $\Pi_2$.
\item
Let $R$ be the $R$ matrix from a QR decomposition of $\Pi_1 X$.
\item
Compute and return the leverage scores of the matrix $X R^{-1} \Pi_2$.
\end{itemize}
For appropriate choices of $r_1$ and $r_2$, if one chooses $\Pi_1$ to be a 
Hadamard-based random projection matrix, then this algorithm runs in 
$o(np^2)$ time, and it returns $1\pm\epsilon$ approximations to all the 
leverage scores of $X$~\cite{DMMW12_JMLR}.
In addition, with a high-quality implementation of the Hadamard-based random 
projection, this algorithm runs faster than traditional deterministic 
algorithms based on \textsc{Lapack} for matrices as small as several 
thousand by several hundred~\cite{AMT10,GM13_TR}.

We have implemented in the software environment R two variants of this fast 
algorithm of \cite{DMMW12_JMLR}, and we have compared it with QR-based 
deterministic algorithms also supported in R for computing the leverage 
scores exactly.
In particular, the following results were obtained on a PC with Intel Core 
i7 Processor and 6 Gbytes RAM running Windows 7, on which we used the 
software package R, version 2.15.2.
In the following, we refer to the above algorithm as BFast (the Binary Fast 
algorithm) when (up to normalization) each element of $\Pi_1$ and $\Pi_2$ is 
generated i.i.d. from $\{-1, 1\}$ with equal sampling probabilities; and we 
refer to the above algorithm as GFast (the Gaussian Fast algorithm) when 
each element of $\Pi_1$ is generated i.i.d. from a Gaussian distribution with 
mean zero and variance $1/n$ and each element of $\Pi_2$ is generated i.i.d. 
from a Gaussian distribution with mean zero and variance $1/p$.
(In particular, note that here we do not consider Hadamard-based projections 
for $\Pi_1$ or more sophisticated parallel and distributed implementations 
of these algorithms~\cite{AMT10,MSM11_TR,GM13_TR,YMM13_TR}.)

\begin{figure}[t]
  \centering
   \makebox{\includegraphics[scale=0.8]{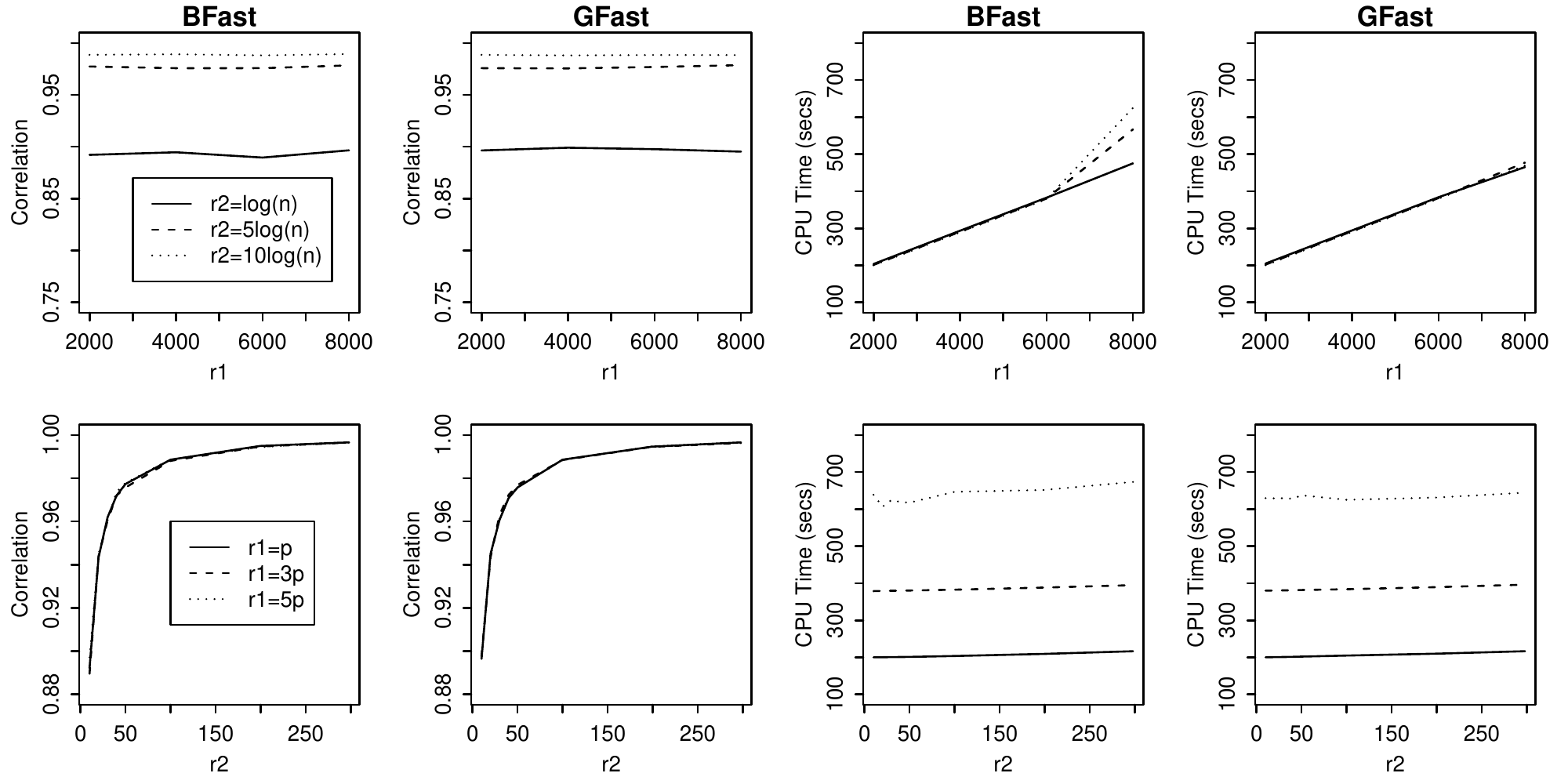}}
\vspace{-1em}
\caption{
(Fast Leveraging Algorithm subsection.)
Effect of approximating leverage scores using BFast and GFast for varying 
parameters.
Upper panels: Varying parameter $r_1$ for fixed $r_2$, where 
$r_2= \log(n) $ (black lines), 
$r_2=5\log(n) $ (dashed lines), and
$r_2=10\log(n)$ (dotted lines).
Lower panels: Varying parameter $r_2$ for fixed $r_1$, where 
$r_1= p$ (black lines), 
$r_1=3p$ (dashed lines), and
$r_1=5p$ (dotted lines).
Left two panels:  Correlation between exact leverage scores and leverage 
scores approximated using BFast and GFast, for varying $r_1$ and $r_2$.
Right two panels: CPU time for varying $r_1$ and $r_2$, using BFast and GFast.
}
\label{fig:fastcor}
\end{figure}

To illustrate the behavior of this algorithm as a function of its 
parameters, we considered synthetic data where the $20,000 \times 1,000$ 
design matrix $X$ is generated from  $T_1$ distribution. 
All the other parameters are set to be the same as before, except
$\Sigma_{ij}= 0.1$, for $i \neq j$, and $\Sigma_{ii}= 2$.  
We then applied BFast and GFast with varying $r_1$ and $r_2$ to the data.  
In particular, we set $r_1=p, 1.5p, 2p, 3p, 5p$, where $p=1,000$, and we set
$r_2= \kappa\log(n)$, for $\kappa=1, 2, 3, 4, 5, 10, 20$,
where $n=20,000$.
See Figure~\ref{fig:fastcor}, which presents both a summary of the 
correlation between the approximate and exact leverage scores as well as a
summary of the running time for computing the approximate leverage scores, 
as $r_1$ and $r_2$ are varied for both BFast and GFast.
We can see that the correlations between approximated and exact leverage 
scores are not very sensitive to varying $r_1$, whereas the running time 
increases roughly linearly for increasing $r_1$.  
In contrast, the correlations between approximated and exact leverage scores 
increases rapidly for increasing $r_2$, whereas the running time does not 
increase much when $r_2$ increases. 
These observations suggest that we may use a combination of small $r_1$ and 
large $r_2$ to achieve high-quality approximation and short running~time.

\begin{figure}[t]
  \centering
   \makebox{\includegraphics[scale=1]{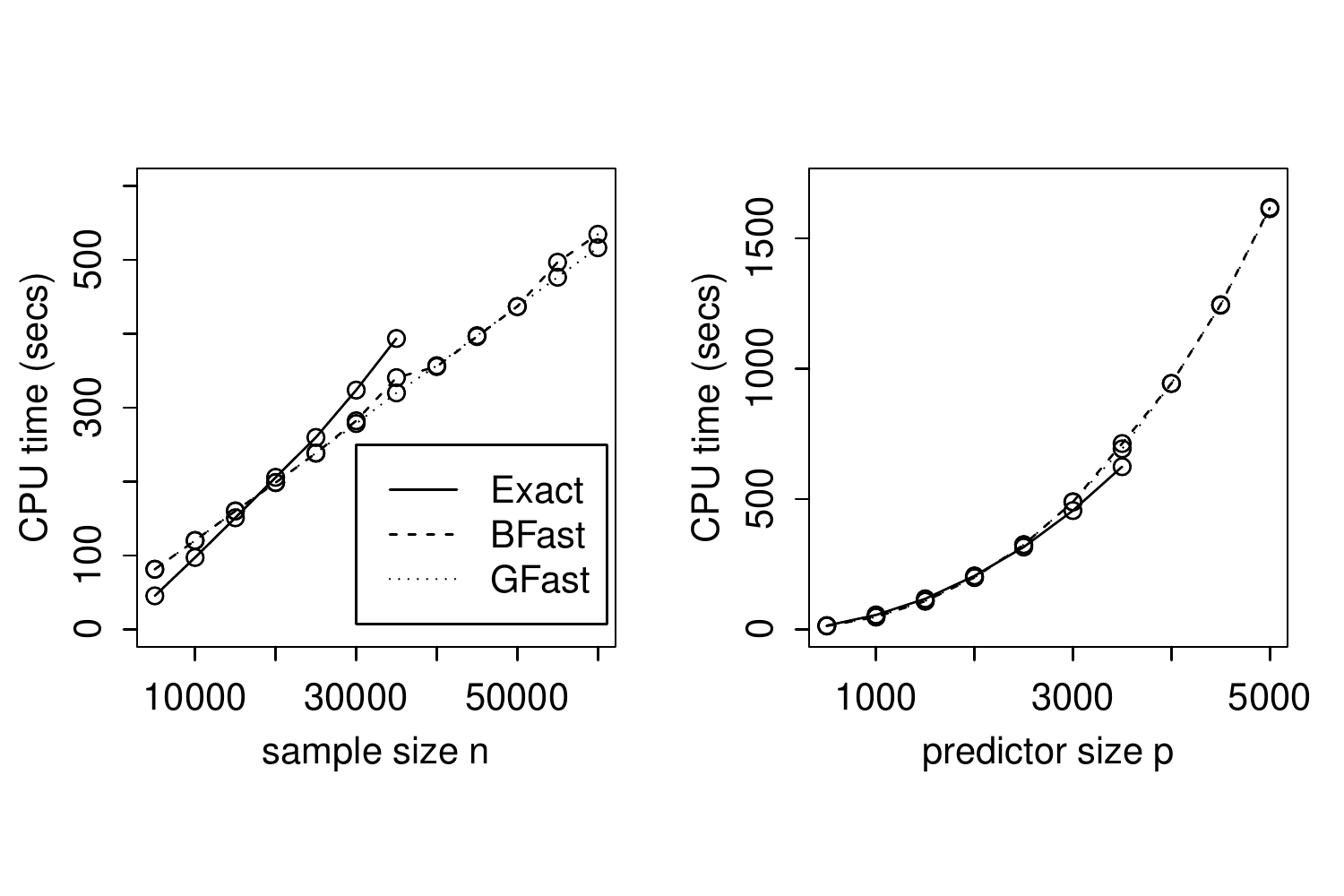}}
\vspace{-1em}
\caption{
(Fast Leveraging Algorithm subsection.)
CPU time for calculating exact leverage scores and approximate leverage 
scores using the BFast and GFast versions of the fast algorithm 
of~\cite{DMMW12_JMLR}.
Left panel is CPU time for varying sample size $n$ for fixed predictor size 
$p=500$;
Right panel is CPU time for varying predictor size $p$ for fixed sample size 
$n=2000$.
Black lines connect the CPU time for calculating exact leverage scores; 
dash lines connect the CPU time for using GFast to approximate the leverage 
scores;
dotted lines connect the CPU time for using BFast to approximate the 
leverage scores.
}
\label{fig:fasttime}
\end{figure}

Next, we examine the running time of the approximation algorithms for 
computing the leverage scores. 
Our results for running times are summarized in Figure~\ref{fig:fasttime}.
In that figure, we plot the running time as sample size $n$ and predictor 
size $p$ are varied for BFast and GFast. 
We can see that when the sample size is very small, the computation time of 
the fast algorithms is slightly worse than that of the exact algorithm.   
(This phenomenon is primarily since the fast algorithm requires additional 
projection and matrix multiplication steps, which dominate the running time 
for very small matrices.)
On the other hand, when the sample size is larger than ca. $20,000$, the 
computation time of the fast approximation algorithms becomes slightly less 
expensive than that of exact algorithm.
Much more significantly, when the sample size is larger than roughly $35,000$, 
the exact algorithm requires more memory than our standard R environment can 
provide, and thus it fails to run at all. 
In contrast, the fast algorithms can work with sample size up to 
roughly~$60,000$.

That is, the use of this randomized algorithm to approximate the leverage 
scores permits us to work with data that are roughly $1.5$ times larger in 
$n$ or $p$, even when a simple vanilla implementation is provided in the
R environment.
(If one is interested in much larger inputs, e.g., with $n=10^6$ or more, 
then one should probably not work within R and instead use Hadamard-based 
random projections for $\Pi_1$ and/or the use of more sophisticated 
methods, such as those described in~\cite{AMT10,MSM11_TR,GM13_TR,YMM13_TR}; 
here we simply evaluate an implementation of these methods in R.)
The reason that BFast and GFast can run for much larger input is likely that the computational bottleneck
for the exact algorithm is a QR decomposition, while the computational 
bottleneck for the fast randomized algorithms is the matrix-matrix 
multiplication step.

\begin{figure}[t]
  \centering
   \makebox{\includegraphics[scale=1]{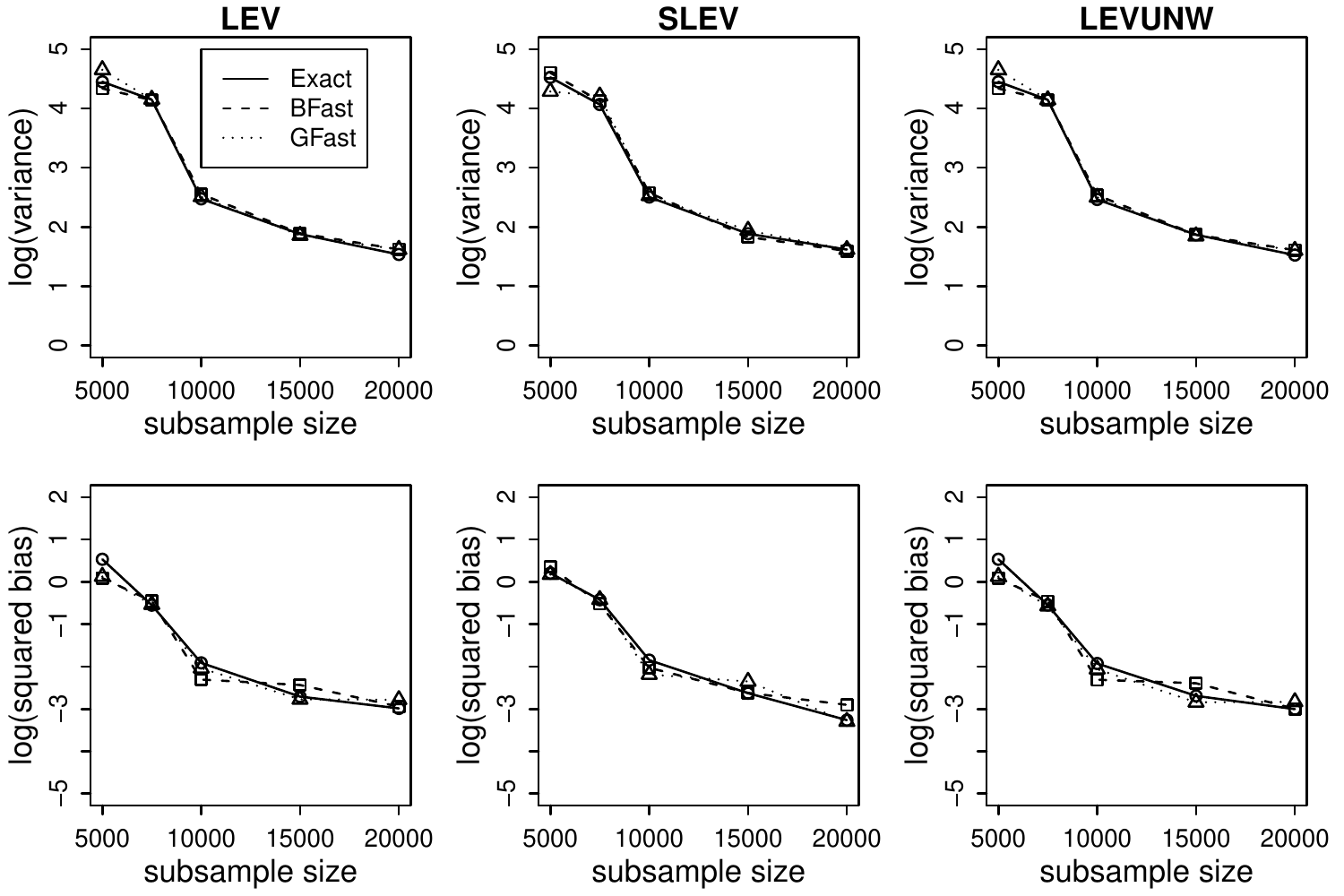}}
\vspace{-1em}
\caption{
(Fast Leveraging Algorithm subsection.)
Comparison of variances and squared biases of the LEV, SLEV, and LEVUNW
estimators in $T_3$ data sets  for $n=20000$ and $p=5000$ using BFast and 
GFast versions of the fast algorithm of~\cite{DMMW12_JMLR}.
Left panels are LEV estimates;
Middle panels are SLEV estimates;
Right panels are LEVUNW estimates.
Black lines are exact algorithm;
dash lines are BFast;
dotted lines are GFast.
}
\label{fig:fastBV2}
\end{figure}

Finally, we evaluate the bias and variance of LEV, SLEV and LEVUNW 
estimates where the leverage scores are calculated using exact algorithm, 
BFast, and GFast.
In Figure~\ref{fig:fastBV2}, we plot the variance and squared bias for 
$T_3$ data sets. 
(We have observed similar but slightly smoother results for the Gaussian 
data sets and similar but slightly choppier results for the $T_1$ data sets.) 
Observe that the variances of LEV estimates where the leverage scores are 
calculated using exact algorithm, BFast, and GFast are almost identical; and 
this observation is also true for SLEV and LEVUNW estimates.
All in all, using the fast approximation algorithm of~\cite{DMMW12_JMLR} 
to compute approximations to the leverage scores for use in LEV, SLEV, and 
LEVUNW leads to improved algorithmic performance, while achieving nearly 
identical statistical results as LEV, SLEV, and LEVUNW when the exact 
leverage scores are~used.

\subsection{Illustration of the Method on Real Data}
\label{sxn:empirical-real}

Here, we provide an illustration of our methods on two real data sets drawn
from two problems in genetics with which we have prior
experience~\cite{Dalpiaz:13,CUR_PNAS}.
The first data set has relatively uniform leverage scores, 
while the second data set has somewhat more nonuniform leverage scores.
These two examples simply illustrate that observations we made on the 
synthetic data also hold for more realistic data that we have studied 
previously.
For more information on the application of these ideas in genetics, see 
previous work on PCA-correlated SNPs~\cite{Paschou07b,Paschou10b}.

\subsubsection{Linear model for bias correction in RNA-Seq data}

In order to illustrate how our methods perform on a real data set with nearly
uniform leverage scores, we consider an RNA-Seq data set containing 
$n=51,751$ read counts from embryonic mouse stem cells \cite{Cloonan:08s}.
Recall that RNA-Seq is becoming the major tool for transcriptome analysis;
it produces digital signals by obtaining tens of millions of short reads; 
and after being mapped to the genome, RNA-Seq data can be summarized by a 
sequence of short-read counts.
Recent work found that short-read counts have significant sequence 
bias~\cite{Li:10}. 
Here, we consider a simplified linear model of \cite{Dalpiaz:13} for 
correcting sequence bias in RNA-Seq.
Let $n_{ij}$ denote the counts of reads that are mapped to the genome 
starting at the $j$th nucleotide of the $i$th gene, where
$i=1, 2, \ldots, 100$ and $ j=1, \ldots, L_i $.
We assume that the log transformed count of reads, 
$y_{ij}=\log(n_{ij}+0.5)$, depends on $40$ nucleotides in the neighborhood, 
denoted as $b_{ij,-20}, b_{ij,-19}, \ldots,  b_{ij,18}, b_{ij,19}$ through 
the following linear model:
$
y_{ij}=\alpha  + \sum_{k=-20}^{19}\sum_{h\in \cal{H}}\beta_{kh}I(b_{ij,k}=h) +\epsilon_{ij},
$
where ${\cal{H}}=\{A, C, G\}$, where $T$ is used as the baseline level, 
$\alpha$ is the grand mean, $I(b_{ij,k} = h)$ equals to 1 if the $k$th 
nucleotide of the surrounding sequence is $h$, and $0$ otherwise, 
$\beta_{kh}$ is the coefficient of the effect of nucleotide $h$ occurring in 
the $k$th position, and $\epsilon_{ij}\sim N(0,\sigma^2)$.
This linear model uses $p=121$ parameters to model the sequence bias of read
counts. 
For $n=51,751$, model-fitting via LS is time-consuming.

\begin{figure}[t]
  \centering
   \makebox{\includegraphics[scale=1]{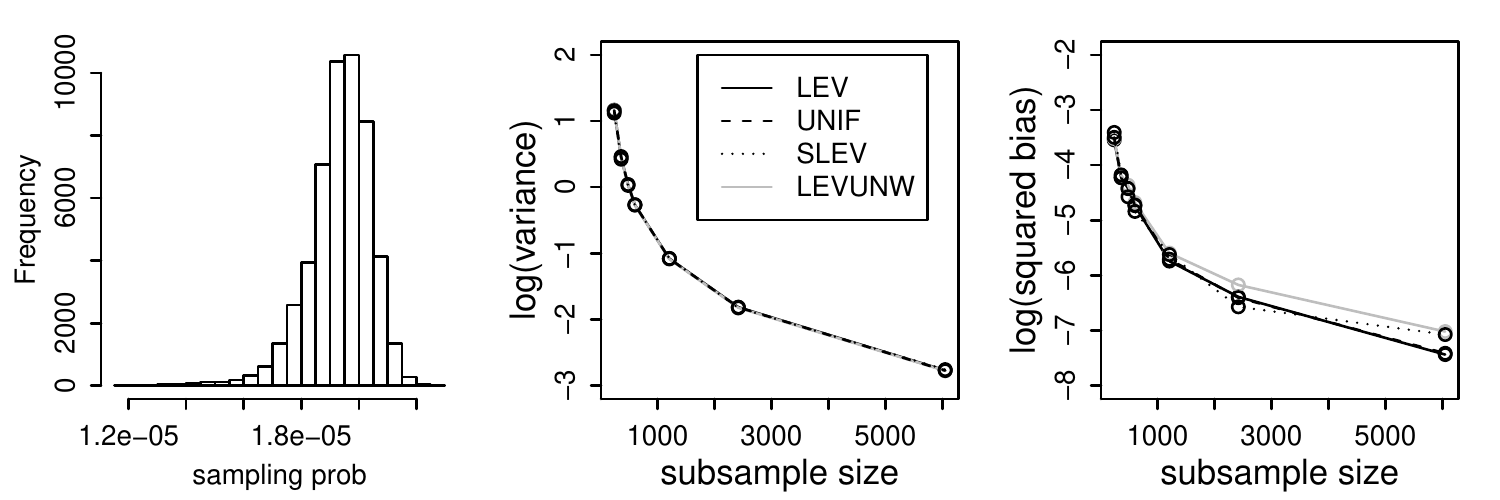}}
\vspace{-1em}
\caption{Empirical results for real data.
Left panel is the histogram of the leverage score sampling probabilities for 
the RNA-Seq data 
(the largest leverage score is $2.25 \times 10^{-5}$, and the mean is 
$1.93 \times 10^{-5}$, i.e., the largest is only slightly larger than the 
mean);
Middle panel is the empirical \emph{conditional} variances
of the LEV, UNIF, LEVUNW, and SLEV estimates;
Right panel is the empirical \emph{conditional} biases.
Black lines for LEV;
dash lines for UNIF;
grey lines for LEVUNW;
dotted lines for SLEV with $\alpha=0.9$.
}
\label{fig:real1}
\end{figure}

Coefficient estimates were obtained using three subsampling algorithms
for seven different subsample sizes: $2p, 3p, 4p, 5p, 10p, 20p, 50p$.
We compare the estimates using the sample bias and variances; and,
for each subsample size, we repeat our sampling $100$ times to get $100$
estimates.
(At each subsample size, we take one hundred subsamples and calculate all 
the estimates; we then calculate the bias of the estimates with respect to 
the full sample least squares estimate and their variance.) 
See Figure~\ref{fig:real1} for a summary of our results.
In the left panel of Figure~\ref{fig:real1}, we plot the histogram of the 
leverage score sampling probabilities.  
Observe that the distribution is quite uniform, suggesting that 
leverage-based sampling methods will perform similarly to uniform sampling.
To demonstrate this, the middle and right panels of Figure~\ref{fig:real1} 
present the (conditional) empirical variances and biases of each of the four 
estimates, for seven different subsample sizes.
Observe that LEV, LEVUNW, SLEV, and UNIF all have comparable sample 
variances.
When the subsample size is very small, all four methods have comparable 
sample bias; but when the subsample size is larger, then LEVUNW has a 
slightly larger bias than the other three estimates.

\subsubsection{Linear model for predicting gene expressions of cancer patient}

\begin{figure}[t]
  \centering
   \makebox{\includegraphics[scale=1]{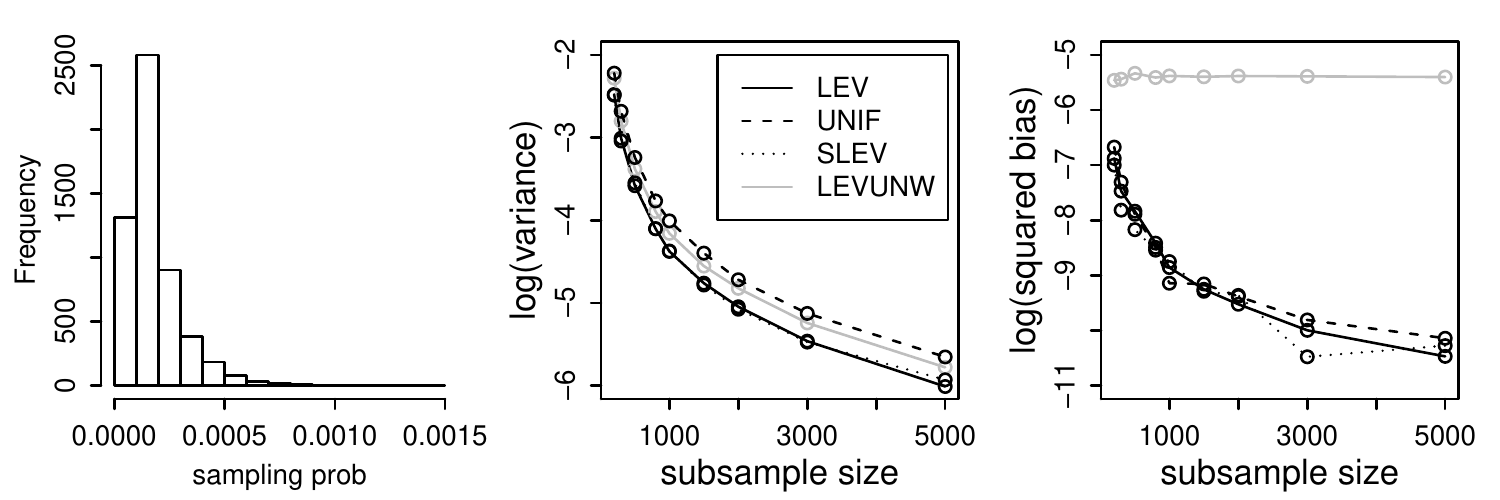}}
\vspace{-1em}
\caption{Empirical results for real data.
Left panel is the histogram of the leverage score sampling probabilities for 
the microarray data
(the largest leverage score is $0.00124$, and the mean is $0.00018$, i.e., 
the largest is $7$ times the mean);
Middle panel is the empirical \emph{conditional} variances
of the LEV, UNIF, LEVUNW, and SLEV estimates;
Right panel is the empirical \emph{conditional} biases.
Black lines for LEV;
dash lines for UNIF;
grey lines for LEVUNW;
dotted lines for SLEV with $\alpha=0.9$.
}
\label{fig:real2}
\end{figure}

In order to illustrate how our methods perform on real data with moderately
nonuniform leverage scores, we consider a microarray data set that was
presented in~\cite{LancetCancer02} (and also considered in~\cite{CUR_PNAS})
for $46$ cancer patients with respect to $n = 5,520$ genes.
Here, we randomly select one patient's gene expression as the response 
$\sbf{y}$ and use the remaining patients' gene expressions as the predictors 
(so $p=45$); and we predict the selected patient's gene expression using 
other patients gene expressions through a linear model.
We fit the linear model using subsampling algorithms  with nine different subsample sizes.
See Figure~\ref{fig:real2} for a summary of our results.
In the left panel of Figure~\ref{fig:real2}, we plot the histogram of the 
leverage score sampling probabilities.  
Observe that the distribution is highly skewed and quite a number of
probabilities are significantly larger than the average probability.
Thus, one might expect that leveraging estimates will have an advantage
over the uniform sampling estimate.
To demonstrate this, the middle and right panels of Figure~\ref{fig:real2} 
present the (conditional) empirical variances and biases of each of the four 
estimates, for nine different subsample sizes.
Observe that SLEV and LEV have smaller sample variance than 
LEVUNW and that UNIF consistently has the largest variance.
Interestingly, since LEVUNW is approximately unbiased to the weighted least 
squares estimate, here we observe that LEVUNW has by far the largest bias 
and that the bias does not decrease as the subsample size increases.  
In addition, when the subsample size is less than 2000, the biases of LEV, 
SLEV and UNIF are comparable; but when the subsample size is greater than 
2000, LEV and SLEV have slightly smaller bias than~UNIF.

\section{Discussion and Conclusion}
\label{sxn:conc}

Algorithmic leveraging---a recently-popular framework for solving large 
least-squares regression and other related matrix problems via sampling 
based on the empirical statistical leverage scores of the data---has been 
shown to have many desirable \emph{algorithmic} properties.
In this paper, we have adopted a \emph{statistical} perspective on 
algorithmic leveraging, and we have demonstrated how this leads to improved 
performance of this paradigm on real and synthetic data.
In particular, from the algorithmic perspective of worst-case analysis, 
leverage-based sampling provides uniformly superior worst-case algorithmic 
results, when compared with uniform sampling.
Our statistical analysis, however, reveals that, from the statistical 
perspective of bias and variance, neither leverage-based sampling nor 
uniform sampling dominates the other.
Based on this, we have developed new statistically-inspired leveraging 
algorithms that achieve improved statistical performance, while maintaining 
the algorithmic benefits of the usual leverage-based method.
Our empirical evaluation demonstrates that our theory is a good predictor of 
the practical performance of both existing as well as our newly-proposed 
leverage-based algorithms.
In addition, our empirical evaluation demonstrates that, by using a 
recently-developed algorithm to approximate the leverage scores, we can 
compute improved approximate solutions for much larger least-squares 
problems than we can compute the exact solutions with traditional 
deterministic~algorithms.

Finally, we should note that, while our results are straightforward and 
intuitive, obtaining them was not easy, in large part due to seemingly-minor 
differences between problem formulations in statistics, computer science, 
machine learning, and numerical linear algebra.
Now that we have ``bridged the gap'' by providing a statistical perspective 
on a recently-popular algorithmic framework, we expect that one can ask even 
more refined statistical questions of this and other related algorithmic 
frameworks for large-scale computation.

\appendix

\section{Asymptotic Analysis and Toy Data}
\label{sxn:interlude}

In this section, we will relate our analytic methods to the notion of 
asymptotic relative efficiency, and we will consider several toy data sets 
that illustrate various aspects of algorithmic leveraging.
Although the results of this section are not used elsewhere, and thus some
readers may prefer skip this section, we include it in order to relate our 
approach to ideas that may be more familiar to certain readers.

\subsection{Asymptotic Relative Efficiency Analysis}
\label{sxn:interlude-asymptotic}

Here, we present an asymptotic analysis comparing UNIF with LEV, SLEV, and 
LEVUNW in terms of their relative efficiency.
Recall that one natural way to compare two procedures is to compare the 
sample sizes at which the two procedures meet a given standard of 
performance.
One such standard is efficiency, which addresses how ``spread out'' about 
$\sbf{\beta}_0$ is the estimator. 
In this case, the smaller the variance, the more ``efficient'' is the 
estimator \cite{serfling10}.   
Since $\sbf{\beta}_0$ is a $p$-dimensional vector, to determine the relative 
efficiency of two estimators, we consider the linear combination of 
$\sbf{\beta}_0$, i.e., $c^{T}\sbf{\beta}_0$, where $c$ is the linear 
combination coefficient.
In somewhat more detail, when $\hat{\sbf{\beta}}$ and $\tilde{\sbf{\beta}}$ 
are two one-dimensional estimates, their relative efficiency can be defined~as
\begin{equation*}
\label{eqn:re}
    e(\hat{\sbf{\beta}}, \tilde{\sbf{\beta}})=\frac{\text{Var}(\tilde{\sbf{\beta}})}{\text{Var}(\hat{\sbf{\beta}})}  ,
\end{equation*}
and when $\hat{\sbf{\beta}}$ and $\tilde{\sbf{\beta}}$ are two 
$p$-dimensional estimates, we can take their linear combinations
$c^{T}\hat{\sbf{\beta}}$ and $c^{T}\tilde{\sbf{\beta}}$, where $c$ is the 
linear combination coefficient vector, and define their relative 
efficiency~as
\begin{equation*}
\label{eqn:re1}
    e(c^{T}\hat{\sbf{\beta}}, c^{T}\tilde{\sbf{\beta}})=\frac{\text{Var}(c^{T}\tilde{\sbf{\beta}})}{\text{Var}(c^{T}\hat{\sbf{\beta}})}  .
\end{equation*}

In order to discuss asymptotic relative efficiency, we start with the 
following seemingly-technical observation.

\begin{definition}
A $k \times k$ matrix A is said to be $A=O(\alpha_n)$ if and only if  every element of $A$ satisfies $A_{ij}=O(\alpha_n)$ for $i,j = 1,\ldots, k$.
\end{definition}

\begin{assumption}
\label{ass:ass1}
$X^{T}X=\sum_{i=1}^{n}\sbf{x}_i\sbf{x}_i^{T}$ is positive definite and
$ (X^{T}X)^{-1}=O(\alpha_n^{-1}) $.
\end{assumption}

\noindent
\textbf{Remark.}
Assuming $X^{T}X$ is nonsingular, for a LS estimator 
$\hat{\sbf{\beta}}_{ols}$ to converge to true value $\sbf{\beta}_0$ in 
probability, it is sufficient and necessary that 
$(X^{T}X)^{-1}  \rightarrow 0$ as 
$n \rightarrow \infty$~\cite{anderson:76,lai:78}.

\noindent
\textbf{Remark.}
Although we have stated this as an assumption, one typically assumes 
an $n$-dependence for $\alpha_n$~\cite{anderson:76}.
Since the form of the $n$-dependence is unspecified, we can alternatively 
view Assumption~\ref{ass:ass1} as a definition of $\alpha_n$.
The usual assumption that is made (typically for analytical convenience) is 
that $\alpha_n= n$~\cite{knight:00}.
We will provide examples of toy data for which $\alpha_n= n$, as well as
examples for which $\alpha_n \neq n$.
In light of our empirical results in 
Section~\ref{sxn:empirical} and the empirical observation that leverage
scores are often very nonuniform~\cite{CUR_PNAS,GM13_TR}, it is an 
interesting question to ask whether the common assumption that 
$\alpha_n= n$ is too restrictive, e.g., whether it excludes interesting 
matrices $X$ with very heterogeneous leveraging~scores.

Under Assumption~\ref{ass:ass1}, i.e., that $ (X^{T}X)^{-1} $ is 
asymptotically parameterized as $ (X^{T}X)^{-1}=O(\alpha_n^{-1}) $, we have 
the following three results to compare the leveraging estimators and the 
uniform sampling estimator.
The expressions in these three lemmas are complicated; and, since they 
are expressed in terms of $\alpha_n$, they are not easy to evaluate on 
real or synthetic data.
(It is partly for this reason that our empirical evaluation is in terms of 
the bias and variance of the subsampling estimators.)
We start by stating a lemma characterizing the relative efficiency of LEV 
and UNIF; the proof of this lemma may be found in 
Appendix~\ref{sxn:app-proofs-ratio-lev}.

\begin{lemma}
\label{variance-ratio}
To leading order,
the asymptotic relative efficiency of $c^{T}\tilde{\sbf{\beta}}_{LEV}$ and
$c^{T}\tilde{\sbf{\beta}}_{UNIF}$ is
\begin{equation}
\label{re-variance-compare}
e(c^{T}\tilde{\sbf{\beta}}_{LEV},c^{T}\tilde{\sbf{\beta}}_{UNIF})
    \simeq O( \frac{\frac{1}{\alpha_n}+ \frac{1}{{r}}\sqrt{\sum_{i}(1-h_{ii})^4 \max (h_{ii})} }{\frac{1}{\alpha_n}+ \frac{1}{{\alpha_n r}}\sqrt{\sum_{i}\frac{(1-h_{ii})^4}{h_{ii}^2} \max (h_{ii})} }),
\end{equation}
where the residual variance is ignored.
\end{lemma}

\noindent
Next, we state a lemma characterizing the relative efficiency of SLEV and
UNIF; the proof of this lemma is similar to that of 
Lemma~\ref{variance-ratio} and is thus omitted.

\begin{lemma}
\label{thm:slev-variance-ratio}
To leading order,
the asymptotic relative efficiency of $c^{T}\tilde{\sbf{\beta}}_{SLEV}$ and
$c^{T}\tilde{\sbf{\beta}}_{UNIF}$ is
\begin{equation*}
\label{eqn:slev-re-variance-compare}
e(c^{T}\tilde{\sbf{\beta}}_{SLEV},c^{T}\tilde{\sbf{\beta}}_{UNIF})
    \simeq O(\frac{\frac{1}{\alpha_n}+ \frac{1}{{r}}\sqrt{\sum_{i}(1-h_{ii})^4 \max (h_{ii})} }{\frac{1}{\alpha_n}+ \frac{1}{{\alpha_n r}}\sqrt{\sum_{i}\frac{(1-h_{ii})^4}{\pi_{i}^2} \max (h_{ii})} })  ,
\end{equation*}
where the residual variance is ignored.
\end{lemma}

\noindent
Finally, we state a lemma characterizing the relative efficiency of LEVUNW and
UNIF; the proof of this lemma may be found in 
Appendix~\ref{sxn:app-proofs-ratio-unweighted}.

\begin{lemma}
\label{levvariance-ratio}
To leading order,
the asymptotic relative efficiency of $c^{T}\tilde{\sbf{\beta}}_{LEVUNW}$ and $c^{T}\tilde{\sbf{\beta}}_{UNIF}$ is
\begin{equation*}
\label{levre-variance-compare}
e(c^{T}\tilde{\sbf{\beta}}_{LEVUNW},c^{T}\tilde{\sbf{\beta}}_{UNIF})
 \simeq O(\frac{\frac{1}{\alpha_n}+ \frac{1}{{r}}\sqrt{\sum_{i}(1-h_{ii})^4 \max (h_{ii})} }{\frac{\max (h_{ii})}{\alpha_n \min (h_{ii}) }+ \frac{1}{{\alpha_n \min (h_{ii}) r}}\sqrt{\sum_{i}(1-g_{ii})^4 \max (g_{ii})} }),
\end{equation*}
where the residual variance is ignored and $g_{ii}=h_{ii}\sbf{x}_i^{T} (X^{T}\Diagnl{h_{ii}}X)^{-1}\sbf{x}_i$.
\end{lemma}

\noindent
Of course, in an analogous manner, one could derive expressions for the 
asymptotic relative efficiencies
$e(c^{T}\tilde{\sbf{\beta}}_{SLEV},c^{T}\tilde{\sbf{\beta}}_{LEV})$,
$e(c^{T}\tilde{\sbf{\beta}}_{LEVUNW},c^{T}\tilde{\sbf{\beta}}_{LEV})$, 
and $e(c^{T}\tilde{\sbf{\beta}}_{LEVUNW},c^{T}\tilde{\sbf{\beta}}_{SLEV})$.

\subsection{Illustration of the Method on Toy Data}
\label{sxn:interlude-toy}

Here, we will consider several toy data sets that illustrate various
aspects of algorithmic leveraging, including various extreme cases of the
method.
While some of these toy data may seem artificial or contrived, they will
highlight properties that manifest themselves in less extreme forms in the
more realistic data in Section~\ref{sxn:empirical}.
Since the leverage score structure of the matrix $X$ is crucial for the
behavior of the method, we will focus primarily on that structure.
To do so, consider the two extreme cases.
At one extreme, when the leverage scores are all equal, i.e., $h_{ii}=p/n$,
for all $i\in [n]$, the first two variance terms in Eqn.~(\ref{uniform-Var1})
are equal to the first two variance terms in Eqn.~(\ref{lsq-wls-var-r-d}).
In this case, LEV simply reduces to UNIF.
At the other extreme, the leverage scores can be very nonuniform---e.g.,
there can be a small number of leverage scores that are much larger than
the rest and/or there can be some leverage scores that are much smaller than
the mean score.
Dealing with these two cases properly is crucial for the method of
algorithmic leveraging, but these two cases highlight important differences
between the more common algorithmic perspective and our more novel
statistical perspective.

The former problem (of a small number of very large leverage scores) is of
particular importance from an algorithmic perspective.
The reason is that in that case one wants to compare the output of the 
sampling algorithm with the optimum based on the empirical data (as opposed 
to the ``ground truth'' solution).
Thus, dealing with large leverage scores was a main issue in the development
of the leveraging paradigm~\cite{DMM06,Mah-mat-rev_BOOK,DMMW12_JMLR}.
On the other hand, the latter problem (of some very small leverage scores)
is also an important concern if we are interested in statistical properties
of algorithmic leveraging.
To see why, consider, e.g., the extreme case that a few data points have
very very small leverage scores, e.g. $h_{ii}=1/n^4$ for some $i$.
In this case, e.g., the second variance term in Eqn.~(\ref{lsq-wls-var-r-d}) 
will be much larger than the second variance term in 
Eqn.~(\ref{uniform-Var1}).

\begin{figure}[t]
  \centering
   \makebox{\includegraphics[scale=1]{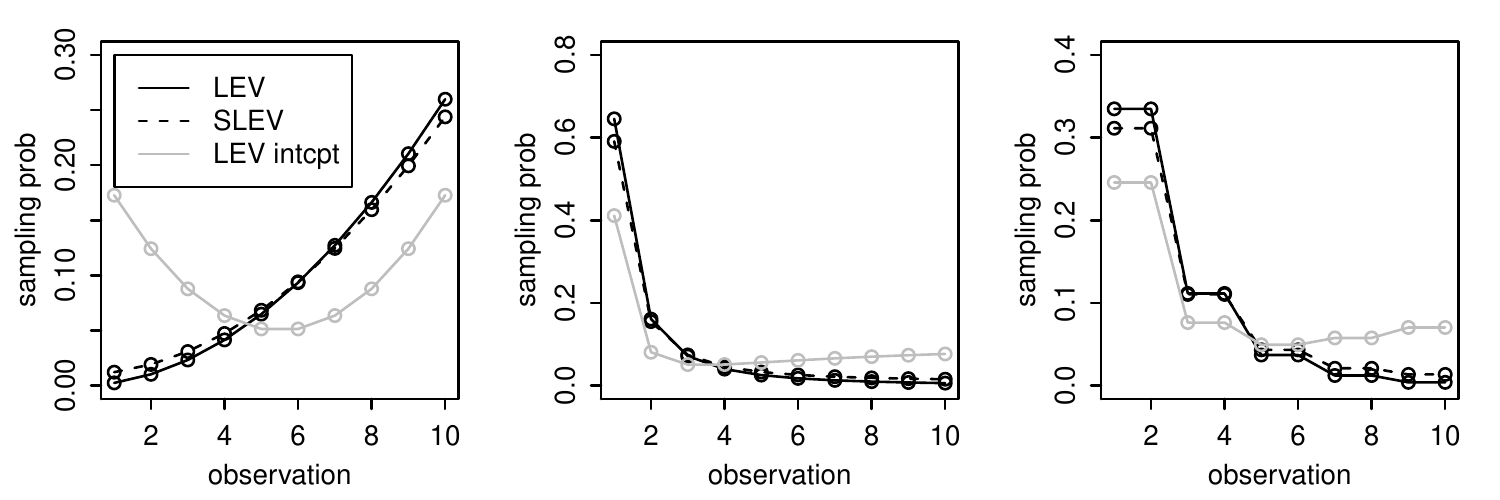}}
\vspace{-1em}
\caption{Leverage score-based sampling probabilities for three toy examples 
(Example 2, Example 3, and Example 4).
Left panel is Inflated Regression Line (Example 2);
Middle panel is In-fill Regression Line (Example 3);
Right panel is Regression Surface (Example 4).
In this example, we set $n=10$.
Black lines connect the sampling probability for each data points for LEV; 
dash lines (below black) connect sampling probability for SLEV; and 
grey lines connect sampling probability for LEV after we add an 
intercept (i.e., the sample mean) as a second column to $X$.  
}
\label{fig:theorem}
\end{figure}

In light of this discussion, here are several toy examples to consider.
We will start with several examples where $p=1$ that illustrate things in
the simplest setting.
\begin{itemize}
\item
\textbf{Example 1A: Sample Mean.}
Let $n$ be arbitrary, $p=1$, and let the $n \times p$ matrix $X$ be such
that $X_i = 1$, for all $i\in[n]$, i.e., let $X$ be the all-ones vector.
In this case,
$X^{T}X=n$ and $h_{ii}=1/n$, for all $i\in[n]$, i.e., the leverage scores
are uniform, and thus algorithmic leveraging reduces to uniform sampling.
Also, in this case, $\alpha_n=n$ in Assumption~\ref{ass:ass1}.  
All three asymptotic efficiencies are equal to $O(1)$.
\item
\textbf{Example 1B: Simple Linear Combination.}
Let $n$ be arbitrary, $p=1$, and let the $n \times p$ matrix $X$ be such
that $X_i = \pm1$, for all $i\in[n]$, either uniformly at random, or such
that $X_i = +1$ if $i$ is odd and $X_i=-1$ if $i$ is even.
In this case, $X^{T}X=n$ and $h_{ii}=1/n$, for all $i\in[n]$, i.e., the 
leverage scores are uniform; and,
in addition, $\alpha_n=n$ in Assumption~\ref{ass:ass1}.
For all four estimators, all four unconditional variances are equal to 
$\sigma^2\{\frac{1}{n} + \frac{(1-1/n)^2}{r}\}$.
In addition, for all four estimators, all three relative efficiencies are 
equal to $O(1)$.
\item
\textbf{Example 2: ``Inflated'' Regression Line Through Origin.}
Let $n$ be arbitrary, $p=1$, and let the $n \times p$ matrix $X$ be such
that $X_i = i$, i.e., they are evenly spaced and increase without
limit with increasing $i$. (We thus refer the $X$ as ``inflated.'')
In this case,
$$ X^{T}X=n(n+1)(2n+1)/6 ,$$ 
and the leverage scores equal
$$ h_{ii}=\frac{6i^2}{n(n+1)(2n+1)} ,$$
i.e., the leverage scores $h_{ii}$ are very nonuniform.
This is illustrated in the left panel of Figure~\ref{fig:theorem}.
Also, in this case, $\alpha_n=n^3$ in Assumption~\ref{ass:ass1}.
It is easy to see that the first variance components of UNIF, LEV, SLEV are 
the same, i.e., they equal
$$ (X^{T}X)^{-1}=\frac{6}{n(n+1)(2n+1)} .$$ 
It is also easy to see that variances of LEV, SLEV and UNIF are dominated 
by their second variance component. 
The leading terms of the second variance component of LEV and UNIF are the 
same, and we expect to see the similar performance based on their variance. 
The leading term of the second variance component of SLEV is smaller than 
that of LEV and UNIF; and thus SLEV has smaller variance than LEV and UNIF.
Simple calculation shows that LEVUNW has a smaller leading term for the 
second variance component than those of LEV, UNIF and SLEV.
\item
\textbf{Example 3: ``In-fill'' Regression Line Through Origin.}
Let $n$ be arbitrary, $p=1$, and let the $n \times p$ matrix $X$ be such
that $X_i = 1/i$.  
This is different than the evenly spaced data points in the ``inflated'' 
toy example since the unevenly spaced data points this this example get 
denser in the interval $(0,1]$.
The asymptotic properties of such design matrix are so-called ``in-fill'' 
asymptotics~\cite{cressie:91}.
In this case,
$$ X^{T}X=\pi^2/6-\psi ^{(1)}(n+1) ,$$
where $\psi^{(k)}$ is the $k^{th}$ derivative of digamma function, and the 
leverage scores equal
$$ h_{ii}=\frac{1}{i^2(\pi^2/6-\psi ^{(1)}(n+1)) } ,$$
i.e., the leverage scores $h_{ii}$ are very nonuniform.
This is illustrated in the middle panel of Figure~\ref{fig:theorem}.
Also, in this case, $\alpha_n=1$ in Assumption~\ref{ass:ass1}.
\end{itemize}

\begin{figure}[t]
  \centering
     \makebox{\includegraphics[scale=1]{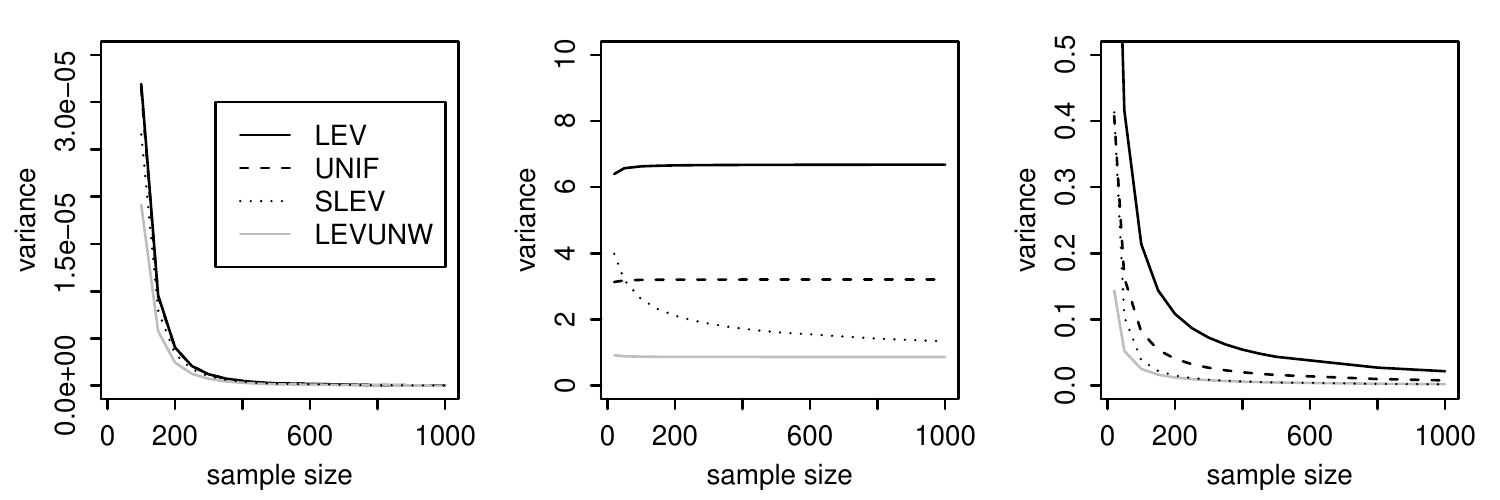}}
\vspace{-1em}
\caption{Theoretical variances for three toy examples (Example 2, Example 3, 
and Example 4) for various sample sizes $n$.
Left panel is Inflated Regression Line (Example 2);
Middle panel is In-fill Regression Line (Example 3);
Right panel is Regression Surface (Example 4).
In this example, we set $\sigma^2=1$ and $r=0.1n$, for varying $n$ from $100$ 
to $1000$.
Black line for LEV (Equation~\ref{lsq-wls-var-r-d}); 
dash line for UNIF (Equation~\ref{uniform-Var1});
dotted line (below black) for SLEV;
and grey line for LEVUNW (Equation~\ref{levnoweightvar11}).
}
\label{fig:theorem2}
\end{figure}

\noindent
To obtain an improved understanding of these examples, consider the first
two panels of Figures~\ref{fig:theorem} and~\ref{fig:theorem2}.
Figure~\ref{fig:theorem} shows the sampling probabilities for the 
Inflated Regression Line and the In-fill Regression Line.
Both the Inflated Regression Line and the In-fill Regression Line have very 
nonuniform leverage scores, and by construction there is a natural ordering 
such that the leverage scores increase or decrease respectively.
For the Inflated Regression Line, the minimum, mean, and maximum leverage
scores are $6/(n(n+1)(2n+1))$, $1/n$, and $6n/(n+1)(2n+1)$, respectively; and
for the In-fill Regression Line, the minimum, mean, and maximum leverage
scores are 
$1/(n^2(\pi^2/6-\psi ^{(1)}(n+1)))$, $1/n$, and $1/(\pi^2/6-\psi ^{(1)}(n+1))$,
respectively.
For reference, note that for the Sample Mean (as well as for the Simple
Linear Combination) all of the the leverage scores are equal to $1/n$, which
equals $0.1$ for the value of $n=10$ used in Figure~\ref{fig:theorem}.

Figure~\ref{fig:theorem2} illustrates the theoretical variances for the same
examples for particular values of $\sigma^2$ and $r$.
In particular, observe that for the Inflated Regression Line, all three
sampling methods tend to have smaller variance as $n$ is increased for a
fixed value of $p$.
This is intuitive, and it is a common phenomenon that we observe in most of 
the synthetic and real data sets.
The property of the In-fill Regression Line where the variances are roughly
flat (actually, they increase slightly) is more uncommon, but it illustrates
that other possibilities exist.
The reason is that leverage scores of most data points are relatively 
homogeneous (as long as $i$ is greater than $\sqrt{6n/\pi^2}$, the leverage 
score of $i$th observation is less than mean $1/n$ but greater than 
$1/n^2(\pi^2/6)$). 
When subsample size $r$ is reasonably large, we have high probabilities to 
sample these data points, whose sample probabilities inflate the variance.
These curves also illustrate that LEV and UNIF can be better or worse with 
respect to each other, depending on the problem parameters; and that SLEV and 
LEVUNW can be better than either, for certain parameter values.

From these examples, we can see that the variance for the leveraging
estimate can be inflated by very small leverage scores.
That is, since the variances involve terms that depend on the inverse
of $h_{ii}$, they can be large if $h_{ii}$ is very small.
Here, we note that the common practice of adding an intercept, i.e., a 
sample mean or all-ones vector \emph{tends} to uniformize the leverage 
scores.
That is, in statistical model building applications, we usually have
intercept---which is an all-ones vector, called the Sample Mean above---in
the model, i.e., the first column of $X$ is $\sbf{1}$ vector; and, in this
case, the $h_{ii}$s are bounded below by $1/n$ and above by 
$1/w_i$~\cite{weisberg:05}.
This is also illustrated in Figure~\ref{fig:theorem}, which shows the
the leverage scores for when an intercept is included.
Interestingly, for the Inflated Regression Line, the scores for elements that
originally had very small score actually increase to be on par with the
largest scores.
In our experience, it is much more common for the small leverage scores to 
simply be increased a bit, as is illustrated with the modified scores for 
the In-fill Regression Line.

We continue the toy examples with an example for $p=2$; this is the
simplest case that allows us to look at what is behind
Assumption~\ref{ass:ass1}.
\begin{itemize}
\item
\textbf{Example 4: Regression Surface Through Origin.}
Let $p=2$ and $n=2k$ be even.
Let the elements of $X$ be defined as
$\sbf{x}_{2j-1,n}=\left(
                    \begin{array}{cc}
                      \sqrt{\frac{n}{3^{j}}} & 0 \\
                    \end{array}
                  \right)
$,
and
$\sbf{x}_{2j,n}=\left(
                    \begin{array}{cc}
                     0 &  \sqrt{\frac{n}{3^{j}}}  \\
                    \end{array}
                  \right)
$.
In this case,
$$ X^{T}X=(n\sum_{j=1}^{n}\frac{1}{3^{j}})I_2=k\frac{3^k-1}{3^k}I_2 =O(n) ,$$ 
and the leverage scores equal 
$$ h_{2j-1,2j-1}=h_{2j,2j}=\frac{2\times 3^k}{3^j(3^k-1)} .$$
Here, $\alpha_n=n$ in Assumption~\ref{ass:ass1}, and the largest leverage 
score does \emph{not} converge to zero.
\end{itemize}

\noindent
To see the leverage scores and the (theoretically-determined) variance for 
the Regression Surface of Example 4, see the third panel of 
Figures~\ref{fig:theorem} and~\ref{fig:theorem2}.
In particular, the third panel of Figure~\ref{fig:theorem} demonstrates 
what we saw with the $p=1$ examples, i.e., that adding an intercept tends 
to increase the small leverage scores; and Figure~\ref{fig:theorem2} 
illustrates  that the variances of all four estimates are getting close as 
sample size $n$ becomes larger.

\noindent
\textbf{Remark.}
It is worth noting that \cite{miller:74} showed $\alpha_n=n$ in 
Assumption~\ref{ass:ass1} implies that $\max h_{ii} \rightarrow 0$. 
In his proof, Miller essentially assumed that $\sbf{x}_i$, $i=1, \ldots, n$ 
is a single sequence.
Example 4 shows that Miller's theorem does not hold for triangular array 
(with one pattern for even numbered observations and the other pattern for 
odd numbered observations) \cite{shao:thesis}.

Finally, we consider several toy data sets with larger values of $p$.
In this case, there starts to be a nontrivial interaction between the
singular value structure and the singular vector structure of the matrix $X$.
\begin{itemize}
\item
\textbf{Example 5: Truncated Hadamard Matrix.}
An $n \times p$ matrix consisting of $p$ columns from a Hadamard Matrix
(which is an orthogonal matrix) has uniform leverage scores---all are equal.
Similarly, for an $n \times p$ matrix with entries i.i.d. from Gaussian 
distribution---that is, unless the aspect ratio of the matrix is 
extremely rectangular, e.g., $p=1$, the leverage scores of a random 
Gaussian matrix are very close to uniform.
(In particular, as our empirical results demonstrate, using nonuniform 
sampling probabilities is not necessary for data generated from Gaussian 
random matrices.)
\item
\textbf{Example 6: Truncated Identity Matrix.}
An $n \times p$ matrix consisting of the first $p$ columns from an Identity
Matrix (which is an orthogonal matrix) has very nonuniform leverage
scores---the first $p$ are large, and the remainder are zero.
(Since one could presumably remove the all-zeros rows, this example might
seem trivial, but it is useful as a worst-case thought experiment.)
\item
\textbf{Example 7: Worst-case Matrix.}
An $n \times p$ matrix consisting of $n-1$ rows all pointing in the same
direction and $1$ row pointing in some other direction.
This has one leverage score---the one corresponding to the row pointing in
the other direction---that is large, and the rest are mediumly-small.
(This is an even better worst-case matrix than Example 6; and in the main 
text we have an even less trivial example of this.)
\end{itemize}

\noindent
Example 5 is ``nice'' from an algorithmic perspective and, as seen in 
Section~\ref{sxn:empirical}, from a statistical perspective as well.
Since they have nonuniform leverage scores; Example 6 and Example 7 are 
worse from an algorithmic perspective.
As our empirical results will demonstrate, they are also problematic from a 
statistical perspective, but for slightly different reasons.

\section{Appendix: Proofs of our main results}
\label{sxn:app-proofs}

In this section, we will provide proofs of several of our main results.

\subsection{Proof of Lemma~\ref{lem:taylor}}
\label{sxn:app-proofs-lev-taylor}

Recall that the matrix $W=S_X D^2S_X^{T}$ encodes information about the
sampling/rescaling process; in particular, this includes UNIF, LEV, and 
SLEV, although our results hold more generally.

By performing a Taylor expansion of $\tilde{\sbf{\beta}}_{W}(\sbf{w})$
around the point $\sbf{w}_0=\sbf{1}$, we have
\begin{equation*}
\label{lsq-wls-taylor-10}
\tilde{\sbf{\beta}}_{W}(\sbf{w})= \tilde{\sbf{\beta}}_{W}(\sbf{w}_0)+\frac{\partial \tilde{\sbf{\beta}}_{W}(\sbf{w})}{\partial \sbf{w}^{T}}|_{\sbf{w}=\sbf{w}_0}(\sbf{w}-\sbf{w}_0) +R_{W}  ,
\end{equation*}
where the second order remainder $R_{W}=o_p(||\sbf{w}-\sbf{w}_0||)$ when
$\sbf{w}$ is close to $\sbf{w}_0$.
By setting $\sbf{w}_0$ as the all-one vector, i.e., $\sbf{w}_0=\sbf{1}$,
$\tilde{\sbf{\beta}}_{W}(\sbf{w}_0)$
is expanded around the full sample ordinary
LS estimate $\hat{\sbf{\beta}}_{ols}$, i.e.,
$\tilde{\sbf{\beta}}_{W}(\sbf{1})= \hat{\sbf{\beta}}_{ols}$.
That is,
\begin{align*}
\tilde{\sbf{\beta}}_{W}(\sbf{w})
 = \hat{\sbf{\beta}}_{ols}+\frac{\partial (X^{T}\Diagnl{\sbf{w}}X)^{-1}X^{T}\Diagnl{\sbf{w}}\sbf{y}}{\partial \sbf{w}^{T}}|_{\sbf{w}=\sbf{1}}(\sbf{w}-\sbf{1})+ R_{W}.
\end{align*}
By differentiation by parts, we obtain
\begin{align}
\frac{\partial ( X^{T}\Diagnl{\sbf{w}}X )^{-1}X^{T}\Diagnl{\sbf{w}}\sbf{y}}{\partial \sbf{w}^{T}}&=\frac{\partial \text{Vec}[(X^{T}\Diagnl{\sbf{w}}X)^{-1}X^{T}\Diagnl{\sbf{w}}\sbf{y}]}{\partial \sbf{w}^{T}}\notag\\
&=(\sbf{1}\otimes (X^{T}\Diagnl{\sbf{w}}X)^{-1})\frac{\partial \text{Vec}[X^{T}\Diagnl{\sbf{w}}\sbf{y}]}{\partial \sbf{w}^{T}}\label{lsq-wls-taylor-2a1}\\
&+(\sbf{y}^{T}\Diagnl{\sbf{w}}X\otimes I_{p})\frac{\partial \text{Vec}[(X^{T}\Diagnl{\sbf{w}}X)^{-1}]}{\partial \sbf{w}^{T}}\label{lsq-wls-taylor-2b1}
\end{align}
where $\text{Vec}$ is Vec operator, which stacks the columns of a matrix 
into a vector, and $\otimes$ is the Kronecker product.
The Kronecker product is defined as follows:
suppose $A=\{a_{ij}\}$ is an $m \times n$ matrix and $B=\{b_{ij}\}$ is a 
$p \times q$ matrix; then, $A \otimes B$ is a $mp \times nq$ matrix, 
comprising $m$ rows and $n$ columns of $p \times q$ blocks, the $ij$th of 
which is $a_{ij}B$.

To simplify (\ref{lsq-wls-taylor-2a1}), note that is easy to show that 
(\ref{lsq-wls-taylor-2a1}) can be seen as
\begin{equation}\label{lsq-wls-taylor-2aa}
(\sbf{1}\otimes (X^{T}\Diagnl{\sbf{w}}X)^{-1})(\sbf{y}^{T}\otimes X^{T}) \frac{\partial \text{Vec}[\Diagnl{\sbf{w}}]}{\partial \sbf{w}^{T}}.
\end{equation}
To simplify (\ref{lsq-wls-taylor-2b1}), we need the following two results of 
matrix differentiation,
\begin{align}\label{harville}
\frac{\partial \text{Vec}[X^{-1}]}{\partial (\text{Vec}X)^{T}}& = -(X^{-1})^{T}\otimes X^{-1}, \mbox{ and}\notag\\
\frac{\partial \text{Vec}[AWB]}{\partial \sbf{w}^{T}} &= (B^{T}\otimes A)\frac{\partial \text{Vec}[W]}{\partial \sbf{w}^{T}},
\end{align}
where the details on these two results can be found on page 366-367 
of~\cite{harville:97}.
By combining the two results in (\ref{harville}), by the chain rule, we have
\begin{eqnarray*}
& & \hspace{-20mm}
\frac{\partial \text{Vec}[(X^{T}\Diagnl{\sbf{w}}X)^{-1}]}{\partial \sbf{w}^{T}} \\
   &=& \frac{\partial \text{Vec}[(X^{T}\Diagnl{\sbf{w}}X)^{-1}]}{\partial \text{Vec}[(X^{T}\Diagnl{\sbf{w}}X)]^{T}}\frac{\partial \text{Vec}[(X^{T}\Diagnl{\sbf{w}}X)]}{\partial \sbf{w}^{T}}\\
   &=& -(X^{T}\Diagnl{\sbf{w}}X)^{-1} \otimes (X^{T}\Diagnl{\sbf{w}}X)^{-1}(X^{T}\otimes X^{T})\frac{\partial \text{Vec}[\Diagnl{\sbf{w}}]}{\partial \sbf{w}^{T}}
\end{eqnarray*}
By simple but tedious algebra, (\ref{lsq-wls-taylor-2a1}) and 
(\ref{lsq-wls-taylor-2b1}) give rise to
\begin{multline}\label{lsq-wls-taylor-4}
\{(\sbf{y}^{T}-\sbf{y}^{T}\Diagnl{\sbf{w}}X(X^{T}\Diagnl{\sbf{w}}X)^{-1}X^{T})\otimes (X^{T}\Diagnl{\sbf{w}}X)^{-1}X^{T}\}\frac{\partial \text{Vec}[\Diagnl{\sbf{w}}]}{\partial \sbf{w}^{T}}\\
=\{(\sbf{y}-X\tilde{\sbf{\beta}}_{W}(\sbf{w}))^{T}\otimes (X^{T}\Diagnl{\sbf{w}}X)^{-1}X^{T}\}\frac{\partial \text{Vec}[\Diagnl{\sbf{w}}]}{\partial\sbf{w}^{T}}
\end{multline}
By combining these results, we thus have,
\begin{align*}
\tilde{\sbf{\beta}}_{W}& = \hat{\sbf{\beta}}_{ols}+ \{(\sbf{y}-X\hat{\sbf{\beta}}_{ols})^{T}\otimes (X^{T}X)^{-1}X^{T} \}\frac{\partial \text{Vec}(\Diagnl{\sbf{w}} )}{\partial \sbf{w}^{T}}(\sbf{w}-\sbf{1}) +R_{W}\\
                   &=\hat{\sbf{\beta}}_{ols}+ \{ \hat{\sbf{e}}^{T}\otimes (X^{T}X)^{-1}X^{T}\}\begin{pmatrix} \mbf{e}_1\mbf{e}_1^{T}\\ \mbf{e}_2\mbf{e}_2^{T}\\ \\ \\ \mbf{e}_n\mbf{e}_n^{T}\end{pmatrix}(\sbf{w}-\sbf{1}) +R_{W}\\
                   &=\hat{\sbf{\beta}}_{ols}+ (X^{T}X)^{-1}X^{T}\Diagnl{\hat{\sbf{e}}}(\sbf{w}-\sbf{1}) +R_{W}
\end{align*}
where $\hat{\sbf{e}}=\sbf{y}-X\hat{\sbf{\beta}}_{ols}$ is the LS residual 
vector, $\mbf{e}_i$ is a length $n$ vector with $i^{th}$ element equal to 
one and all other elements equal to zero, from which the lemma follows.

\subsection{Proof of Lemma~\ref{lem:bv}}
\label{sxn:app-proofs-lev-bv}

Recall that we will use $W$ to refer to the sampling process.

We start by establishing the conditional result.
Since $\Expect{\sbf{w}}=\sbf{1}$, 
it is straightforward to calculate conditional expectation
of $\tilde{\sbf{\beta}}_{W}$.
Then, it is easy to see that
\begin{align*}
\Expect{(w_i-1)(w_j-1)} 
   &=\frac{1}{r \pi_i}-\frac{1}{r} \quad \text{for}\quad i=j\\
   &=-\frac{1}{r}  \quad\text{for}\quad i \neq j  .
\end{align*}
We rewrite it in matrix form,
\begin{equation*}
\label{lsq-wls-var-p-2}
\Varnce{\sbf{w}}=\Expect{(\sbf{w}-\sbf{1})(\sbf{w}-\sbf{1})^T}=\Diagnl{\frac{1}{r\sbf{\pi}}}-\frac{1}{r}J_n  ,
\end{equation*}
where $\sbf{\pi}=(\pi_1, \pi_2, \ldots, \pi_n)^{T}$ and $J_n$ is a 
$n \times n $ matrix of ones.
Some additional algebra yields that the variance of 
$\tilde{\sbf{\beta}}_{W}$ is
\begin{align*}
\CVarnce{\tilde{\sbf{\beta}}_{W} -\hat{\sbf{\beta}}|\sbf{y} } &= \Varnce{ (X^{T}X)^{-1}X^{T}\Diagnl{\hat{\sbf{e}}} (\sbf{w}-\sbf{1})|\sbf{y}} + \CVarnce{R_{W}}\\
 &=(X^{T}X)^{-1}X^{T}\Diagnl{\hat{\sbf{e}}}(\Diagnl{\frac{1}{r\sbf{\pi}}}-\frac{1}{r}J_n) \Diagnl{\hat{\sbf{e}}} X(X^{T}X)^{-1}  + \CVarnce{R_{W}}\\
 &=(X^{T}X)^{-1}X^{T}[\Diagnl{\hat{\sbf{e}}}\Diagnl{\frac{1}{r\sbf{\pi}}} \Diagnl{\hat{\sbf{e}}}]X(X^{T}X)^{-1}
  + \Varnce{R_{W}}\\
 &=(X^{T}X)^{-1}X^{T}\Diagnl{\frac{1}{r\sbf{\pi}}\hat{\sbf{e}}^2}X(X^{T}X)^{-1}
  +  \CVarnce{R_{W}}.
\end{align*}
Setting $\pi_i=h_{ii}/p$ in above equations, we thus prove the conditional 
result.

We next establish the unconditional result as follows.
If we take one more expectation of the expectation result in 
Lemma \ref{lem:lev-bv} with respect to the response, then
we have the expectation result.
By rule of double expectations, we have the variance of
$\tilde{\sbf{\beta}}_{W}$ result, from which the lemma follows.

\subsection{Proof of Lemma~\ref{lem:unwl-taylor}}
\label{sxn:app-proofs-unwl-taylor}

First note that the unweighted leveraging estimate 
$\tilde{\sbf{\beta}}_{LEVUNW}$ can be written as
\begin{equation*}
\label{wlsq-sample-expr}
\tilde{\sbf{\beta}}_{LEVUNW}= (X^{T}S_XS_X^{T}X)^{-1}X^{T}S_XS_X^{T}\sbf{y}= (X^{T}W_{LEVUNW}X)^{-1}X^{T}W_{LEVUNW}\sbf{y}  ,
\end{equation*}
where $W_{LEVUNW}=S_XS_X^{T}=\Diagnl{\sbf{w}_{LEVUNW}}$, and where 
$\sbf{w}_{LEVUNW}$ has a multinomial distribution $Multi(r, \sbf{\pi})$.
The proof of this lemma is analogous to the proof of Lemma~\ref{lem:taylor};
and so here we provide only some details on the differences.
By employing a Taylor expansion, we have
\begin{equation*}
\label{lsq-wls-taylor-11}
\tilde{\sbf{\beta}}_{LEVUNW}(\sbf{w}_{LEVUNW})= \tilde{\sbf{\beta}}_{LEVUNW}(\sbf{w}_0)+\frac{\partial \tilde{\sbf{\beta}}_{LEVUNW}(\sbf{w})}{\partial \sbf{w}^{T}}|_{\sbf{w}=\sbf{w}_0}(\sbf{w}_{LEVUNW}-\sbf{w}_0) +R_{LEVUNW}  ,
\end{equation*}
where $R_{LEVUNW}=o_p(||\sbf{w}_{LEVUNW}-\sbf{w}_0||)$.
Following the proof of the previous lemma, we have~that
\begin{align*}
\label{ulsq-wls-taylor-1-simple}
\tilde{\sbf{\beta}}_{LEVUNW}
   & = \hat{\sbf{\beta}}_{wls} + \{(\sbf{y}-X\hat{\sbf{\beta}}_{wls})^{T}\otimes (X^{T}W_0X)^{-1}X^{T} \}\frac{\partial \text{vec}(\Diagnl{\sbf{w}_{LEVUNW}} )}{\partial \sbf{w}_{LEVUNW}^{T}}(\sbf{w}_{LEVUNW}-\sbf{w}_0) \\
   &+R_{LEVUNW}\\
   &=\hat{\sbf{\beta}}_{wls}+ \{ \hat{\sbf{e}}_{w}^{T}\otimes (X^{T}W_0X)^{-1}X^{T}\}\begin{pmatrix} \mbf{e}_1\mbf{e}_1^{T}\\ \mbf{e}_2\mbf{e}_2^{T}\\ \\ \\ \mbf{e}_n\mbf{e}_n^{T}\end{pmatrix}(\sbf{w}_{LEVUNW}-\sbf{w}_0) +R_{LEVUNW}\\
   &=\hat{\sbf{\beta}}_{wls}+ (X^{T}W_0X)^{-1}X^{T}\Diagnl{\hat{\sbf{e}}_{w_0}}(\sbf{w}_{LEVUNW}-\sbf{w}_0) +R_{LEVUNW}  , 
\end{align*}
where $W_0=\Diagnl{\sbf{w}_0}=\Diagnl{r\sbf{\pi}}$, $\hat{\sbf{\beta}}_{wls}=(X^{T}W_0X)^{-1}X^{T}W_0\sbf{y}$,
$\hat{\sbf{e}}_{w}=\sbf{y}-X\hat{\sbf{\beta}}_{wls}$ is the weighted LS residual vector, $\mbf{e}_i$ is a length $n$ vector with $i^{th}$ element equal to one and all other elements equal to zero.
From this the lemma follows.

\subsection{Proof of Lemma~\ref{lem:unwl-bv}}
\label{sxn:app-proofs-unwl-bv}

By taking the conditional expectation of Taylor expansion of the LEVUNW estimate
$\tilde{\sbf{\beta}}_{LEVUNW}$ in Lemma~\ref{lem:unwl-taylor}, we have that  
\begin{align*}
\CExpect{\tilde{\sbf{\beta}}_{LEVUNW}|\sbf{y}}   = \hat{\sbf{\beta}}_{wls} + (X^{T}W_0 X)^{-1}X^{T}\Diagnl{\hat{\sbf{e}}_w} \CExpect{\sbf{w}-r\sbf{\pi}} +\CExpect{R_{LEVUNW}}  .
\end{align*}
Since $\CExpect{\sbf{w}_{LEVUNW}}=r\sbf{\pi}$, the conditional expectation is 
thus obtained.
Since $\sbf{w}_{LEVUNW}$ is multinomial distributed, we have
\begin{equation*}
\label{wl-lsq-wls-var-p-2}
\Varnce{\sbf{w}_{LEVUNW}}=\Expect{(\sbf{w}_{LEVUNW}-r\sbf{\pi})(\sbf{w}_{LEVUNW}-r\sbf{\pi})^T}=\Diagnl{r\sbf{\pi}}-{r}\sbf{\pi}\sbf{\pi}^{T}  .
\end{equation*}
Some algebra yields that the conditional variance of $\tilde{\sbf{\beta}}_{LEVUNW}$ is
\begin{eqnarray*}
& & \hspace{-20mm}
\CVarnce{ \tilde{\sbf{\beta}}_{LEVUNW} -\hat{\sbf{\beta}}_{wls}|\sbf{y} } \\
   &=& \CVarnce{ (X^{T}W_0 X)^{-1}X^{T}\Diagnl{\hat{\sbf{e}}_w}(\sbf{w}_{LEVUNW}-r\sbf{\pi})|\sbf{y} } +\CVarnce{R_{LEVUNW}}\\
   &=& (X^{T}W_0X)^{-1}X^{T}\Diagnl{\hat{\sbf{e}}_w}W_0\Diagnl{\hat{\sbf{e}}_w} X(X^{T}W_0 X)^{-1}
  +  \CVarnce{R_{LEVUNW}}.
\end{eqnarray*}
Finally, note that
$$
\Expect{\hat{\sbf{\beta}}_{wls}}= (X^{T}W_0X)^{-1}XW_0\Expect{\sbf{y}}= (X^{T}W_0X)^{-1}XW_0X\sbf{\beta}_{0}= \sbf{\beta}_{0}.
$$
From this the lemma follows.

\subsection{Proof of Lemma~\ref{variance-ratio}}
\label{sxn:app-proofs-ratio-lev}

Since 
$\text{Var}(c^{T}\tilde{\sbf{\beta}}_{LEV})
   =c^{T}\text{Var}(\tilde{\sbf{\beta}}_{LEV})c$, 
we shall the derive the asymptotic order of 
$\text{Var}(\tilde{\sbf{\beta}}_{LEV})$.
The second variance component of 
$\tilde{\sbf{\beta}}_{LEV}$ in (\ref{lsq-wls-var-r-d}) is seen to be
\begin{eqnarray*}
& & \hspace{-20mm}
\frac{p\sigma^2}{r}(X^{T}X)^{-1}X^{T}\Diagnl{\frac{(1-h_{ii})^2}{h_{ii}}}X(X^{T}X)^{-1}  \\
   &=&\frac{p\sigma^2}{r}\sum_{i}\frac{(1-h_{ii})^2}{h_{ii}}(X^{T}X)^{-1}\sbf{x}_i\sbf{x}_i^{T} (X^{T}X)^{-1}\\
   &\le& \frac{p\sigma^2}{r} \sqrt{\sum_{i}\frac{(1-h_{ii})^4}{h_{ii}^2}\sum_{i}((X^{T}X)^{-1}\sbf{x}_i\sbf{x}_i^{T} (X^{T}X)^{-1})^2},
\end{eqnarray*}
where Cauchy-Schwartz inequality has been used.
Next, we show that 
$$\sum_{i}((X^{T}X)^{-1}\sbf{x}_i\sbf{x}_i^{T} (X^{T}X)^{-1})^2
=O( \max (h_{ii}) \alpha_n^{-2}) .$$
To see this, observe that
\begin{align*}
\sum_{i}((X^{T}X)^{-1}\sbf{x}_i\sbf{x}_i^{T} (X^{T}X)^{-1})^2
&\le \max((X^{T}X)^{-1}\sbf{x}_i\sbf{x}_i^{T} )  \sum_{i}(X^{T}X)^{-2}\sbf{x}_i\sbf{x}_i^{T} (X^{T}X)^{-1}\\
&\le \max (\sbf{x}_i^{T} (X^{T}X)^{-1}\sbf{x}_i) \sum_{i}(X^{T}X)^{-2}\sbf{x}_i\sbf{x}_i^{T} (X^{T}X)^{-1}\\
&= \max (\sbf{x}_i^{T} (X^{T}X)^{-1}\sbf{x}_i) (X^{T}X)^{-2}\\
&= O( \max (h_{ii}) \alpha_n^{-2})
\end{align*}
Thus, the second variance component of $\tilde{\sbf{\beta}}_{LEV}$ in 
(\ref{lsq-wls-var-r-d}) is of the order of 
$$O(\frac{1}{{\alpha_n r}}\sqrt{\sum_{i}\frac{(1-h_{ii})^4}{h_{ii}^2} \max (h_{ii})}).$$
Analogously, the second variance component of $\tilde{\sbf{\beta}}_{UNIF}$ in
(\ref{uniform-Var1}) is of the order of
$$O(\frac{1}{{r}}\sqrt{\sum_{i}(1-h_{ii})^4 \max (h_{ii})}).$$
The lemma then follows immediately.

\subsection{Proof of Lemma~\ref{levvariance-ratio}}
\label{sxn:app-proofs-ratio-unweighted}

It is easy to see that
$(X^{T}\Diagnl{h_{ii}} X)^{-1}= O(1/(\min (h_{ii}) \alpha_n)) $.
The second variance component of $\tilde{\sbf{\beta}}_{LEVUNW}$ in 
(\ref{levnoweightvar11}) is seen to be
\begin{eqnarray*}
& & \hspace{-20mm}
\frac{p\sigma^2}{r}(X^{T}\Diagnl{h_{ii}}X)^{-1}X^{T}\Diagnl{(1-g_{ii})^2h_{ii}}X(X^{T}\Diagnl{h_{ii}}X)^{-1}\\
   &=& \frac{p\sigma^2}{r}\sum_{i}(1-g_{ii})^2 h_{ii}(X^{T}\Diagnl{h_{ii}}X)^{-1}\sbf{x}_i\sbf{x}_i^{T} (X^{T}\Diagnl{h_{ii}}X)^{-1}\\
   &\le& \frac{p\sigma^2}{r} \sqrt{\sum_{i} (1-g_{ii})^4 \sum_{i}(h_{ii}(X^{T}\Diagnl{h_{ii}}X)^{-1}\sbf{x}_i\sbf{x}_i^{T} (X^{T}\Diagnl{h_{ii}}X)^{-1})^2}  ,
\end{eqnarray*}
where Cauchy-Schwartz inequality has used.
Next, we show that
$$\sum_{i}(h_{ii}(X^{T}\Diagnl{h_{ii}}X)^{-1}\sbf{x}_i\sbf{x}_i^{T} (X^{T}\Diagnl{h_{ii}}X)^{-1})^2=O( \max (g_{ii}) (\min(h_{ii}) \alpha_n)^{-2}).$$
To see this, observe that
\begin{eqnarray*}
& & \hspace{-10mm}
\sum_{i}(h_{ii}(X^{T}\Diagnl{h_{ii}}X)^{-1}\sbf{x}_i\sbf{x}_i^{T} (X^{T}\Diagnl{h_{ii}}X)^{-1})^2  \\
   &\le& \max(h_{ii}(X^{T}\Diagnl{h_{ii}}X)^{-1}\sbf{x}_i\sbf{x}_i^{T})  \sum_{i}h_{ii}(X^{T}\Diagnl{h_{ii}}X)^{-2}\sbf{x}_i\sbf{x}_i^{T} (X^{T}\Diagnl{h_{ii}}X)^{-1}\\
   &\le& \max (h_{ii}\sbf{x}_i^{T} (X^{T}\Diagnl{h_{ii}}X)^{-1}\sbf{x}_i) \sum_{i}h_{ii}(X^{T}\Diagnl{h_{ii}}X)^{-2}\sbf{x}_i\sbf{x}_i^{T} (X^{T}\Diagnl{h_{ii}}X)^{-1}\\
   &=& \max (h_{ii}\sbf{x}_i^{T} (X^{T}\Diagnl{h_{ii}}X)^{-1}\sbf{x}_i) (X^{T}\Diagnl{h_{ii}}X)^{-2}
= O( \max (g_{ii}) (\min(h_{ii}) \alpha_n)^{-2}  ).
\end{eqnarray*}
Thus, the second variance component of $\tilde{\sbf{\beta}}_{LEVUNW}$ in 
(\ref{levnoweightvar11}) is of the order of 
$$O(\frac{1}{{\alpha_n \min(h_{ii}) r }}\sqrt{\sum_{i}(1-g_{ii})^4 \max (g_{ii})})  . $$
The lemma then follows immediately.

\vspace{0.2in}
\noindent
\textbf{Acknowledgments.}
This research was partially supported by a grant from the U.S. National Science Foundation.

\end{document}